\documentclass[12pt,aps,preprintnumbers,superscriptaddress,nofootinbib,prd,onecolumn,floatfix]{revtex4-2}
\pdfoutput=1

\usepackage{comment}
\usepackage{multirow,array}
\usepackage{amsmath,amssymb}
\usepackage{graphicx}
\usepackage{color}
\usepackage{url}
\usepackage{slashed}
\usepackage{multirow}
\usepackage{footmisc}
\usepackage{hyperref}
\usepackage{braket}
\usepackage{colortbl}
\usepackage{orcidlink}
\usepackage{physics}
\usepackage{mathrsfs}
\usepackage{feynmp-auto,expdlist}
\usepackage[compat=1.1.0]{tikz-feynman}
\usepackage{booktabs}
\usepackage{makecell}
\usepackage{tablefootnote}
\usepackage[flushleft]{threeparttable}
\usepackage[normalem]{ulem}
\usepackage[T1]{fontenc}

\newcommand{\bea}{\begin{eqnarray}}
\newcommand{\eea}{\end{eqnarray}}
\newcommand{\be}{\begin{equation}}
\newcommand{\ee}{\end{equation}}
\newcommand{\bma}{\begin{pmatrix}}
\newcommand{\ema}{\end{pmatrix}}

\definecolor{nicered}{rgb}{0.5,0.,0.}
\definecolor{nicegreen}{rgb}{0.,0.5,0.}
\definecolor{niceblue}{rgb}{0.,0.,0.5}
\hypersetup{colorlinks,citecolor=nicegreen,linkcolor=nicered,urlcolor=niceblue}

\begin{document}

\title{Probing \boldmath{$\tau$} lepton dipole moments at FCC-ee}

\author{Dario Buttazzo\orcidlink{0000-0002-4678-7975}}
\affiliation{Istituto Nazionale di Fisica Nucleare, Sezione di Pisa, I-56127 Pisa, Italy}
\author{Gabriele Levati\orcidlink{0009-0000-7693-2152}}
\affiliation{Albert Einstein Center for Fundamental Physics, Institute for
Theoretical Physics, University of Bern, Sidlerstrasse 5, 3012 Bern,
Switzerland}
\author{Yang Ma\orcidlink{0000-0002-9419-6598}}
\affiliation{Centre for Cosmology, Particle Physics and Phenomenology (CP3),
UCLouvain, 1348 Louvain-la-Neuve, Belgium}
\author{Fabio Maltoni\orcidlink{0000-0003-4890-0676}}
\affiliation{Centre for Cosmology, Particle Physics and Phenomenology (CP3),
UCLouvain, 1348 Louvain-la-Neuve, Belgium}
\affiliation{Dipartimento di Fisica e Astronomia, Università di Bologna and INFN,
Sezione di Bologna, via Irnerio 46, 40126 Bologna, Italy}
\author{Paride Paradisi\orcidlink{0000-0003-0556-7797}}
\affiliation{Dipartimento di Fisica e Astronomia `G. Galilei', Universit\`a di Padova, Italy}
\affiliation{Istituto Nazionale di Fisica Nucleare, Sezione di Padova, I--35131 Padova, Italy}
\author{ZeQiang Wang\orcidlink{0000-0002-1643-2727}}
\affiliation{Centre for Cosmology, Particle Physics and Phenomenology (CP3),
UCLouvain, 1348 Louvain-la-Neuve, Belgium}

\preprint{IRMP-CP3-26-08,
COMETA-2026-07}

\newpage

\begin{abstract}
We study the sensitivity of future high-luminosity $e^+e^-$ colliders to the electromagnetic and weak dipole moments of the $\tau$ lepton within the Standard Model Effective Field Theory. We analyse several complementary processes across the FCC-ee operating stages and show that the exceptional luminosity and clean environment of future $e^+e^-$ colliders allow for substantial improvements over current limits from LEP and the LHC. The $e^+e^- \to \tau^+\tau^-$ process provides excellent sensitivity to weak dipole operators through resonant $Z$ production and $Z$-pole energy scans, while $\gamma\gamma \to \tau^+\tau^-$ yields the strongest constraint on the anomalous magnetic moment owing to the dominance of linear interference effects in the near-threshold region. Higher-energy stages improve the sensitivity to the $\tau$ electric dipole moment. The radiative processes $e^+e^- \to \tau^+\tau^-\gamma$ and $H\to\tau^+\tau^-\gamma$ provide complementary, though subleading, probes of $\tau$ dipole operators. Combining all channels, we estimate the statistical reach at the 95\% CL to be $|\Delta a_\tau|<2.8\times10^{-5}$, $|d_\tau|<2.5\times10^{-18}\,e\,{\rm cm}$, $|\Delta a_\tau^Z|<1.0\times10^{-5}$, and $|d_\tau^Z|<4.4\times10^{-19}\,e\,{\rm cm}$, improving upon the current bounds by several orders of magnitude. These projections quantify the potential statistical reach of the FCC-ee program, motivating dedicated studies incorporating detector effects, systematic uncertainties, and higher-order theoretical predictions.
\end{abstract}

\maketitle
\newpage
\tableofcontents
\section{Introduction}

Lepton dipole moments provide some of the most stringent tests of the Standard Model (SM) and constitute powerful indirect probes of physics beyond the SM (BSM). The anomalous magnetic moments $a_\ell$ and electric dipole moments (EDM) $d_\ell$ of the electron and muon have been measured with extraordinary precision, reaching the level of $0.13$ parts per trillion for $a_e$~\cite{Fan:2022eto} and $124$ parts per billion for $a_\mu$~\cite{Muong-2:2023cdq,Muong-2:2025xyk}, while the corresponding bounds on the electric dipole moments constrain $|d_e|<4.1\times10^{-30}\,e\,{\rm cm}$~\cite{Roussy:2022cmp} and $|d_\mu|<2\times10^{-19}\,e\,{\rm cm}$~\cite{Muong-2:2008ebm}. 

In the SM, lepton EDMs are many orders of magnitude below present experimental sensitivities, so that any non-vanishing signal would unambiguously imply new sources of CP violation. The situation is different for anomalous magnetic moments, where meaningful NP searches require a simultaneous improvement of both experimental measurements and theoretical predictions. In the electron case, the dominant limitation arises from the determination of the fine-structure constant through atom-interferometry measurements in Rb and Cs~\cite{Morel:2020dww,Parker:2018vye}, while theory uncertainties are already smaller than the direct experimental uncertainty~\cite{Aoyama:2019ryr,Volkov:2019phy,Volkov:2024yzc,Aoyama:2024aly,DiLuzio:2024sps,Hoferichter:2025fea}. For the muon case, recent progress in lattice-QCD determinations of the hadronic vacuum polarization contribution has substantially reshaped the SM prediction~\cite{Aliberti:2025beg}. The latest White Paper result, $a_\mu^{\rm SM}=116\,592\,033(62)\times10^{-11}$ ~\cite{Aliberti:2025beg,Aoyama:2012wk,Volkov:2019phy,Volkov:2024yzc,Aoyama:2024aly,Parker:2018vye,Morel:2020dww,Fan:2022eto,Czarnecki:2002nt,Gnendiger:2013pva,Ludtke:2024ase,Hoferichter:2025yih,RBC:2018dos,Giusti:2019xct,Borsanyi:2020mff,Lehner:2020crt,Wang:2022lkq,Aubin:2022hgm,Ce:2022kxy,ExtendedTwistedMass:2022jpw,RBC:2023pvn,Kuberski:2024bcj,Boccaletti:2024guq,Spiegel:2024dec,RBC:2024fic,Djukanovic:2024cmq,ExtendedTwistedMass:2024nyi,MILC:2024ryz,FermilabLatticeHPQCD:2024ppc,Keshavarzi:2019abf,DiLuzio:2024sps,Kurz:2014wya,Colangelo:2015ama,Masjuan:2017tvw,Colangelo:2017qdm,Colangelo:2017fiz,Hoferichter:2018dmo,Hoferichter:2018kwz,Eichmann:2019tjk,Bijnens:2019ghy,Leutgeb:2019gbz,Cappiello:2019hwh,Masjuan:2020jsf,Bijnens:2020xnl,Bijnens:2021jqo,Danilkin:2021icn,Stamen:2022uqh,Leutgeb:2022lqw,Hoferichter:2023tgp,Hoferichter:2024fsj,Estrada:2024cfy,Deineka:2024mzt,Eichmann:2024glq,Bijnens:2024jgh,Hoferichter:2024vbu,Hoferichter:2024bae,Holz:2024lom,Holz:2024diw,Cappiello:2025fyf,Colangelo:2014qya,Blum:2019ugy,Chao:2021tvp,Chao:2022xzg,Blum:2023vlm,Fodor:2024jyn}, is consistent with the current experimental world average, $a_\mu^{\rm exp}=116\,592\,071.5(14.5)\times10^{-11}$, leading to $\Delta a_\mu = 38(63)\times10^{-11}$ and therefore showing no significant discrepancy at present~\cite{Aliberti:2025beg}. Considerable efforts are nevertheless ongoing to further reduce both theoretical and experimental uncertainties and to maximize the sensitivity to BSM effects~\cite{Hertzog:2025ssc,Aliberti:2025beg}.

The case of the $\tau$ lepton is qualitatively different. Owing to its short lifetime, the $\tau$ dipole moments cannot be accessed through the interaction of polarized leptons with external electromagnetic fields, as in the electron and muon cases. Instead, they must be extracted from precise measurements of production and decay amplitudes involving $\tau$ leptons. 
At the same time, $\tau$ dipole moments provide a particularly promising window into BSM physics. Indeed, naive dimensional analysis suggests the scaling $a_\ell \propto m_\ell^2/\Lambda^2$, implying enhanced sensitivity to heavy new physics in the $\tau$ sector compared with the lighter charged leptons. This scaling can be modified in specific UV scenarios, and sizable enhancements can arise in models featuring additional sources of chiral-symmetry breaking or preferential couplings to third-generation fermions, as commonly predicted in theories addressing the flavor structure of the SM~\cite{Giudice:2012ms,Crivellin:2021spu,Athron:2025ets,Barbieri:1995uv,Barbieri:1996ae,Barbieri:2012uh,Matsedonskyi:2014iha,Panico:2016ull,Glioti:2024hye,Froggatt:1978nt,Berezhiani:1983hm,Bordone:2017bld,Fuentes-Martin:2022xnb,Davighi:2023iks,Nakai:2025dmp}. 

The SM prediction for the $\tau$ anomalous magnetic moment,
$a_\tau^{\rm SM}=117\,720(4)\times10^{-8}$, is known with an uncertainty far below any foreseeable experimental sensitivity~\cite{Eidelman:2007sb,Eidelman:2016aih,DiLuzio:2024sps,Hoferichter:2025fea,Wittig:2025azm}. Consequently, precision measurements of $a_\tau$ and $d_\tau$ represent exceptionally clean probes of NP. Deviations from the SM can be systematically parameterized within the Standard Model Effective Field Theory (SMEFT), where the SM Lagrangian is supplemented by higher-dimensional gauge-invariant operators suppressed by powers of the NP scale $\Lambda$~\cite{Buchmuller:1985jz}. 

Experimentally, the most stringent bounds on $a_\tau$ are currently provided by the ATLAS and CMS Collaborations at the LHC. The CMS Collaboration reported a best-fit value of $a_\tau = 0.9^{+3.2}_{-3.1}\times10^{-3}$ at $68\%$ C.L. from measurements of $\gamma\gamma\to\tau^+\tau^-$ production in proton--proton collisions at $\sqrt{s}=13~{\rm TeV}$, corresponding to $-0.0042 < a_\tau^{\rm CMS} < 0.0062$ at $95\%$ C.L.~\cite{CMS:2024qjo}. Subsequently, the ATLAS Collaboration obtained a stronger constraint, $a_\tau^{\rm ATLAS} = -0.5^{+2.6}_{-0.9}\times10^{-3}$, from high-mass $\tau^+\tau^-$ production in proton--proton collisions, corresponding to $a_\tau \in [-2.4,\,4.7]\times10^{-3}$ at $95\%$ C.L.~\cite{ATLAS:2025oiy}. 
The most stringent current constraint on the $\tau$ electric dipole moment has been reported by the Belle collaboration, $d_\tau \in [-1.85,\,0.61]\times10^{-17}\,e\,{\rm cm}$ at $95\%$ C.L.~\cite{Belle:2021ybo}, while no comparably competitive bound on $a_\tau$ is presently available from $B$ factories. 
Future Belle II analyses are expected to significantly improve the sensitivity to both dipole moments by exploiting suitably constructed asymmetries in $e^+e^-\to\tau^+\tau^-$ processes~\cite{Bernabeu:2006wf,Bernabeu:2007rr,Bernabeu:2008ii}. Assuming an integrated luminosity of $50~{\rm ab}^{-1}$, projected sensitivities at the level of $|a_\tau| \lesssim 10^{-5}$ and $|d_\tau| \lesssim 10^{-19}\,e\,{\rm cm}$ have been estimated in Refs.~\cite{Crivellin:2021spu,USBelleIIGroup:2022qro,Aihara:2024zds,Gogniat:2025eom,Hoferichter:2025zjp,Hoferichter:2025ijh,Belle-II:2018jsg, Gogniat:2026zvf}.

In this work, we study the sensitivity of the Future Circular Collider (FCC-ee) to electromagnetic and weak dipole moments of the $\tau$ lepton within the SMEFT framework, providing not only all the technical details but also additional results with respect to Ref.~\cite{Buttazzo:2026amk}.
Section~\ref{sec:current_status} reviews the current experimental and theoretical status of the $\tau$ dipole moments. Section~\ref{sec:Theory} introduces the SMEFT framework and the matching onto electromagnetic and weak dipole operators. Section~\ref{sec:pheno} presents the phenomenological analysis of the relevant production and decay channels, and Sec.~\ref{sec:summary} summarizes our results.

%
\section{Current status of the tau lepton dipole moments}
\label{sec:current_status}
The most general electromagnetic interaction between an on-shell fermion and an off-shell photon is described by the vertex function $\Gamma^\mu(q^2)$, constrained by Lorentz and $U(1)_{\rm EM}$ gauge invariance. For a fermion with incoming and outgoing momenta $p$ and $p'$, respectively, the corresponding vertex reads
\begin{equation}
\bar u(p')\Gamma_\mu(q^2)u(p),
\end{equation}
where $q =p'-p$ is the incoming momentum of the off-shell photon
and $\Gamma^\mu (q^2)$ reads
\begin{equation}
\Gamma^\mu(q)
=
-ieQ_f
\left[
F_1(q^2)\gamma^\mu
+\frac{i\sigma^{\mu\nu}q_\nu}{2m_f}F_2(q^2)
+\frac{\sigma^{\mu\nu}q_\nu}{2m_f}\gamma_5 F_3(q^2)
+\frac{q^2\gamma^\mu-q^\mu\slashed q}{m_f^2}
\gamma_5 F_A(q^2)
\right],\label{eq:FormFactor}
\end{equation}
where $e = \sqrt{4\pi\alpha}$, $\alpha$ is the fine-structure constant, $m_f$ is the fermion mass, $Q_f$ is the fermion's electric charge in units of $e$, and $\sigma^{\mu\nu} = \frac{i}{2} [\gamma^\mu, \gamma^\nu]$. 
The functions $F_1(q^2)$, $F_2(q^2)$, $F_3(q^2)$, and $F_A(q^2)$ are the Dirac, Pauli, electric-dipole, and anapole form factors, respectively.
While these form factors depend on the momentum transfer $q^2$, their values in the static limit $q^2 \to 0$ correspond to physical observables.\footnote{The anapole form factor defined through the off-shell electromagnetic vertex of a fundamental charged particle is generally gauge dependent and therefore cannot be regarded as a physical observable on its own.}

In particular, the Dirac form factor $F_1(q^2)$ is normalized to $F_1(0) = 1$ by charge conservation, while $F_{2}(q^2)$ and $F_{3}(q^2)$ are related to the fermion's anomalous magnetic moment $a_f$ and the electric dipole moment (EDM) $d_f$:
\begin{equation}
a_f = F_2(0), \qquad d_f = \frac{eQ_f}{2 m_f} F_3(0).
\label{eq:af_df}
\end{equation}
The anomalous magnetic moment of a lepton $\ell$ is defined as $a_\ell = (g_\ell - 2)/2$, where $g_\ell$ is the gyromagnetic ratio relating the lepton's spin to its magnetic moment. This quantity is commonly referred to as the ``$g-2$'', as it characterizes the quantum mechanical deviation of $g_\ell$ from the classical Dirac value of $g_\ell = 2$. 
In the SM, the leading contribution arises from QED radiative effects and is given by the universal Schwinger term $a_\ell^{\rm QED,\,1\text{-loop}}=\alpha/2\pi$~\cite{Schwinger:1948iu}.

Similarly, for the electroweak interaction between an off-shell $Z$ boson with momentum $q$ and two fermions, the vertex function $\Gamma^\mu_Z$ can be written as:
\begin{align}
\Gamma_Z^\mu(q) =
-ie Q_f
\bigg[
&
\gamma^\mu
\bigg(
F_1^Z(q^2)+F_A^Z(q^2)\gamma_5
\bigg)
+\frac{\sigma^{\mu\nu}q_\nu}{2m_f}
\bigg(
iF_2^Z(q^2)+F_3^Z(q^2)\gamma_5
\bigg)
\nonumber\\
&+\frac{q^\mu}{2m_f}
\bigg(
F_4^Z(q^2)+F_5^Z(q^2)\gamma_5
\bigg)
\bigg]\,,\label{eq:FormFactor_Z}
\end{align}
where the first two structures describe the vector and axial-vector interactions, $F_2^Z$ and $F_3^Z$ encode the weak magnetic dipole moment (WMDM) and the weak electric dipole moment (WEDM), while $F_4^Z$ and $F_5^Z$ parameterize longitudinal interactions.
At the Z-pole mass, these observables are defined as:
\begin{equation}
a_f^Z = F_2^Z(M_Z^2), \qquad d_f^Z = \frac{eQ_f}{2 m_f} F^Z_3(m_Z^2).\label{eq:af_df_Z_refined}
\end{equation}
While $a_f^{(Z)}$ is CP conserving, $d_f^{(Z)}$ violates the CP symmetry. As a result, the two observables probe different sectors of the underlying NP theory. In particular, $a_f^{(Z)}$ is sensitive to CP-even effects, such as contributions from heavy states that preserve the SM symmetries, whereas a non-vanishing $d_f^{(Z)}$ would provide an unambiguous signal of new sources of CP violation beyond the SM.

To quantify possible NP contributions to the $\tau$ dipole moments, we decompose the magnetic and weak magnetic dipole moments as
\begin{equation}
    \begin{aligned}
    a_\tau = a_\tau^{\rm SM} + \Delta a_\tau,\qquad\qquad
    a^Z_\tau = a_\tau^{Z,\,\rm SM} + \Delta a^{Z}_\tau,
    \end{aligned}
\end{equation}
where the SM predictions for $a_\tau^{\text{SM}}$~\cite{Eidelman:2007sb,Eidelman:2016aih,DiLuzio:2024sps,Hoferichter:2025fea,Wittig:2025azm} and $a_\tau^{Z,\,\rm SM}$~\cite{Bernabeu:1994wh,Kosnik:2026cfq} are given by
\begin{equation}
    \begin{aligned}
    a_\tau^{\text{SM}} = 117\,720\,(4) \times 10^{-8},\qquad\qquad
    a_\tau^{Z,\,\rm SM} = -(2.10 + 0.61 \,i )\times 10^{-6}\,.
    \end{aligned}
\end{equation}
The absorptive (imaginary) part of $a_\tau^{Z,\rm SM}$ arises from virtual particles lighter than the $Z$ boson going on shell. By contrast, the SM predictions for the electric and weak electric dipole moments are negligibly small~\cite{Hollik:1998wk,Yamaguchi:2020eub,Yamaguchi:2020dsy} and can be safely neglected. Consequently, throughout this work $d_\tau$ and $d^Z_\tau$ are identified with the corresponding NP contributions to the $\tau$ electric and weak electric dipole moments.

A summary of the present experimental constraints on the $\tau$ anomalous magnetic moment and electric dipole moments is reported in Table~\ref{tab:dipole_exp}. 

\begin{table}[ht]
\centering
\begin{tabular}{l c c c c}
\toprule
\textbf{Exp} & \textbf{Collision} & \textbf{$a_\tau$} [$10^{-3}$] & 
\textbf{$d_\tau~[10^{-17}\,e \cdot {\rm cm}]$} & \textbf{C.L.} \\
\midrule
L3~\cite{L3:1998gov}         & $e^+e^-$ & $[-52, 58]$  & $[-31,31]$&95\%\\
OPAL~\cite{OPAL:1998dsa}     & $e^+e^-$ & $[-68, 65]$  & $[-38,36]$&95\%\\
DELPHI~\cite{DELPHI:2003nah} & $e^+e^-$ & $[-52, 13]$  & $[-37,37]$&95\%\\
Belle~\cite{Belle:2021ybo}   & $e^+e^-$ & $-$          & 
  $[-1.85,0.61]$\tablefootnote{The interval $[-1.85, 0.61]\times 10^{-17}\,e\, {\rm cm}$ corresponds to 
  the constraint on $\mathrm{Re}(d_\tau)$; the constraint on $\mathrm{Im}(d_\tau)$ 
  is $[-1.03,0.23]\times 10^{-17}\,e\, {\rm cm}$ at 95\% C.L.} & 95\% \\
\midrule
ATLAS~\cite{ATLAS:2022ryk} & Pb+Pb & $[-57, 24]$ & $-$ & 95\% \\
CMS~\cite{CMS:2022arf}     & Pb+Pb & $[-51,21]$  & $-$ & 95\% \\
\midrule
CMS~\cite{CMS:2024qjo}     & $pp$ & $[-4.2, 6.2]$ & $[-2.9,2.9]$ &95\% \\
ATLAS~\cite{ATLAS:2025oiy} & $pp$ & $[-2.4, 4.7]$ & $[-2.0,2.0]$ &95\% \\
\bottomrule
\end{tabular}
\caption{Summary of current experimental constraints on the anomalous magnetic moment $a_\tau$ and electric dipole moment $d_\tau$ of the $\tau$ lepton.}
\label{tab:dipole_exp}
\end{table}

The first direct constraints on the $\tau$ dipole moments were obtained at LEP by the OPAL and L3 collaborations through the radiative process $Z\to\tau^+\tau^-\gamma$~\cite{Grifols:1990ha,L3:1998gov,OPAL:1998dsa}.
These limits were subsequently improved by the DELPHI collaboration using
the photon-induced process $\gamma\gamma\to\tau^+\tau^-$, leading to the
bounds $-0.052<a_\tau<0.013$ and $|d_\tau|<3.7\times10^{-16}\,e\,{\rm cm}$, at the $95\%$ C.L.~\cite{DELPHI:2003nah}. 

In recent years, substantial progress has been achieved at the LHC through measurements of exclusive $\gamma\gamma\to\tau^+\tau^-$ production. In ultra-peripheral Pb+Pb collisions, the ATLAS collaboration obtained the limit $-0.057 < a_\tau < 0.024$ at $95\%$ C.L.~\cite{ATLAS:2022ryk}, while CMS reported the model-dependent determination $a_\tau^{\rm CMS}=0.001^{+0.055}_{-0.089}$ through the same channel~\cite{CMS:2022arf}. These results are broadly comparable to the LEP sensitivity.
The current best constraint on $a_\tau$ is provided by the ATLAS \cite{ATLAS:2025oiy} and CMS \cite{CMS:2024qjo} collaborations, based on measurements of $\gamma\gamma\to\tau^+\tau^-$ production in proton--proton collisions at $\sqrt{s}=13~{\rm TeV}$ while the most stringent limit on the electric dipole moment is still obtained by the Belle experiment through CP-odd observables in $e^+e^- \to \tau^+\tau^-$ (see Table~\ref{tab:dipole_exp}) \cite{Belle:2021ybo,ATLAS:2022ryk}. All existing measurements are fully consistent with the SM expectation.

Despite the remarkable experimental progress achieved over the last decades, current experimental sensitivities remain many orders of magnitude above the SM predictions for the EDM and are still far from probing the SM contribution to the anomalous magnetic moment. Furthermore, while the electromagnetic dipole moments have been the subject of extensive experimental investigations, direct constraints on the weak magnetic and weak electric dipole moments are still
comparatively weak and largely model-dependent. This situation provides a strong motivation for future high-luminosity $e^+e^-$ colliders, where the clean experimental environment, the possibility of operating at and around the $Z$ pole, and the enormous samples of $\tau$ leptons can significantly improve the sensitivity to both electromagnetic and weak dipole interactions.

\section{Tau dipole moments in the SMEFT}
\label{sec:Theory}

Heavy NP contributions to the $\tau$ dipole moments can be described in a model-independent way within the SMEFT framework. At dimension six, the relevant operators are~\cite{Allwicher:2021jkr,Bresciani:2026acy}
\begin{equation}
    \label{eq:SMEFT}
    \mathscr{L} \supset \frac{\mathcal{C}_{\ell B}}{\Lambda^2} \, \bar{\ell}_L \sigma^{\mu\nu} e_R \varphi B_{\mu\nu} + \frac{\mathcal{C}_{\ell W}}{\Lambda^2} \, \bar{\ell}_L \sigma^{\mu\nu} \,\tau_I e_R \,  \varphi \, W^I_{\mu\nu} + {\rm h.c.}, 
\end{equation}
where $B_{\mu\nu}$ and $W^I_{\mu\nu}$ denote the field-strength tensors of the $U(1)_Y$ and $SU(2)_L$ gauge groups, $\varphi$ is the $SU(2)_L$ Higgs doublet, $\ell_L$ denotes the left-handed lepton $SU(2)_L$ doublet, and $e_R$ is the corresponding right-handed charged-lepton singlet. The matrices $\tau^I$ are the Pauli matrices, and $\sigma^{\mu\nu} = \frac{i}{2}[\gamma^\mu,\gamma^\nu]$ is the antisymmetric tensor that encodes the dipole structure.
After spontaneous symmetry breaking, the Higgs field acquires a vacuum expectation value (VEV), $\langle \varphi \rangle = (0, v/\sqrt{2})^T$ with $v=246$ GeV. This triggers the matching between the gauge-invariant SMEFT operators and the anomalous dipole form factors. From Eq.~\eqref{eq:SMEFT}, the effective Lagrangian in the broken phase is given by
\begin{equation}\label{eq:SMEFT_az}
    \mathscr{L} \supset \frac{v+H}{\sqrt{2} \Lambda^2} \big(\mathcal{C}_{\ell \gamma}\, \bar{\ell} \sigma^{\mu\nu}P_R\, \ell F_{\mu\nu} + \mathcal{C}_{\ell Z}\, \bar{\ell} \sigma^{\mu\nu} P_R\,\ell Z_{\mu\nu} \big) + \text{h.c.},
\end{equation}
where $\ell$ is the spinor for the lepton, $P_R=(1+\gamma^5)/2$ is the right-handed projection operator, and we have defined the Wilson coefficients in the $(\gamma, Z)$ basis as 
\begin{equation}\label{eq:CgammaCz}
    \begin{aligned}
    &\mathcal{C}_{\ell \gamma} \equiv c_W \mathcal{C}_{\ell B} - s_W \mathcal{C}_{\ell W}, \\
    &\mathcal{C}_{ \ell Z} \equiv -c_W \mathcal{C}_{\ell W} - s_W \mathcal{C}_{\ell B} \,. 
    \end{aligned}
\end{equation}

Matching Eq.~\eqref{eq:SMEFT_az} onto the dipole form factors for the $\tau$ lepton defined in  Eqs.~\eqref{eq:FormFactor} and \eqref{eq:FormFactor_Z}, one obtains the 
tree-level relations
\begin{equation}
\label{eq:adtau_final}
\begin{aligned}
\Delta a_\tau &= 
\frac{2\sqrt{2} m_\tau v}{e \Lambda^2} 
\text{Re}\,\mathcal{C}_{\tau\gamma}(m_\tau),
\quad 
d_\tau = 
\frac{\sqrt{2} v}{\Lambda^2} 
\text{Im}\,\mathcal{C}_{\tau\gamma}(m_\tau), 
\\
\Delta a_\tau^{Z} &= 
\frac{2\sqrt{2} m_\tau v}{e \Lambda^2} 
\text{Re}\,\mathcal{C}_{\tau Z}(M_Z),
\quad 
d_\tau^{Z} = 
\frac{\sqrt{2} v}{\Lambda^2} 
\text{Im}\,\mathcal{C}_{\tau Z}(M_Z).
\end{aligned}
\end{equation}
At tree level, the dimension-6 dipole operators generate momentum-independent contributions to the dipole form factors; a non-trivial $q^2$ dependence would require higher-dimensional operators and lies beyond the dimension-6 approximation adopted here.

The SMEFT parameterization of the dipole interactions are implemented using the Universal \textsc{FeynRules} Output (UFO)~\cite{Degrande:2011ua} model files into \textsc{LEPTONdipole}~\cite{Ma:2026LEPTONdipole}, which is cross checked with \textsc{SMEFTsim}~\cite{Brivio:2020onw}.

The Wilson coefficients entering Eq.~\eqref{eq:adtau_final} are evaluated at the scales $m_\tau$ and $M_Z$, respectively. When probing the dipole interactions at a characteristic scale $Q$, the corresponding coefficients should in principle be evolved from $Q$ to these scales using the renormalization-group (RG) equations. Using the RG evolution of Refs.~\cite{Jenkins:2013zja,Jenkins:2013wua,Alonso:2013hga,Jenkins:2017jig,Jenkins:2017dyc}, we estimate that running effects modify the coefficients by at most $\mathcal{O}(15\%)$, even for $Q=100~\mathrm{TeV}$. At FCC-ee, where the highest center-of-mass energy considered in our analysis is $\sqrt{s}=365~\mathrm{GeV}$, these effects are well below the accuracy relevant for our analysis. We therefore neglect RG running in the numerical results
and, for simplicity, omit the scale dependence of
$\mathcal{C}_{\tau\gamma}$ and $\mathcal{C}_{\tau Z}$.

Within the SMEFT framework, perturbative unitarity imposes an upper bound on the energy range over which scattering amplitudes remain perturbatively
calculable. For the dipole operators in Eq.~\eqref{eq:SMEFT}, one finds
the consistency conditions~\cite{Allwicher:2021jkr,Bresciani:2025toe,Bresciani:2026acy}
\begin{align}
s \,\frac{|{\mathcal C}_{\ell B}|}{\Lambda^2} < 4\pi\,, 
\qquad
s \,\frac{|{\mathcal C}_{\ell W}|}{\Lambda^2}  < 4\pi\, \sqrt{2/3}\,.
\label{eq:perturbative}
\end{align}
Considering one operator at a time, it is straightforward to check that perturbative unitarity requires
\begin{equation}
\sqrt{s} \lesssim
\begin{cases}
\displaystyle
2.1~\mathrm{TeV}
\left(\frac{10^{-2}}{|\Delta a_\tau|}\right)^{1/2},
& \mathcal C_{\tau W}=0\,,
\\[3mm]
\displaystyle
1.4~\mathrm{TeV}
\left(\frac{10^{-2}}{|\Delta a_\tau|}\right)^{1/2},
& \mathcal C_{\tau B}=0\,.
\end{cases}
\label{eq:unitarity_atau_numerical}
\end{equation}
Thus, for $|\Delta a_\tau|=10^{-2}$, perturbative unitarity is preserved up to center-of-mass energies of approximately $2.1$~TeV and $1.4$~TeV in the $\mathcal C_{\tau B}$-only and $\mathcal C_{\tau W}$-only scenarios, respectively.

\section{Probing tau dipole moments at future \boldmath{$e^+e^-$} colliders}
\label{sec:pheno}

In this section we investigate the sensitivity of future high-luminosity $e^+e^-$ colliders to the electromagnetic and weak dipole moments of the $\tau$ lepton. As will become clear in the following, the different production channels probe complementary aspects of the underlying dipole interactions. In particular, resonant $Z$ production provides the best sensitivity to the weak dipole moments, whereas photon-induced $\tau$-pair production is particularly powerful in probing the anomalous magnetic moment. Radiative $\tau$-pair production and radiative Higgs decays offer additional, albeit less competitive, information. After describing the common analysis framework, we will discuss each process separately before combining their sensitivities.

\subsection{Analysis framework}
In this work we adopt the Future Circular Collider (FCC-ee) as
the reference benchmark, while the Circular Electron--Positron Collider
(CEPC) is expected to provide a comparable physics program with similar
center-of-mass energies and integrated luminosities
~\cite{FCC:2018evy,Bernardi:2022hny,FCC:2025lpp,Altmann:2025feg,
CEPCStudyGroup:2018ghi,CEPCStudyGroup:2018rmc,An:2018dwb,
CEPCPhysicsStudyGroup:2022uwl,CEPCStudyGroup:2023quu,
CEPCStudyGroup:2025kmw}. 

\paragraph*{FCC-ee operating points.}
The center-of-mass energies and integrated luminosities assumed in our analysis are summarized in Table~\ref{tab:FCCee}. These benchmarks cover the Tera-$Z$, electroweak, Higgs, and top-factory operating stages of the FCC-ee and provide sensitivity over a broad range of energies.
\begin{table}[htb]
    \centering
    \setlength{\tabcolsep}{10pt}
    \renewcommand{\arraystretch}{1.25}
    \scalebox{0.9}{
    \begin{tabular}{c|ccc|cc|c|c}
        \hline
        Working point & \multicolumn{3}{c|}{$Z$ pole} & \multicolumn{2}{c|}{$WW$} & $ZH$ & $t\bar t$ \\ \hline
        $\sqrt{s}$ [GeV]          & 87.9 & 91.2 & 94.3 & 157.5 & 162.5 & 240 & 365 \\
        $\mathcal{L}$ [ab$^{-1}$] & 40   & 125  & 40   & 9.6   & 9.6   & 10.8 & 2.7 \\ 
        \hline
    \end{tabular}
    }
    \caption{
    Center of mass energies, $\sqrt{s}$, and integrated 
    luminosities, $\mathcal{L}$, at FCC-ee~\cite{janot_2024_yr3v6-dgh16,FCC:2025lpp} 
    in the stages of a Tera-$Z$ ($Z$ pole),  
    electroweak ($WW$), Higgs ($ZH$), and top ($t\bar t$)
    factories.}
    \label{tab:FCCee}
\end{table}

\paragraph*{Event selection.}
To account for detector acceptance, all visible final-state particles are required to satisfy a polar-angle cut
$\theta>5^\circ$ with respect to the beam axis. Furthermore, we impose the isolation requirement $\Delta R_{ij}>0.4$, where
$\Delta R=\sqrt{(\Delta\eta)^2+(\Delta\phi)^2}$. The angular cut reflects the finite detector acceptance, whereas the isolation requirement ensures a reliable reconstruction of photons and $\tau$ leptons and simultaneously regulates the collinear configurations arising from photon emission off charged fermions. Additional kinematic selections specific to each process are introduced in the corresponding subsections.

The numerical results are obtained using \textsc{MadGraph\_aMC@NLO}~\cite{Alwall:2014hca,Frederix:2018nkq} and cross checked with \textsc{Whizard}~\cite{Kilian:2007gr}.

\paragraph*{Statistical procedure.}
The sensitivity to the dipole operators is quantified through the profile-likelihood significance ~\cite{ParticleDataGroup:2020ssz,Cowan:2010js}
\begin{equation}
\label{eq:sen}
\mathcal S=
\sqrt{2(S+B)\ln\!\left(1+\frac{S}{B}\right)-2S},
\end{equation}
where
\begin{equation}
S(\mathcal C_{\tau W},\mathcal C_{\tau B}) =
\epsilon\mathcal L \left[
\sigma(\mathcal C_{\tau W},\mathcal C_{\tau B}) - \sigma(0,0)
\right],
\qquad
B=\epsilon \mathcal L\,\sigma(0,0).
\end{equation}
Here $\mathcal L$ denotes the integrated luminosity, $\epsilon$ denotes the relevant efficiency for the analysis and $\sigma(\mathcal C_{\tau W},\mathcal C_{\tau B})$ the corresponding cross section.
Throughout this work, $\mathcal S=2$ and $\mathcal S=3$ correspond to the 95\% and 99\% confidence levels, respectively.
The statistical significances obtained from independent channels are combined according to
\begin{equation}
\label{eq:s_combine}
\mathcal S_{\rm combined} = \sqrt{\sum_i\mathcal S_i^2}.
\end{equation}
Moreover, we perform the analysis at the $\tau$ production level and assume a constant reconstruction efficiency $\epsilon_\tau=80\%$.

\paragraph*{Scope of the projections.}
The sensitivities presented below should be interpreted as estimates of the intrinsic statistical reach of the different channels under the assumptions specified above, rather than as forecasts of the final experimental performance. A realistic assessment will require the explicit simulation of $\tau$ decays and reconstruction, detector response, reducible backgrounds, and channel-dependent systematic uncertainties, together with sufficiently precise higher-order predictions for both signal and background processes. Our results are therefore primarily intended to identify the most promising channels and kinematic regimes and to motivate dedicated experimental and theoretical studies.

\subsection{\boldmath{$e^+e^-\to \tau^+\tau^-$}}
\label{sec:theory2}
Among the production channels considered in this work,
$e^+e^-\to\tau^+\tau^-$ at Tera-$Z$ provides the most direct probe of the weak dipole operators. The reason is twofold. First, at center-of-mass energies close to the $Z$ pole, the process benefits from the resonant enhancement of the $Z$-exchange amplitude. Second, an energy scan across the resonance changes the relative weight of the photon and $Z$-mediated contributions, thereby helping to disentangle the electromagnetic and weak dipole interactions. This makes $e^+e^-\to\tau^+\tau^-$ a central channel for probing $\Delta a_\tau^Z$ and $d_\tau^Z$ at future high-luminosity $e^+e^-$ colliders. Meanwhile, owing to the $1/s$ suppression of the SM background, $e^+e^-\to\tau^+\tau^-$ also provides the best sensitivity to $d_\tau$ at the higher-energy stages of FCC-ee.

\subsubsection{SMEFT description}
\label{sec:ee2tt_smeft}
\begin{figure}[htb]
\centering
\includegraphics[width=0.25\textwidth]{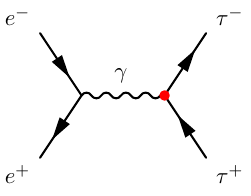}
\hspace{50pt}
\includegraphics[width=0.25\textwidth]{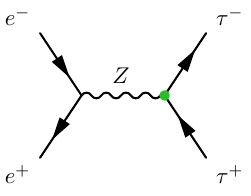}
\caption{
Leading-order Feynman diagrams for $e^+e^- \to \tau^+\tau^-$, including both SM and SMEFT contributions. The red and green dots mark vertices that can receive corrections from the electromagnetic $\mathcal O_{\tau\gamma}$ and weak $\mathcal O_{\tau Z}$ dipole operators, respectively. When these dimension-6 contributions are switched off, the marked vertices reduce to their SM counterparts, and the diagrams reproduce the corresponding SM amplitudes.
}
\label{fig:FD_ee2tt}
\end{figure}

Within the SMEFT, the leading-order diagrams for $e^+e^-\to \tau^+\tau^-$ are shown in Fig.~\ref{fig:FD_ee2tt}. The red node denotes an insertion of the electromagnetic dipole operator $\mathcal O_{\tau\gamma}$, while the green node denotes an insertion of the weak dipole operator $\mathcal O_{\tau Z}$. In the absence of these
insertions, the same diagrams reduce to the SM contribution.

The total cross section can be organized as an expansion in the dipole Wilson coefficients,
\begin{equation}
\label{eq:ee2tata_xs}
\sigma_{\ell^+\ell^-\to\tau^+\tau^-} = \sigma_{\ell\ell}^{\rm SM} + \sigma_{\ell\ell}^{\rm Lin} + \sigma_{\ell\ell}^{\rm Quad},
\end{equation}
where the three terms denote the pure SM contribution, the interference between the SM and dimension-six dipole amplitudes, and the quadratic dimension-six contribution, respectively. Explicitly, 
\begin{subequations}
\label{eq:sigma_ee_all}
\begin{align}
\sigma_{\ell\ell}^{\rm SM} &=
\frac{\beta}{12\pi}
\Bigg[
\frac{e^4(1+2\tau)}{s}
+
\frac{
2e^2g_Z^2g_V^2s\Delta(1+2\tau)
}{
\Delta^2s^2+M_Z^2\Gamma_Z^2
}
\nonumber\\
&\hspace{2.0cm}
+
\frac{
g_Z^4s(g_V^2+g_A^2)
\left[
g_V^2(1+2\tau)+g_A^2(1-4\tau)
\right]
}{
\Delta^2s^2+M_Z^2\Gamma_Z^2
}
\Bigg],
\label{eq:ee_sigma_sm}
\\[2mm]
\sigma_{\ell\ell}^{\rm Lin}
&=
\frac{\beta m_\tau v}{\sqrt{2}\pi s\Lambda^2}
\,
{\rm Re}
\Bigg[
e^3\mathcal C_{\tau\gamma}
+
\frac{
eg_Z^2g_V^2\mathcal C_{\tau\gamma}\Delta s^2
}{
\Delta^2s^2+M_Z^2\Gamma_Z^2
}
\nonumber\\
&\hspace{2.0cm}
-
\frac{
g_Zg_V\mathcal C_{\tau Z}s^2
\left[
e^2\Delta+g_Z^2(g_V^2+g_A^2)
\right]
}{
\Delta^2s^2+M_Z^2\Gamma_Z^2
}
\Bigg],
\label{eq:ee_sigma_lin}
\\[2mm]
\sigma_{\ell\ell}^{\rm Quad}
&=
\frac{\beta v^2(1+8\tau)}{12\pi\Lambda^4}
\Bigg[
e^2\bigl({\rm Re}\,\mathcal C_{\tau\gamma}\bigr)^2
+
\frac{
g_Z^2\bigl({\rm Re}\,\mathcal C_{\tau Z}\bigr)^2
s^2(g_V^2+g_A^2)
}{
\Delta^2s^2+M_Z^2\Gamma_Z^2
}
\nonumber\\
&\hspace{2.0cm}
-
\frac{
2eg_Zg_V\,{\rm Re}\,\mathcal C_{\tau\gamma}\,
{\rm Re}\,\mathcal C_{\tau Z}\,\Delta s^2
}{
\Delta^2s^2+M_Z^2\Gamma_Z^2
}
\Bigg]
\nonumber\\
&\quad
+
\frac{\beta v^2(1-4\tau)}{12\pi\Lambda^4}
\Bigg[
e^2\bigl({\rm Im}\,\mathcal C_{\tau\gamma}\bigr)^2
+
\frac{
g_Z^2\bigl({\rm Im}\,\mathcal C_{\tau Z}\bigr)^2
s^2(g_V^2+g_A^2)
}{
\Delta^2s^2+M_Z^2\Gamma_Z^2
}
\nonumber\\
&\hspace{2.0cm}
-
\frac{
2eg_Zg_V\,{\rm Im}\,\mathcal C_{\tau\gamma}\,
{\rm Im}\,\mathcal C_{\tau Z}\,\Delta s^2
}{
\Delta^2s^2+M_Z^2\Gamma_Z^2
}
\Bigg].
\label{eq:ee_sigma_quad}
\end{align}
\end{subequations}
Here
\begin{equation}
\tau=\frac{m_\tau^2}{s},
\qquad
\beta=\sqrt{1-4\tau},
\qquad
\Delta=1-\frac{M_Z^2}{s},
\qquad
g_Z=\frac{e}{c_Ws_W},
\end{equation}
while $g_V$ ($g_A$) denotes the vector (axial-vector) couplings of charged leptons to the $Z$ boson.

\paragraph*{Threshold behaviour.}

The real and imaginary parts of the dipole Wilson coefficients enter
Eq.~\eqref{eq:ee_sigma_quad} with different kinematic weights. In the
high-energy limit, $\sqrt{s}\gg m_\tau$, the difference between the
coefficients of
$({\rm Re}\,\mathcal C_{\tau V})^2$ and
$({\rm Im}\,\mathcal C_{\tau V})^2$ becomes negligible.
Near the $\tau^+\tau^-$ threshold, however, the contribution
proportional to $({\rm Im}\,\mathcal C_{\tau V})^2$ is
suppressed by an additional factor of $\beta^2$ relative to the real
part. This follows from the fact that the CP-odd dipole amplitude
contains the final-state bilinear
$\bar u_\tau\sigma^{\mu\nu}q_\nu\gamma^5v_\tau$, whose non-relativistic
matrix element is proportional to the three-momentum
$|\vec p_\tau|$. Therefore the corresponding amplitude vanishes
linearly at threshold, and the cross section is suppressed as
$|\vec p_\tau|^2/m_\tau^2\sim\beta^2$.

\paragraph*{Resonant structure and $Z$-pole scan.}

The parameter $\Delta$ measures the detuning from the $Z$ resonance and
controls the interference between photon and $Z$ exchange. Since
$\Delta$ changes sign when crossing the pole, an energy scan around
$\sqrt{s}=M_Z$ modulates the relative contributions of
$\mathcal C_{\tau\gamma}$ and $\mathcal C_{\tau Z}$. This is the key
feature that allows the $Z$-pole program to lift the degeneracy between the electromagnetic and weak dipole operators, as shown in Sec.~\ref{sec:ee2tt}.

\begin{figure}[htb]
\centering
\includegraphics[width=0.6\textwidth]{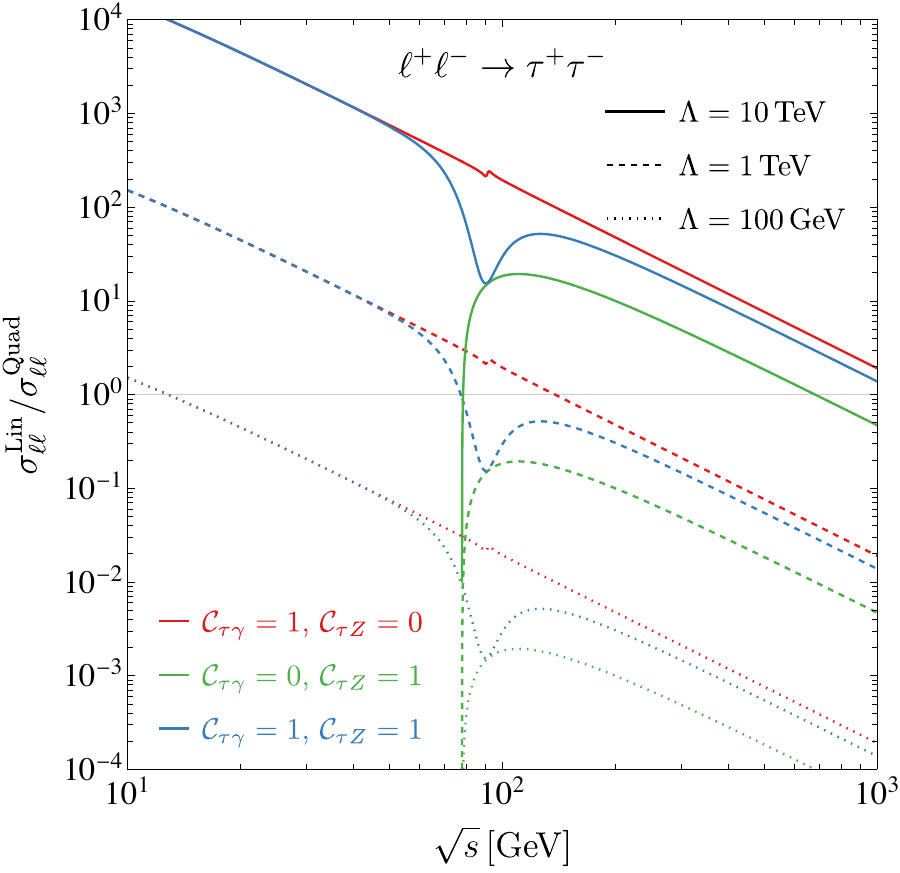}
\caption{
Ratio of the linear interference term to the quadratic BSM contribution,
$\sigma^{\rm Lin}_{\ell\ell}/\sigma^{\rm Quad}_{\ell\ell}$, for
$\ell^+\ell^- \to \tau^+\tau^-$ as a function of $\sqrt{s}$, for
different Wilson-coefficient choices and values of $\Lambda$.}
\label{fig:L2Q_eett}
\end{figure}

\paragraph*{Linear versus quadratic contributions.}
The SM-NP interference term in Eq.~\eqref{eq:ee_sigma_lin} is suppressed by $m_\tau v/s$. This suppression reflects the helicity-flipping nature of the dipole amplitude relative to the SM vector and axial-vector
amplitudes. Parametrically, 
\begin{equation}
\sigma_{\ell\ell}^{\rm Lin} \sim \frac{m_\tau v}{s\Lambda^2}, 
\qquad
\sigma_{\ell\ell}^{\rm Quad} \sim \frac{v^2}{\Lambda^4},
\qquad
\frac{\sigma_{\ell\ell}^{\rm Lin}} {\sigma_{\ell\ell}^{\rm Quad}} \sim \frac{m_\tau\Lambda^2}{sv}.
\end{equation}
The relative importance of the two contributions is illustrated in Fig.~\ref{fig:L2Q_eett}. At fixed Wilson coefficients and fixed $\Lambda$, the linear term is progressively suppressed at larger center-of-mass energies and eventually gives way to the quadratic contribution. Conversely, the high luminosity of the FCC-ee allows one to probe larger values of $\Lambda$, where the linear interference can remain the dominant source of sensitivity to the CP-even dipole moments $a_\tau^{(Z)}$. This feature will be important at the $Z$ pole, while the quadratic contribution becomes relatively more relevant at higher-energy stages.

\paragraph*{EFT interpretation.}

The expressions above retain the complete dependence on the dimension-six dipole amplitudes up to $\mathcal{O}(1/\Lambda^4)$. In particular, they include the interference between the SM and NP amplitudes involving two insertions of dimension-six operators, as well as the squared contributions from NP amplitudes with a single dimension-six insertion. Interference effects from dimension-eight dipole operators also scale as $1/\Lambda^4$, but they inherit the same helicity suppression as the dimension-six interference term. They are therefore parametrically suppressed with respect to the quadratic dimension-six contribution in the high-energy regime. This justifies the dimension-six truncation used in the following phenomenological analysis.

\subsubsection{Phenomenological analysis}
\label{sec:ee2tt}

The exceptional sensitivity of the $e^+e^-\to\tau^+\tau^-$ channel originates from the possibility of operating future $e^+e^-$ colliders at and around the $Z$-boson resonance. Unlike the higher-energy FCC-ee stages, where the production cross section decreases approximately as $1/s$, the resonant enhancement of the $Z$-exchange amplitude, together with the enormous integrated luminosity of the Tera-$Z$ program, provides the largest sample of reconstructed $\tau$ leptons available at any future collider. As discussed in the previous subsection, the energy scan across the resonance also modifies the relative weight of the photon- and $Z$-mediated amplitudes, thereby providing a unique handle to separate the electromagnetic and weak dipole operators.

\paragraph*{Energy scan and initial-state radiation.}

To exploit the full potential of the Tera-$Z$ program, it is essential to account for initial-state radiation (ISR), which lowers the effective partonic center-of-mass energy through photon emission from the incoming electron and positron. The observable cross section is therefore obtained by convolving the partonic prediction of Eq.~\eqref{eq:ee2tata_xs}
with the ISR structure functions,
\begin{equation}
\label{eq:isr_conv_ee}
\sigma_{e^+e^-\to X}(s)
=
\int_{\xi_0}^{1} d\xi
\int_{\xi}^{1}\frac{dx}{x}\,
 f_{e^-/e^-}(x,Q^2)
 f_{e^+/e^+}\!\left(\frac{\xi}{x},Q^2\right)
 \hat\sigma_{e^+e^-\to X}(\hat s),
\qquad
\hat s=\xi s .
\end{equation}
Here $f_{e^\pm/e^\pm}$ denotes the probability density for finding an electron or positron carrying a fraction $x$ of the beam energy after photon radiation. For $X=\tau^+\tau^-$, the partonic cross section $\hat\sigma_{e^+e^-\to X}$ is given by
Eq.~\eqref{eq:ee2tata_xs}. Throughout this work ISR effects are
simulated with \textsc{MadGraph5\_aMC@NLO}~\cite{Alwall:2014hca,Frederix:2018nkq} and cross-checked against \textsc{Whizard}~\cite{Kilian:2007gr}.

\begin{figure}[htb]
\centering
\includegraphics[width=0.45\textwidth]{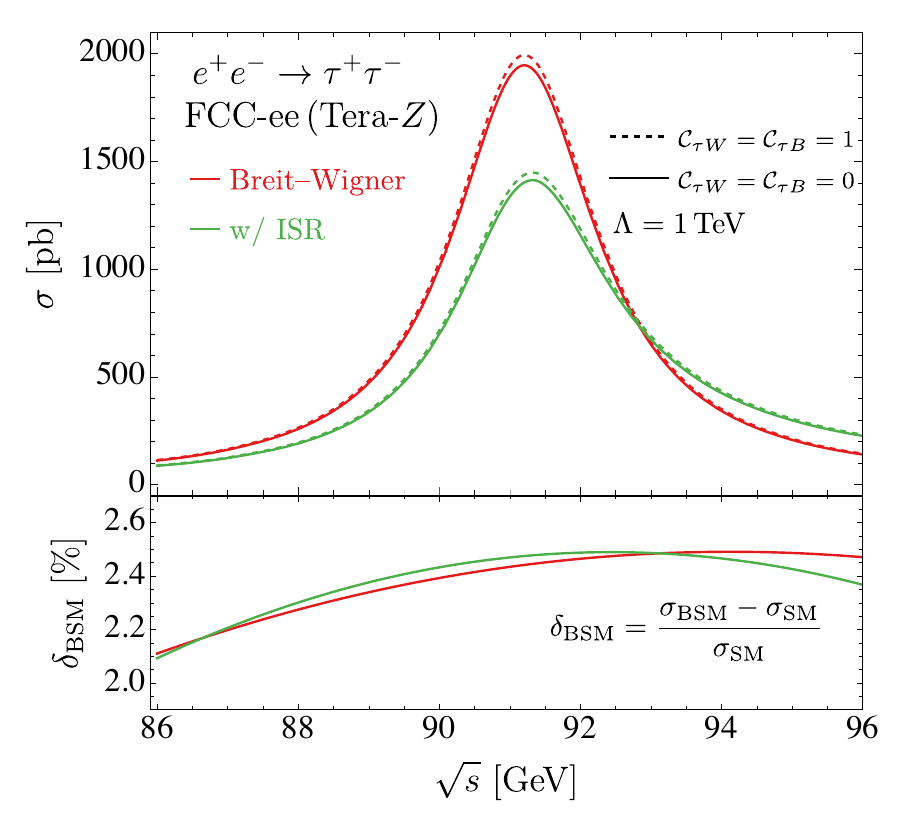}
\hspace{8pt}
\includegraphics[width=0.45\textwidth]{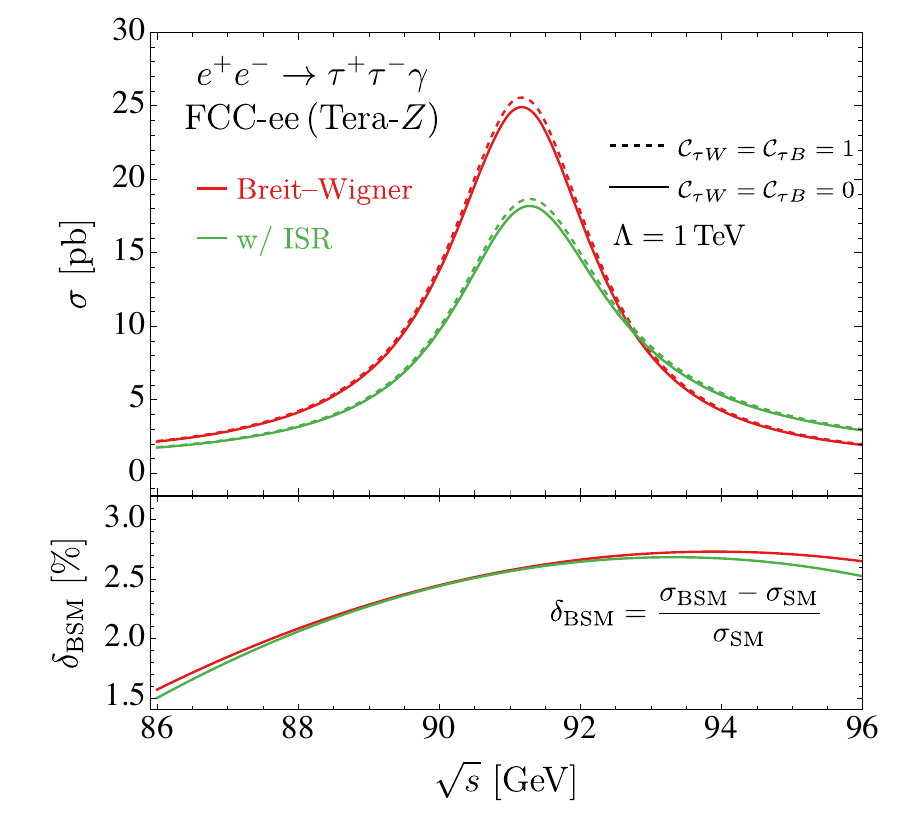}
\caption{
Leading-order cross sections for $e^+e^-\to\tau^+\tau^-$ (left) and $e^+e^-\to\tau^+\tau^-\gamma$ (right) as a function of the nominal center-of-mass energy $\sqrt{s}$ near the $Z$ pole. The red curves show the Breit--Wigner prediction, while the green curves include ISR effects. Solid lines denote the SM prediction, whereas dashed lines correspond to a BSM benchmark with $\mathcal C_{\tau W}=\mathcal C_{\tau B}=1$ and $\Lambda=1~{\rm TeV}$. The lower panels show the relative deviation $\delta_{\rm BSM}=(\sigma_{\rm BSM}-\sigma_{\rm SM})/\sigma_{\rm SM}$.}
\label{fig:ee_ISR}
\end{figure}
The impact of ISR is illustrated in Fig.~\ref{fig:ee_ISR}, where we compare the Breit--Wigner prediction with the complete simulation including photon radiation for both $e^+e^-\to\tau^+\tau^-$ and $e^+e^-\to\tau^+\tau^-\gamma$. The emission of ISR photons induces the well-known radiative-return effect, whereby events produced above the $Z$ pole migrate towards the resonance. This shifts the apparent maximum of the visible cross section to slightly larger nominal center-of-mass energies. The lower panels show that the relative BSM contribution remains stable across the three Tera-$Z$ operating points, ensuring that the sensitivity to the dipole operators is largely preserved once ISR effects are included.

\begin{figure}[!p]
\centering
\includegraphics[width=0.42\textwidth]{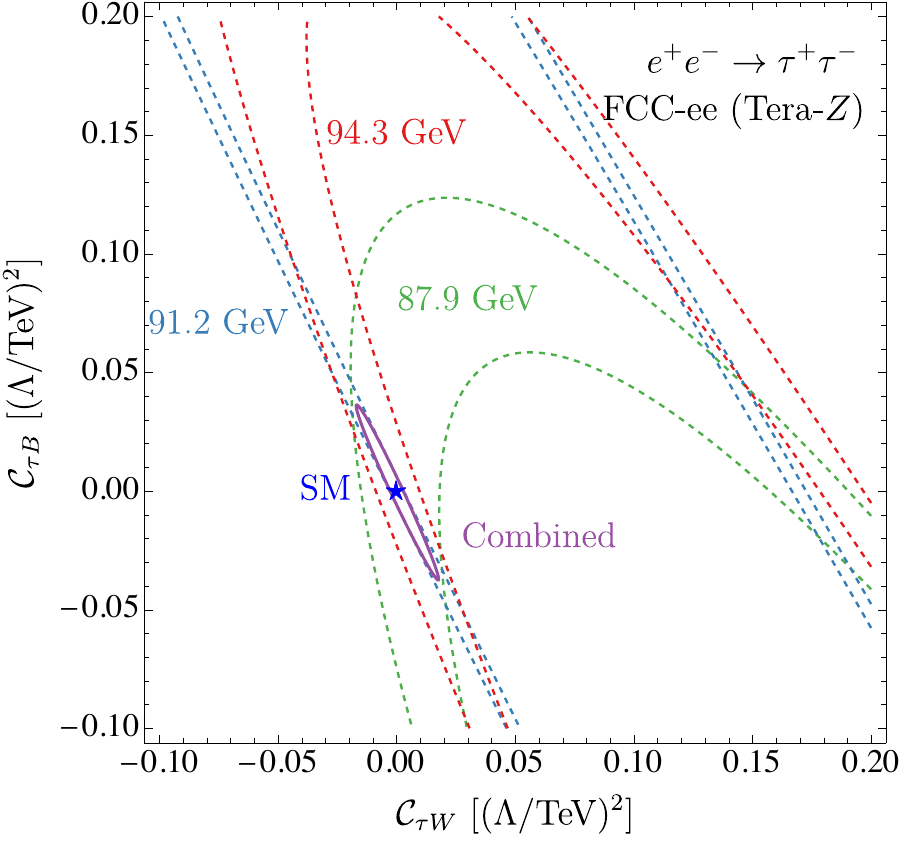}
\includegraphics[width=0.415\textwidth]{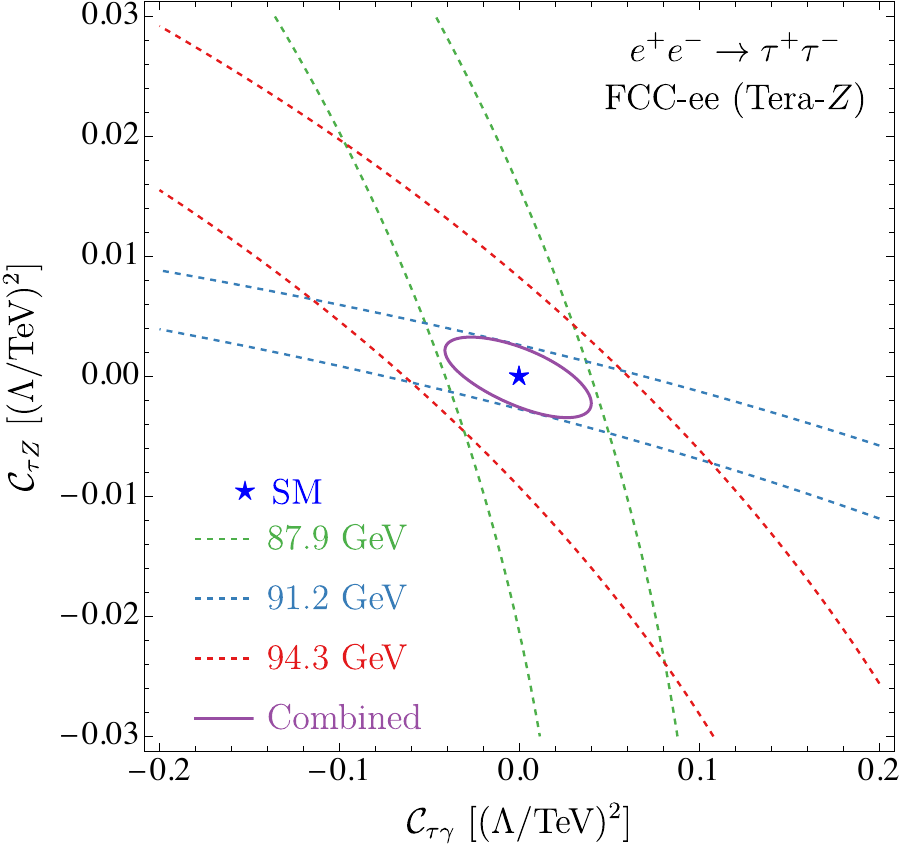}
\includegraphics[width=0.42\textwidth]{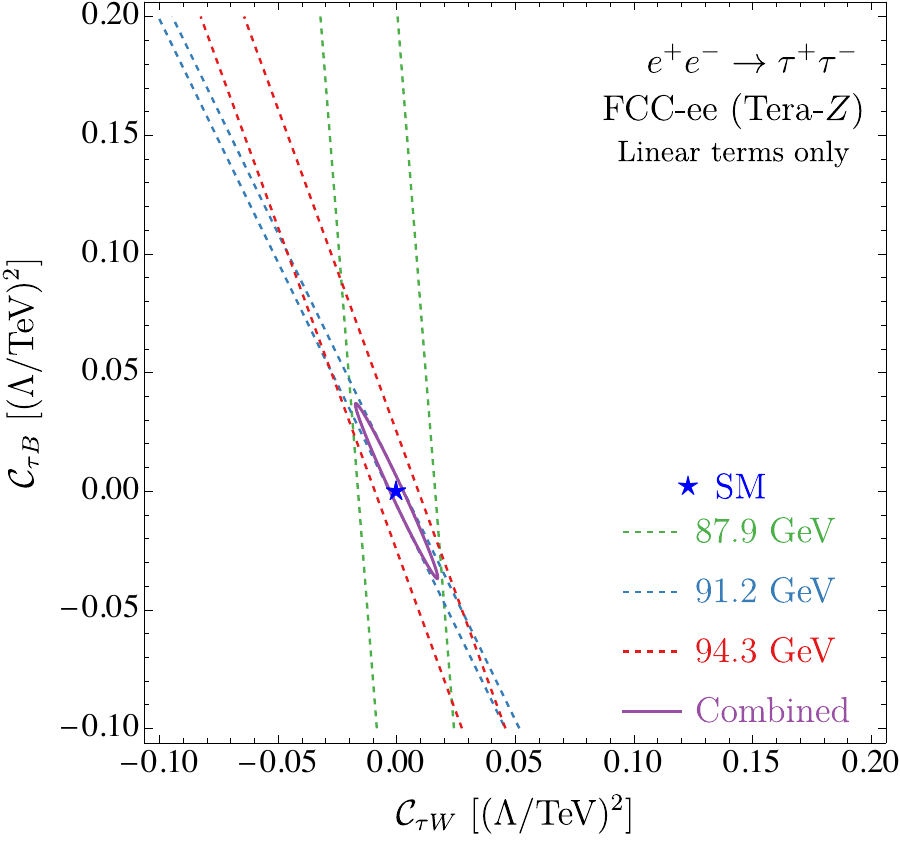}
\includegraphics[width=0.415\textwidth]{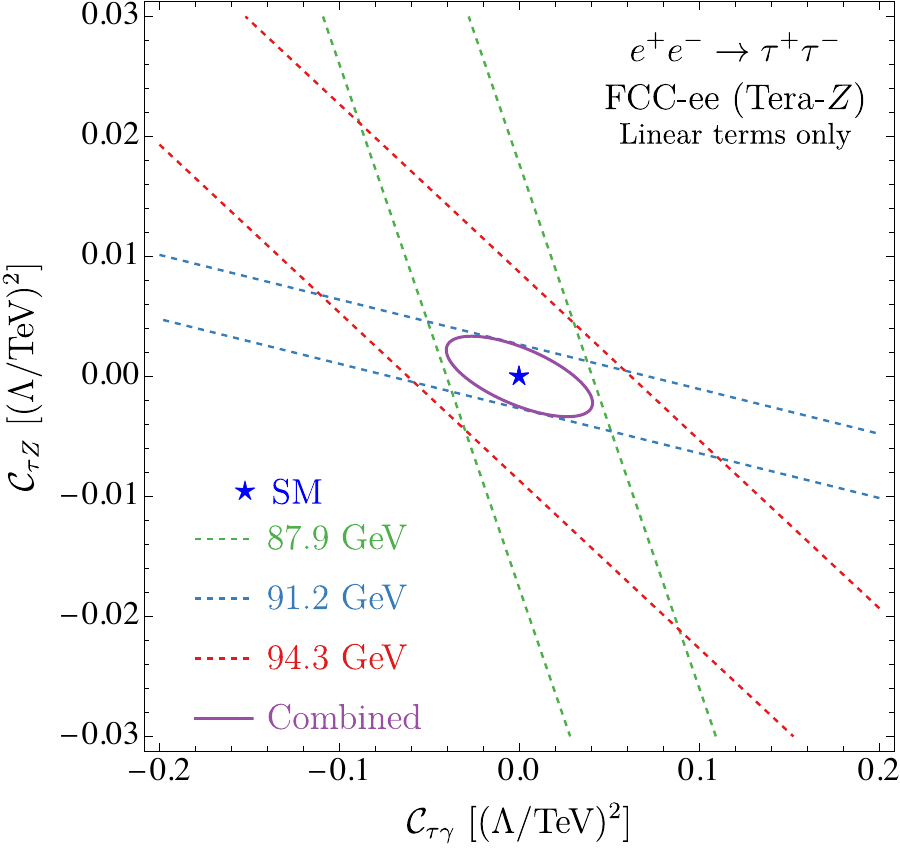}
\includegraphics[width=0.415\textwidth]{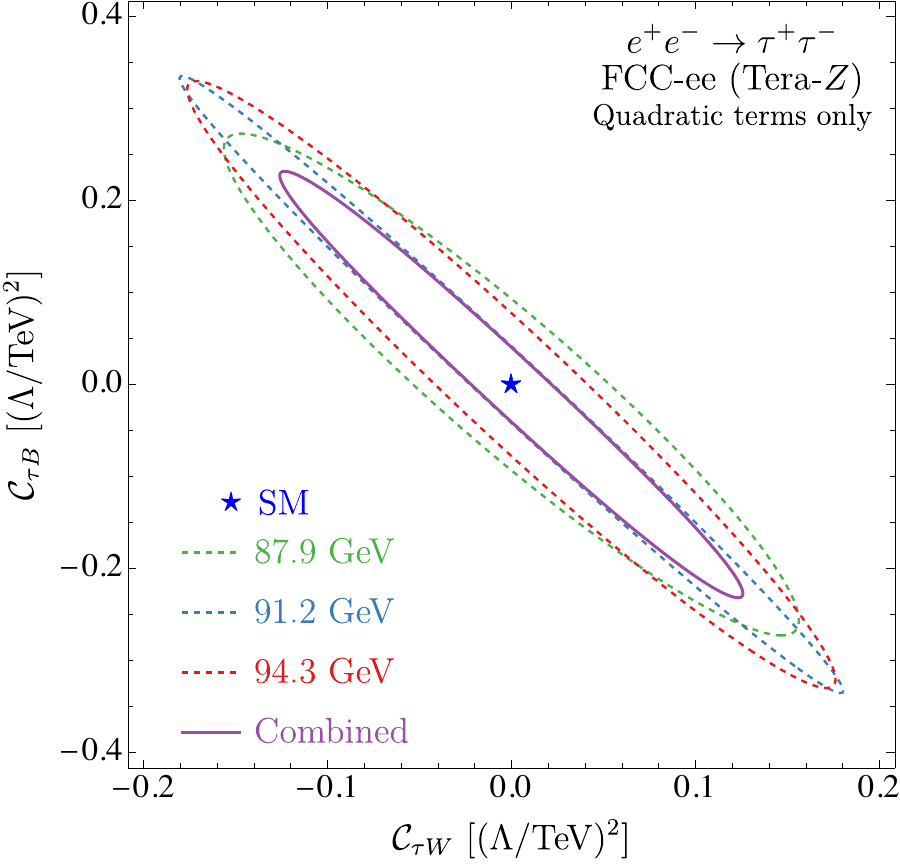}
\includegraphics[width=0.42\textwidth]{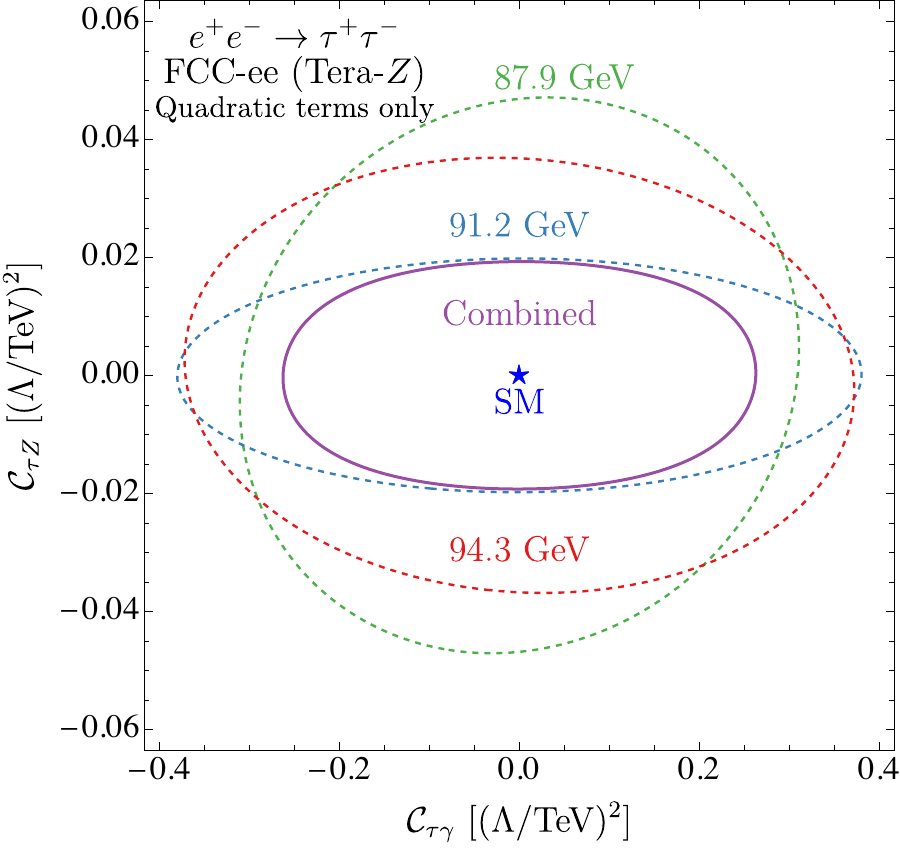}
\caption{
Expected $95\%$ CL sensitivity to the $\tau$ dipole Wilson coefficients from $e^+e^-\to\tau^+\tau^-$ at the FCC-ee Tera-$Z$ stage. The left column shows constraints in the $(\mathcal C_{\tau W},\mathcal C_{\tau B})$ plane, while the right column shows the corresponding constraints in the $(\mathcal C_{\tau\gamma},\mathcal C_{\tau Z})$ plane. From top to bottom, the three rows show the full dimension-six prediction, the linear interference terms only, and the quadratic BSM terms only. Dashed contours denote individual center-of-mass energies, while solid contours show their combination.}
\label{fig:ee_3TeraZ}
\end{figure}

\paragraph*{Sensitivity at the Tera-$Z$ stage.}
We now quantify the sensitivity of the three Tera-$Z$ operating points using the statistical procedure introduced in Sec.~\ref{sec:pheno}. Fig.~\ref{fig:ee_3TeraZ} shows the corresponding $95\%$ CL sensitivity contours in the $(\mathcal C_{\tau W},\mathcal C_{\tau B})$ and $(\mathcal C_{\tau\gamma},\mathcal C_{\tau Z})$ planes. The three rows display, respectively, the complete SMEFT prediction, the isolated linear interference contribution, and the pure quadratic dimension-six contribution.
\begin{itemize}
\item {\it Complete SMEFT prediction.}
The first row of Fig.~\ref{fig:ee_3TeraZ} demonstrates the
complementarity of the three Tera-$Z$ running points. Each individual energy leaves an approximate flat direction in the Wilson-coefficient plane, corresponding to different combinations of $\mathcal C_{\tau\gamma}$ and $\mathcal C_{\tau Z}$ that produce nearly the same deviation from the SM cross section. Combining the three measurements efficiently removes these degeneracies. This improvement is a direct consequence of the energy dependence of the photon--$Z$ interference discussed in the previous subsection.
\item {\it Linear interference.}
The second row isolates the contribution linear in the Wilson coefficients. As anticipated from the analytical discussion in Sec.~\ref{sec:ee2tt_smeft}, the high luminosity of the Tera-$Z$ program allows the SM-NP interference term to dominate the sensitivity to the CP-even dipole moments despite its helicity suppression. This linear contribution therefore provides the leading constraints on the anomalous magnetic and weak magnetic dipole moments.
\item {\it Quadratic contribution.}
The third row shows the quadratic dimension-six contribution. Unlike the interference term, this contribution is insensitive to the sign of the Wilson coefficients and therefore gives rise to the characteristic annular allowed regions. Since the CP-odd dipole moments do not interfere with the SM amplitude in the production rate without spin information, the quadratic contribution also provides the leading sensitivity to the electric and weak electric dipole moments\footnote{
For a recent proposal to constrain $d_\tau$ and $d_\tau^Z$ through spin-sensitive CP-odd asymmetries at future Z factories, see Ref.~\cite{Du:2026ynu}.
}.
\end{itemize}

The marginalized constraints obtained by combining the three Tera-$Z$ energies are
\begin{equation}
\label{eq:ee_TeraZ_bounds}
\Delta a_\tau \in [-1.7,1.6]\times10^{-4},
\qquad
|d_\tau| < 6.0\times10^{-18}\ e\, {\rm cm},
\end{equation}
\begin{equation}
\label{eq:ee_TeraZ_bounds_Z}
\Delta a^Z_\tau \in [-1.4,1.3]\times10^{-5},
\qquad
|d^Z_\tau| < 4.4\times10^{-19}\ e\, {\rm cm}.
\end{equation}
The corresponding one-operator-at-a-time limits are summarized in
Table~\ref{tab:ee_TeraZ_bounds}. The substantial improvement with
respect to present constraints originates from the combination of 
the enormous statistics available at the Tera-$Z$ stage and the complementary information provided by the three operating energies.

\begin{table}[htb]
\centering
\renewcommand{\arraystretch}{1.18}
\setlength{\tabcolsep}{5pt}
\begin{tabular}{c|cccccc}
\hline
$\sqrt{s}$ [GeV]
& Max$(\Delta a_\tau)$
& Min$(\Delta a_\tau)$
& Max$(\Delta a^Z_\tau)$
& Min$(\Delta a^Z_\tau)$
& $|d_\tau|$ [$e\,{\rm cm}$]
& $|d^Z_\tau|$ [$e\,{\rm cm}$]
\\
\hline
87.9     & $1.6\times10^{-4}$ & $-1.7\times10^{-4}$ & $6.3\times10^{-5}$ & $-8.6\times10^{-5}$ & $7.0\times10^{-18}$ & $1.1\times10^{-18}$ \\
91.2     & $2.8\times10^{-4}$ & $-3.0\times10^{-4}$ & $1.1\times10^{-5}$ & $-1.1\times10^{-5}$ & $8.6\times10^{-18}$ & $4.5\times10^{-19}$ \\
94.3     & $2.5\times10^{-4}$ & $-2.6\times10^{-4}$ & $3.3\times10^{-5}$ & $-3.7\times10^{-5}$ & $8.4\times10^{-18}$ & $8.4\times10^{-19}$ \\
Combined & $1.2\times10^{-4}$ & $-1.3\times10^{-4}$ & $1.0\times10^{-5}$ & $-1.0\times10^{-5}$ & $6.0\times10^{-18}$ & $4.4\times10^{-19}$ \\
\hline
\end{tabular}
\caption{
Projected $95\%$ CL one-operator-at-a-time constraints on the $\tau$
dipole moments from $e^+e^-\to\tau^+\tau^-$ at the FCC-ee Tera-$Z$
stage. Bounds on the magnetic dipole moments assume real Wilson
coefficients, while limits on the electric dipole moments assume purely
imaginary coefficients. For $\Delta a_\tau$ and $|d_\tau|$ the
$Z$-dipole coefficient is set to zero, while for $\Delta a^Z_\tau$ and
$|d^Z_\tau|$ the electromagnetic dipole coefficient is set to zero.}
\label{tab:ee_TeraZ_bounds}
\end{table}

\paragraph*{Sensitivity at higher-energy stages.}
Although the Tera-$Z$ program provides the strongest sensitivity to the weak dipole moments, the higher-energy FCC-ee stages remain phenomenologically important. As discussed in Sec.~\ref{sec:ee2tt_smeft}, the interference term decreases as $m_\tau v/(s\Lambda^2)$, whereas the quadratic contribution scales as $v^2/\Lambda^4$.  Consequently, the relative importance of the quadratic contribution increases with the center-of-mass energy, partially compensating for the loss in sensitivity to the $\tau$ EDM arising from the decreasing production cross section and the lower integrated luminosity at the higher-energy stages of FCC-ee.

\begin{figure}[!p]
\centering
\includegraphics[width=0.45\textwidth]{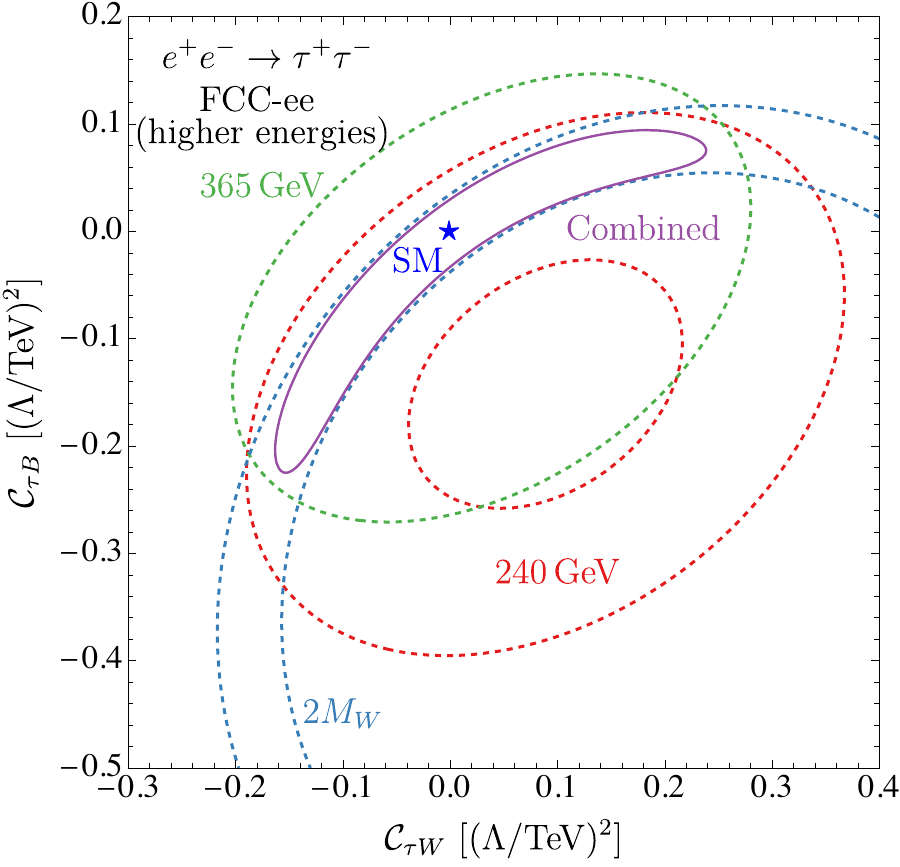}
\includegraphics[width=0.445\textwidth]{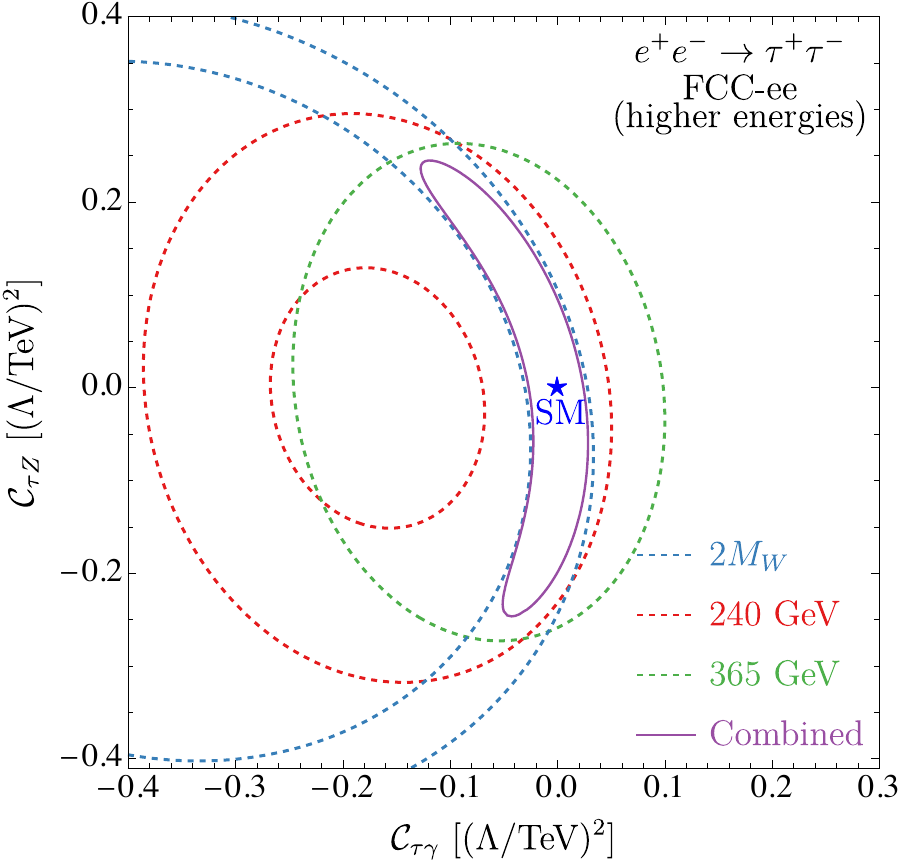}
\includegraphics[width=0.45\textwidth]{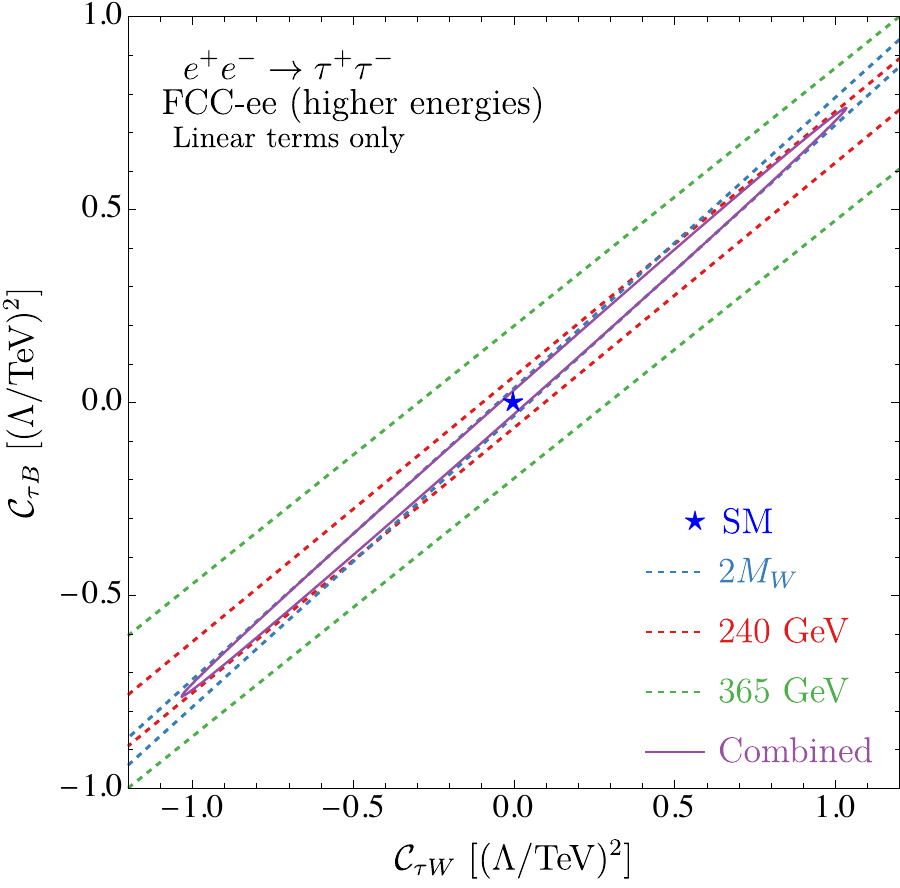}
\includegraphics[width=0.445\textwidth]{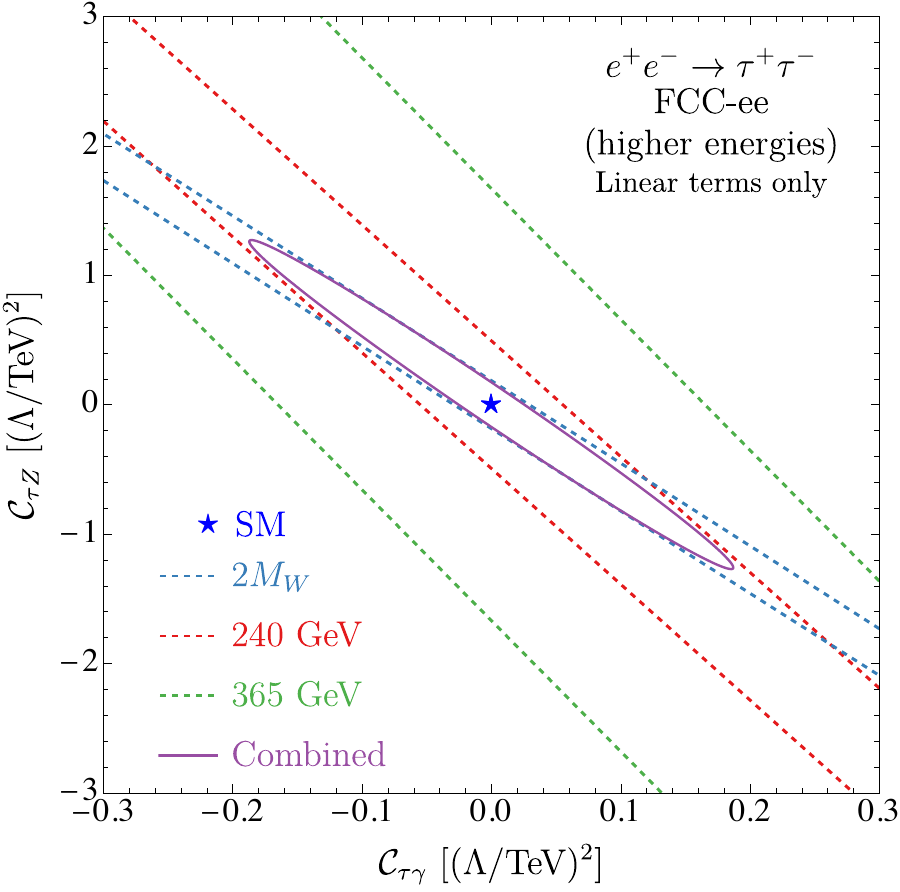}
\includegraphics[width=0.45\textwidth]{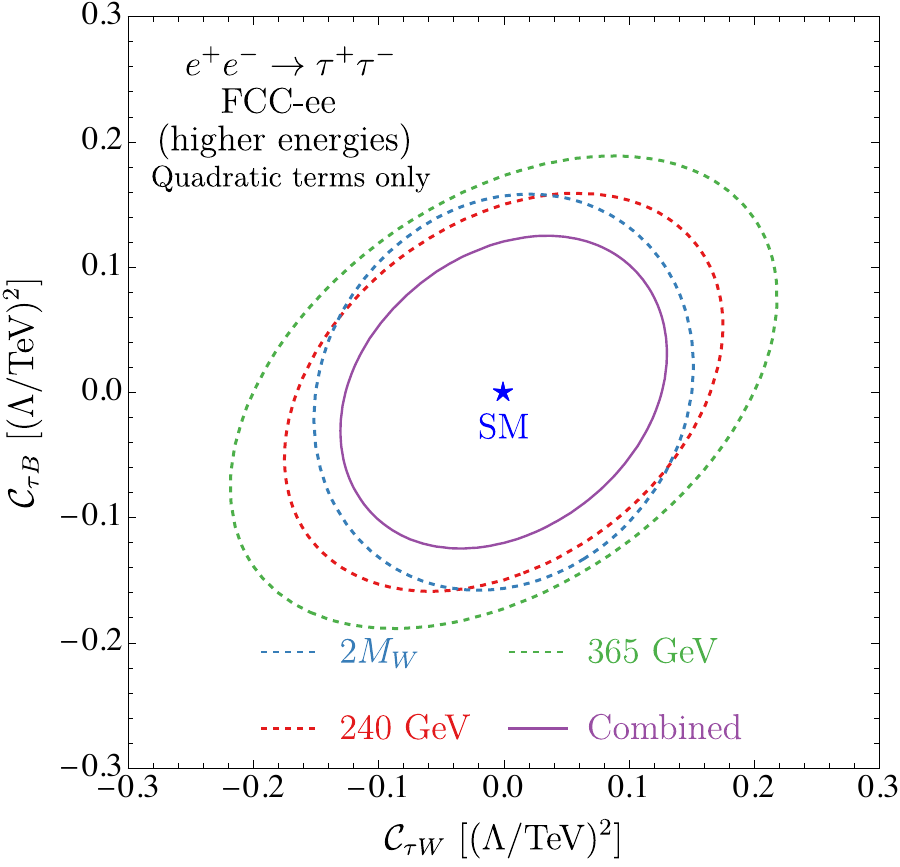}
\includegraphics[width=0.445\textwidth]{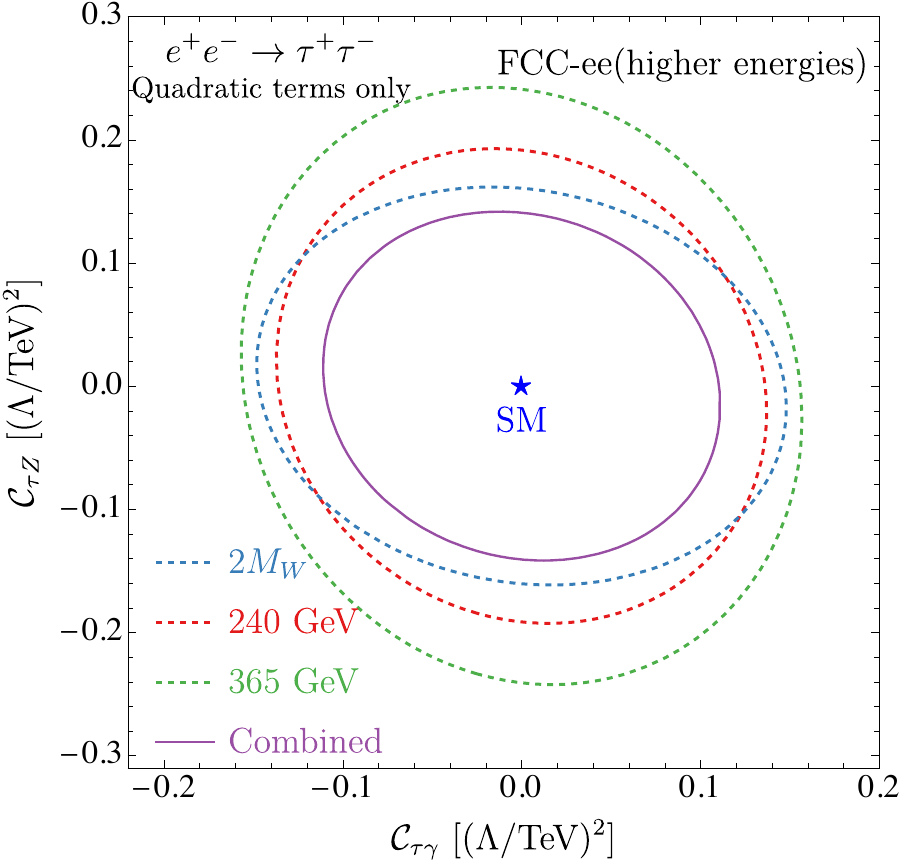}
\caption{
Same as Fig.~\ref{fig:ee_3TeraZ}, but for $e^+e^-\to\tau^+\tau^-$ at the higher-energy FCC-ee stages.}
\label{fig:ee_highE}
\end{figure}
Fig.~\ref{fig:ee_highE} presents the sensitivity contours at the electroweak, Higgs and top operating stages using the same decomposition adopted in Fig.~\ref{fig:ee_3TeraZ}. As expected, the linear contribution gradually loses importance with increasing $\sqrt{s}$, while the quadratic contribution becomes progressively more relevant. The combined contours at high energy differ more visibly from the linear-only approximation than at the $Z$ pole, reflecting the transition towards a regime where quadratic terms play a larger role. Nevertheless, the reduced luminosity of the highest-energy stages prevents these measurements from surpassing the ultimate precision achievable at the Tera-$Z$ stage for the weak dipole moments.

After marginalization over the correlated Wilson coefficients, the combination of all FCC-ee stages gives
\begin{equation}
\label{eq:ee_all_bounds}
\Delta a_\tau \in [-9.1,8.6]\times10^{-5},
\qquad
|d_\tau| < 2.5\times10^{-18}\ e\, {\rm cm},
\end{equation}
\begin{equation}
\label{eq:ee_all_bounds_Z}
\Delta a^Z_\tau \in [-1.2,1.1]\times10^{-5},
\qquad
|d^Z_\tau| < 4.4\times10^{-19}\ e\, {\rm cm}.
\end{equation}
The corresponding one-operator-at-a-time limits are collected in Table~\ref{tab:ee_all_bounds}. The limits on the magnetic dipole moments remain mildly asymmetric because of the linear interference term, whereas the limits on the electric dipole moments are symmetric and are controlled by the quadratic contribution.

\begin{table}[htb]
\centering
\renewcommand{\arraystretch}{1.18}
\setlength{\tabcolsep}{5pt}
\begin{tabular}{c|cccccc}
\hline
$\sqrt{s}$
& Max$(\Delta a_\tau)$
& Min$(\Delta a_\tau)$
& Max$(\Delta a^Z_\tau)$
& Min$(\Delta a^Z_\tau)$
& $|d_\tau|$ [$e\,{\rm cm}$]
& $|d^Z_\tau|$ [$e\,{\rm cm}$]
\\
\hline
$M_Z$        & $1.2\times10^{-4}$ & $-1.3\times10^{-4}$ & $1.0\times10^{-5}$ & $-1.0\times10^{-5}$ & $6.0\times10^{-18}$ & $4.4\times10^{-19}$ \\
$2M_W$       & $1.1\times10^{-4}$ & $-1.2\times10^{-4}$ & $4.2\times10^{-4}$ & $-9.8\times10^{-4}$ & $3.3\times10^{-18}$ & $3.6\times10^{-18}$ \\
$240$ GeV    & $2.0\times10^{-4}$ & $-2.8\times10^{-4}$ & $6.4\times10^{-4}$ & $-9.3\times10^{-4}$ & $3.1\times10^{-18}$ & $4.4\times10^{-18}$ \\
$365$ GeV    & $4.0\times10^{-4}$ & $-1.0\times10^{-3}$ & $9.0\times10^{-4}$ & $-1.0\times10^{-3}$ & $3.5\times10^{-18}$ & $5.5\times10^{-18}$ \\
Combined     & $7.8\times10^{-5}$ & $-8.2\times10^{-5}$ & $1.0\times10^{-5}$ & $-1.0\times10^{-5}$ & $2.5\times10^{-18}$ & $4.4\times10^{-19}$ \\
\hline
\end{tabular}
\caption{
Projected $95\%$ CL one-operator-at-a-time constraints on the $\tau$
dipole moments from $e^+e^-\to\tau^+\tau^-$ at all FCC-ee operating
stages.}
\label{tab:ee_all_bounds}
\end{table}

The analysis of $e^+e^-\to\tau^+\tau^-$ leads to three main conclusions. First, the Tera-$Z$ stage provides the strongest sensitivity to the weak dipole moments owing to the simultaneous resonant enhancement of the $Z$-exchange amplitude and the unprecedented integrated luminosity. Second, the energy scan across the $Z$ resonance plays a crucial role in lifting the approximate degeneracies between the electromagnetic and weak dipole operators through the energy dependence of the photon--$Z$ interference. Finally, although the higher-energy FCC-ee stages are individually less sensitive to the weak dipoles, they probe a complementary regime in which the quadratic dimension-six contribution becomes increasingly important. In addition to the $1/s$ suppression on the SM background, the $e^+e^-\to\tau^+\tau^-$ process provides the best sensitivity to the $\tau$ EDM. The combination of all operating stages therefore provides the most robust determination of the $\tau$ dipole operators within the SMEFT framework. 

\subsection{\boldmath{$\gamma\gamma\to\tau^+\tau^-$}}
\label{sec:gmgm2tt}
Photon-induced $\tau$-pair production provides a qualitatively different probe of the electromagnetic dipole operator with respect to direct $e^+e^-$ annihilation. In this channel, the observed process is $e^+e^-\to e^+e^-\tau^+\tau^-$, dominated by quasi-real photon fusion, $\gamma\gamma\to\tau^+\tau^-$. Since the photon couples directly to the combination $\mathcal C_{\tau\gamma}$, this process is essentially insensitive to the weak dipole coefficient $\mathcal C_{\tau Z}$. The enhancement of the photon luminosity at small momentum fractions favors the near-threshold region, where SM--NP interference dominates the sensitivity to $\Delta a_\tau$.

\subsubsection{SMEFT description}
\label{sec:gmgm2tt_smeft}
We describe photon-induced production in the equivalent-photon approximation (EPA). The cross section for $\ell^+\ell^-\to\ell^+\ell^-+\mathcal F$ can be written as a convolution of the photon-photon luminosity with the partonic cross section,
\begin{equation}
\label{eq:sigma_EPA}
\begin{aligned}
\sigma_{\mathcal F}(s) &=
\int_{\xi_0}^{1} d\xi\,
\frac{d\mathcal L_{\gamma\gamma}}{d\xi}\,
\hat\sigma_{\gamma\gamma}(\hat s),
\qquad
\hat s=\xi s,
\\[1mm]
\frac{d\mathcal L_{\gamma\gamma}}{d\xi} &=
\int_{\xi}^{1}\frac{dx}{x}\,
f_{\gamma/\ell}(x,Q^2)
 f_{\gamma/\ell}\!\left(\frac{\xi}{x},Q^2\right).
\end{aligned}
\end{equation}
Here $\xi_0=\hat s_{\rm min}/s$ is fixed by the minimum invariant mass of the final state, or by the applied kinematic selection cuts. The photon distribution $f_{\gamma/\ell}(x,Q^2)$ gives the probability of finding a quasi-real photon carrying a fraction $x$ of the lepton energy at the factorization scale $Q$. At leading logarithmic accuracy it is described by the Weizs\"acker--Williams approximation (WWA)~\cite{vonWeizsacker:1934nji,Williams:1934ad} 
\begin{equation}
\label{eq:EPA}
f_{\gamma/\ell}^{\rm WWA}(x,Q^2)
=
\frac{\alpha}{2\pi}
\frac{1+(1-x)^2}{x}
\ln\!\left(\frac{Q^2}{m_\ell^2}\right).
\end{equation}
In the numerical analysis we use the improved Weizs\"acker--Williams approximation (IWW)~\cite{Frixione:1993yw}, which retains the constant terms associated with the photon virtuality integration,
\begin{equation}
\label{eq:IWW}
f_{\gamma/\ell}^{\rm IWW}(x,Q^2)
=
\frac{\alpha}{2\pi}
\left[
2m_\ell^2x\left(\frac{1}{Q^2}-\frac{1}{Q_{\rm min}^2}\right)
+
\frac{1+(1-x)^2}{x}\ln\!\frac{Q^2}{Q_{\rm min}^2}
\right],
\qquad
Q_{\rm min}^2=\frac{m_\ell^2x^2}{1-x}.
\end{equation}
Following the LEP-II analysis, we take as central scale the average transverse energy of the final-state $\tau$ leptons in the partonic center-of-mass frame,
\begin{equation}
\label{eq.mg5EPA_Q}
Q=\frac{E_T}{2}
=
\frac{E_{\tau^+}\sin\theta_{\tau^+}
+E_{\tau^-}\sin\theta_{\tau^-}}{2}.
\end{equation}

\begin{figure}[htb]
\centering
\includegraphics[width=0.618\textwidth]{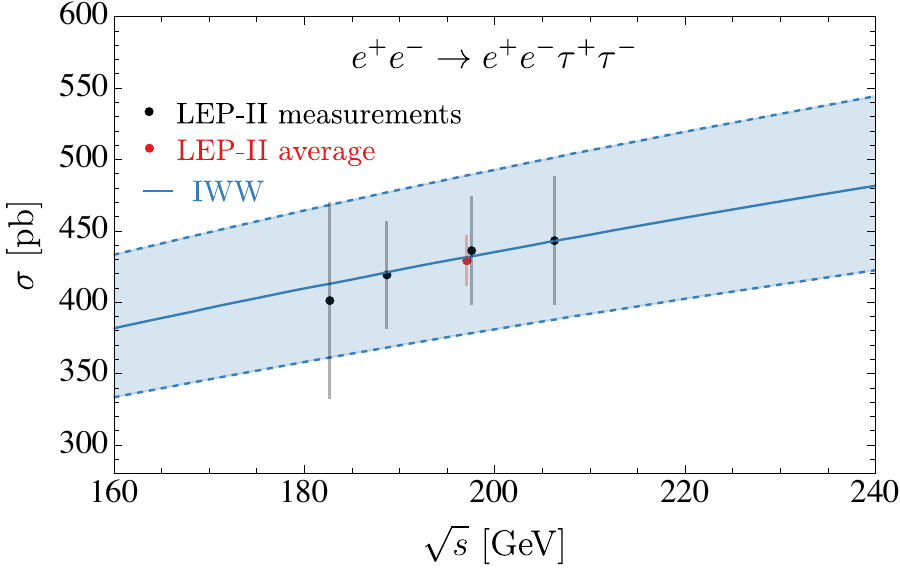}
\caption{
SM prediction for $\sigma(e^+e^-\to e^+e^-\tau^+\tau^-)$ as a function of $\sqrt{s}$ in the improved Weizs\"acker--Williams approximation. The shaded band is obtained by varying the central scale $Q=E_T/2$ in the range $[Q/2,2Q]$. Black markers show the LEP-II measurements from 1998--2000, and the red marker denotes their weighted average as reported by DELPHI~\cite{DELPHI:2003nah}.
}
\label{fig:aa2tt_IWW}
\end{figure}

As a validation of this choice, in Fig.~\ref{fig:aa2tt_IWW} we show the SM prediction for $e^+e^-\to e^+e^-\tau^+\tau^-$ obtained with the IWW photon flux. The uncertainty band is estimated by varying the central scale $Q=E_T/2$ in the range $[Q/2,2Q]$. The agreement with the DELPHI measurements at LEP-II supports the use of this EPA setup for the FCC-ee projections below.

\begin{figure}[htb]
\centering
\includegraphics[width=0.25\textwidth]{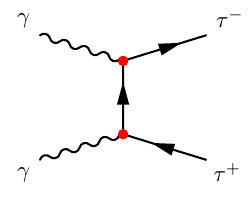}
\hspace{50pt}
\includegraphics[width=0.25\textwidth]{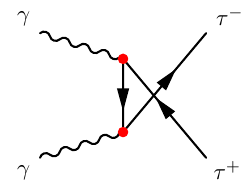}
\caption{
Leading-order Feynman diagrams for $\gamma\gamma\to\tau^+\tau^-$, including both SM and SMEFT contributions. The red dots mark vertices that can receive contributions from the electromagnetic dipole operator $\mathcal O_{\tau\gamma}$. In the absence of these dimension-6 contributions, the marked vertices reduce to their SM counterparts.
}
\label{fig:FD_aa2tt}
\end{figure}

The leading-order diagrams for the partonic subprocess $\gamma\gamma\to\tau^+\tau^-$ are shown in Fig.~\ref{fig:FD_aa2tt}. The red nodes denote insertions of the anomalous electromagnetic dipole operator $\mathcal O_{\tau\gamma}$. Unlike $e^+e^-\to\tau^+\tau^-$, this channel probes only the photon combination of the SMEFT dipoles. 
The partonic cross section can be organized as
\begin{equation}
\label{eq:gmgm2tata_xs}
\hat\sigma_{\gamma\gamma\to\tau^+\tau^-}
=
\hat\sigma_{\gamma\gamma}^{\rm SM}
+
\hat\sigma_{\gamma\gamma}^{\rm Lin}
+
\hat\sigma_{\gamma\gamma}^{\rm Quad}
+\mathcal O\!\left(\frac{1}{\Lambda^6}\right),
\end{equation}
where the second term denotes the interference between the SM amplitude and a single insertion of $\mathcal O_{\tau\gamma}$, while the third term contains the quadratic dimension-six contribution, including both the squared contributions from NP amplitudes with a single insertion of \(\mathcal{O}_{\tau\gamma}\) and the interference between the SM amplitude and NP amplitudes with two insertions of \(\mathcal{O}_{\tau\gamma}\). Explicitly,
\begin{subequations}
\label{eq:gmgm2tata_xs_dc}
\begin{align}
\hat\sigma_{\gamma\gamma}^{\rm SM}
&=
\frac{e^4}{4\pi\hat s}
\left[
(1+4\hat\tau-8\hat\tau^2)
\ln\!\left(\frac{1+\hat\beta}{1-\hat\beta}\right)
-
\hat\beta(1+4\hat\tau)
\right],
\label{eq:gmgm2tata_xs_sm}
\\[1mm]
\hat\sigma_{\gamma\gamma}^{\rm Lin}
&=
\frac{\sqrt{2}e^3m_\tau v}{\pi\hat s}
\frac{{\rm Re}\,\mathcal C_{\tau\gamma}}{\Lambda^2}
\ln\!\left(\frac{1+\hat\beta}{1-\hat\beta}\right),
\label{eq:gmgm2tata_xs_lin}
\\[1mm]
\hat\sigma_{\gamma\gamma}^{\rm Quad}
&=
\frac{e^2v^2}{\pi\Lambda^4}
\left[
\left(
({\rm Re}\,\mathcal C_{\tau\gamma})^2
-
({\rm Im}\,\mathcal C_{\tau\gamma})^2
\right)
\hat\tau
\ln\!\left(\frac{1+\hat\beta}{1-\hat\beta}\right)
+
2|\mathcal C_{\tau\gamma}|^2\hat\beta
\right],
\label{eq:gmgm2tata_xs_quad}
\end{align}
\end{subequations}
with
\begin{equation}
\hat\tau=\frac{m_\tau^2}{\hat s},
\qquad
\hat\beta=\sqrt{1-4\hat\tau}.
\end{equation}

\begin{figure}[htb]
\centering
\includegraphics[width=0.6\textwidth]{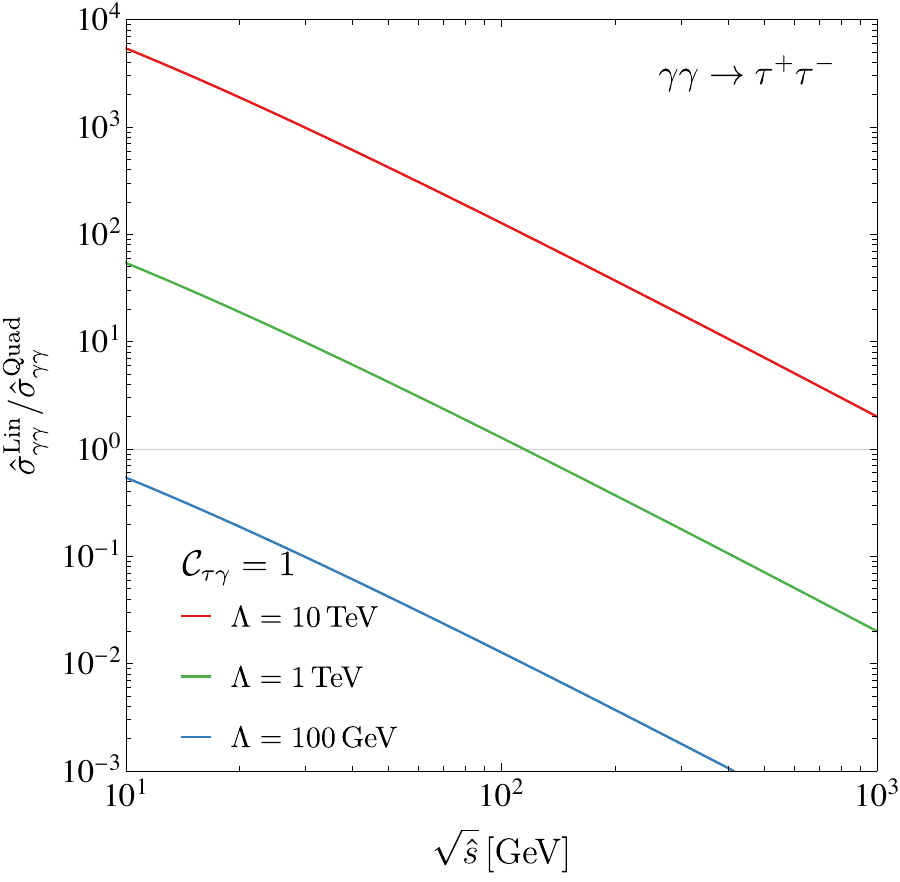}
\caption{
Ratio of the linear interference term to the quadratic BSM contribution, $\hat\sigma^{\rm Lin}_{\gamma\gamma}/\hat\sigma^{\rm Quad}_{\gamma\gamma}$, for the partonic process $\gamma\gamma\to\tau^+\tau^-$ as a function of $\sqrt{\hat s}$.
}
\label{fig:L2Q_aatt}
\end{figure}

\paragraph*{Threshold dominance and interference.}
Although the interference term in Eq.~\eqref{eq:gmgm2tata_xs_lin} is formally helicity suppressed by $m_\tau v/\hat s$, the photon luminosity strongly enhances the low-$\hat s$ region. The convolution in Eq.~\eqref{eq:sigma_EPA} is therefore dominated by $\hat s\gtrsim4m_\tau^2$, where $m_\tau/\sqrt{\hat s}$ is not a small parameter. This makes the linear interference contribution the leading source of sensitivity to $\Delta a_\tau$. The relative size of the interference and quadratic terms is shown in Fig.~\ref{fig:L2Q_aatt}. For values of $\Lambda$ probed by future $e^+e^-$ colliders, the signal in the threshold-dominated region is controlled by the linear term. This feature is the main reason why $\gamma\gamma\to\tau^+\tau^-$ provides the strongest projected bound on the anomalous magnetic moment.
The same threshold dominance makes the electric dipole sensitivity less competitive. Since the EDM contribution is extracted from the quadratic term and the SM background is large in the low-$\hat s$ region, the resulting bound on $d_\tau$ is weaker than the one obtained from direct $e^+e^-\to\tau^+\tau^-$ production at higher energies.

\paragraph*{EFT interpretation.}
The dominance of the linear dimension-six contribution in the relevant phase-space region also ensures the robustness of the dimension-six truncation. Quadratic dimension-six terms and interference effects from dimension-eight dipole operators are both suppressed by an additional power of $1/\Lambda^2$ with respect to the leading interference term in the range of Wilson coefficients probed below.

\subsubsection{Phenomenological analysis}
\label{sec:gmgm2tt_pheno}
We now quantify the FCC-ee sensitivity to the electromagnetic dipole moments from $e^+e^-\to e^+e^-\tau^+\tau^-$. To remain close to the LEP-II strategy used by DELPHI, we adopt the central scale $Q=E_T/2$ and do not impose additional kinematic cuts beyond those implicit in the EPA setup. The expected signal and background yields are computed as

\begin{equation}
\label{eq:gmgm_signal_background}
S=\epsilon_{\rm LEP-II}\,\mathcal L
\left[
\sigma_{\gamma\gamma\to\tau^+\tau^-}(\mathcal C_{\tau\gamma},s)
-
\sigma_{\gamma\gamma\to\tau^+\tau^-}^{\rm SM}(s)
\right],
\qquad
B=\epsilon_{\rm LEP-II}\,\mathcal L\,
\sigma_{\gamma\gamma\to\tau^+\tau^-}^{\rm SM}(s).
\end{equation}
Here $\epsilon_{\rm LEP-II}\simeq1\%$ is obtained from the average DELPHI efficiency in the 1997--1999 data-taking periods. We do not use the lower efficiencies reported in 2000, which were affected by the removal of events near unstable TPC sectors~\cite{DELPHI:2003nah}. This choice provides a conservative benchmark for FCC-ee projections; any improvement in the reconstruction efficiency would directly strengthen the sensitivity.

\begin{figure}[htb]
\centering
\includegraphics[width=0.311\textwidth]{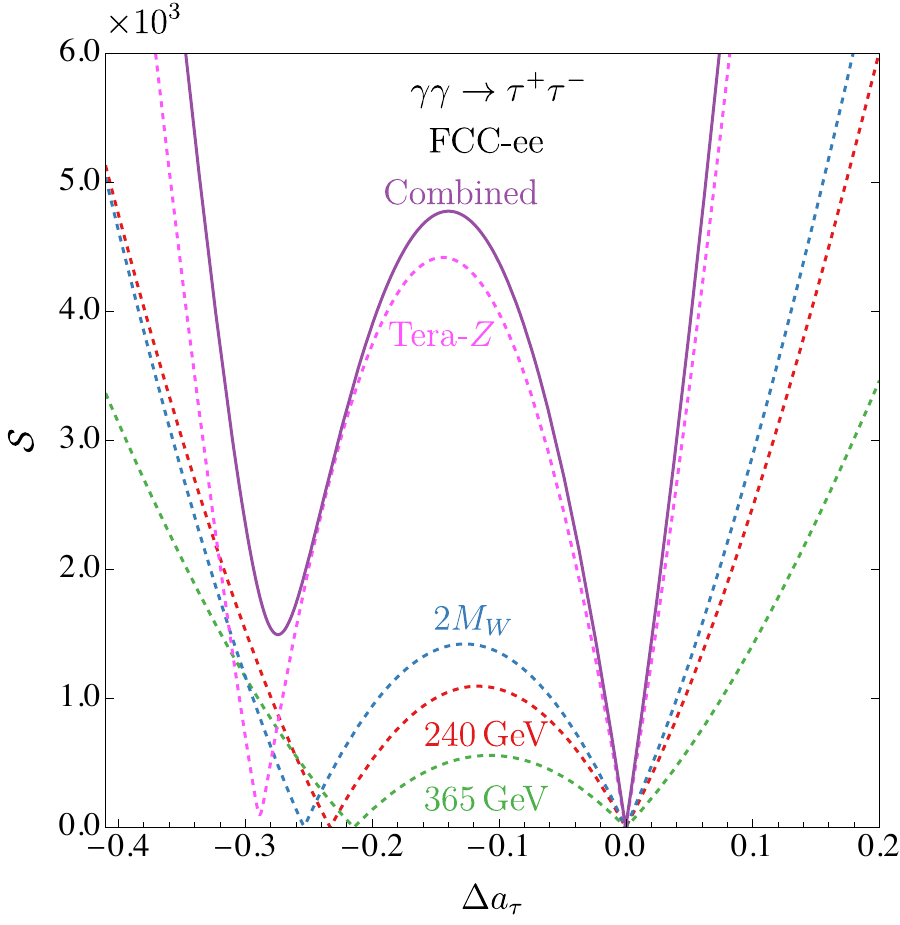}
\includegraphics[width=0.314\textwidth]{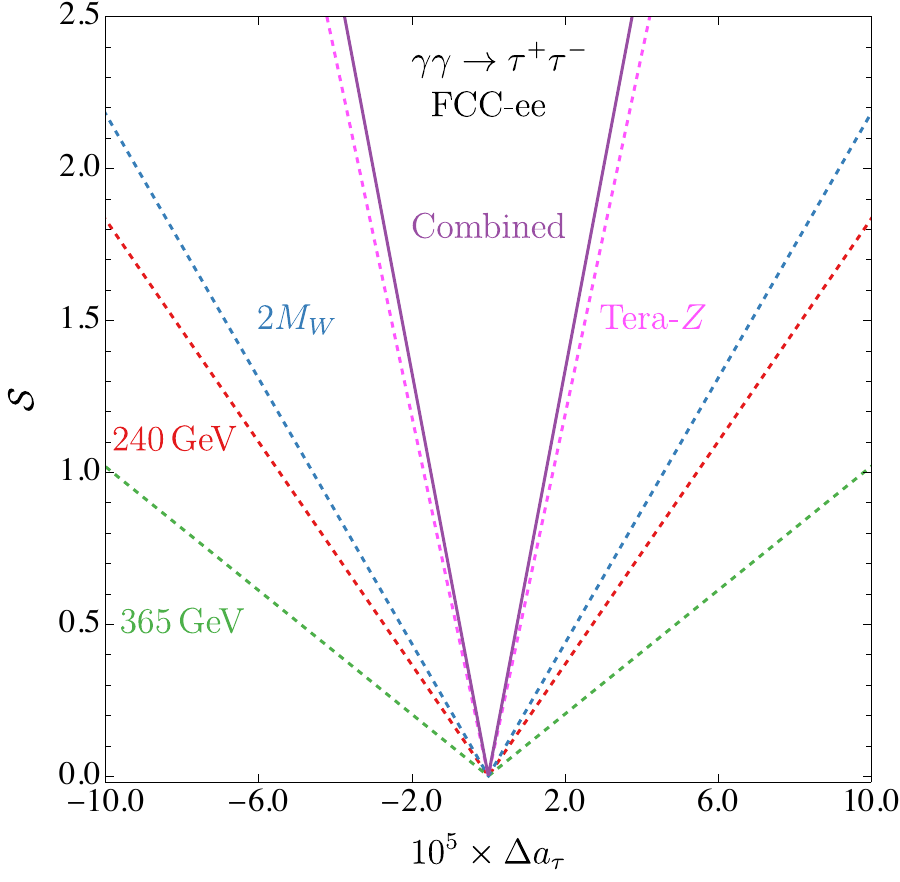}
\includegraphics[width=0.31\textwidth]{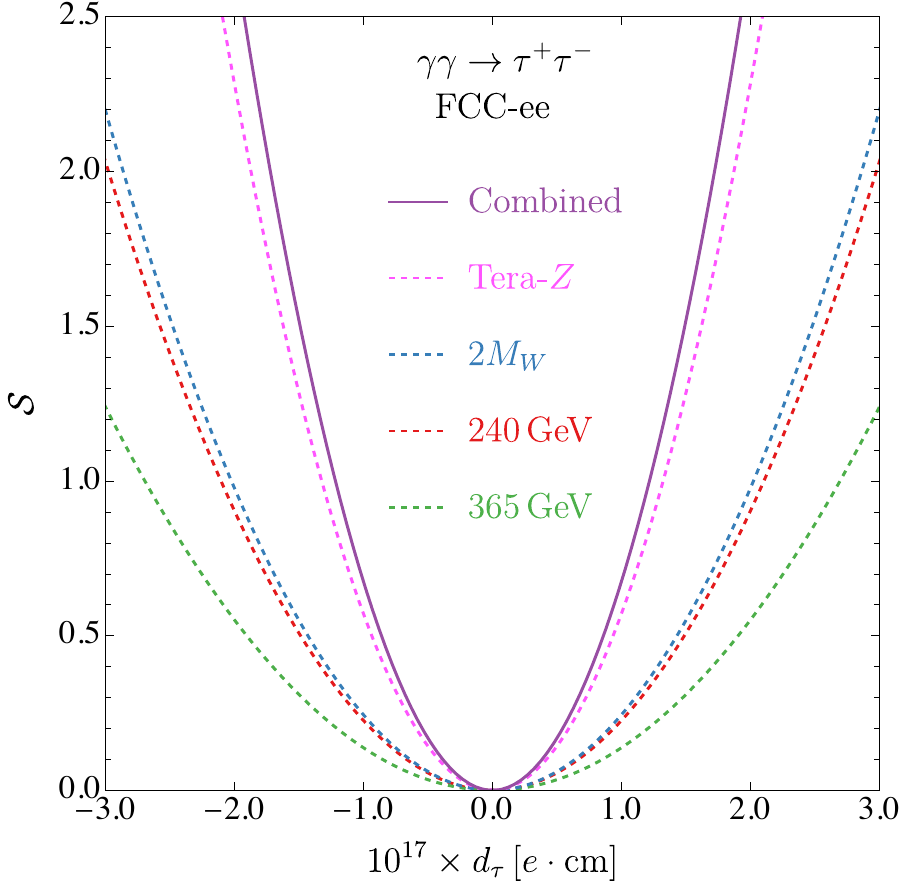}
\caption{
Statistical significance and 95\% CL sensitivity reach for the
$\tau$ electromagnetic dipole moments from photon-induced production at
FCC-ee. The left panel shows the significance over a broad range of
$\Delta a_\tau$, the middle panel zooms into the region close to the SM
point where the linear term dominates, and the right panel shows the
quadratic sensitivity to $d_\tau$. Dashed curves denote individual FCC-ee
stages, while the solid purple curve shows their combination.
}
\label{fig:sig_FCC_aa2tt}
\end{figure}
\paragraph*{Significance as a function of the dipole moments.}
Fig.~\ref{fig:sig_FCC_aa2tt} shows the statistical significance for the different FCC-ee operating stages. The left and middle panels refer to $\Delta a_\tau$ and are obtained from the full cross section, including both the interference and quadratic terms. The right panel shows the sensitivity to $d_\tau$, extracted from the quadratic contribution with a purely imaginary $\mathcal C_{\tau\gamma}$.
The broad scan in the left panel displays the characteristic double-dip
structure: besides the SM point, the significance also vanishes at a
non-zero value of $\Delta a_\tau$ where the linear and quadratic BSM
contributions cancel. 
The location of this cancellation is random and depends on the collision energy, so that combining measurements at different center-of-mass energies effectively lifts this degeneracy.
The middle panel focuses on the phenomenologically relevant region near the SM point. There the interference term dominates, and the significance grows approximately linearly with $|\Delta a_\tau|$. By contrast, the EDM contribution is purely quadratic in the Wilson coefficient, leading to the parabolic behaviour shown in the right panel and to a comparatively weaker sensitivity to $d_\tau$.

\begin{figure}[htb]
\centering
\includegraphics[width=0.45\textwidth]{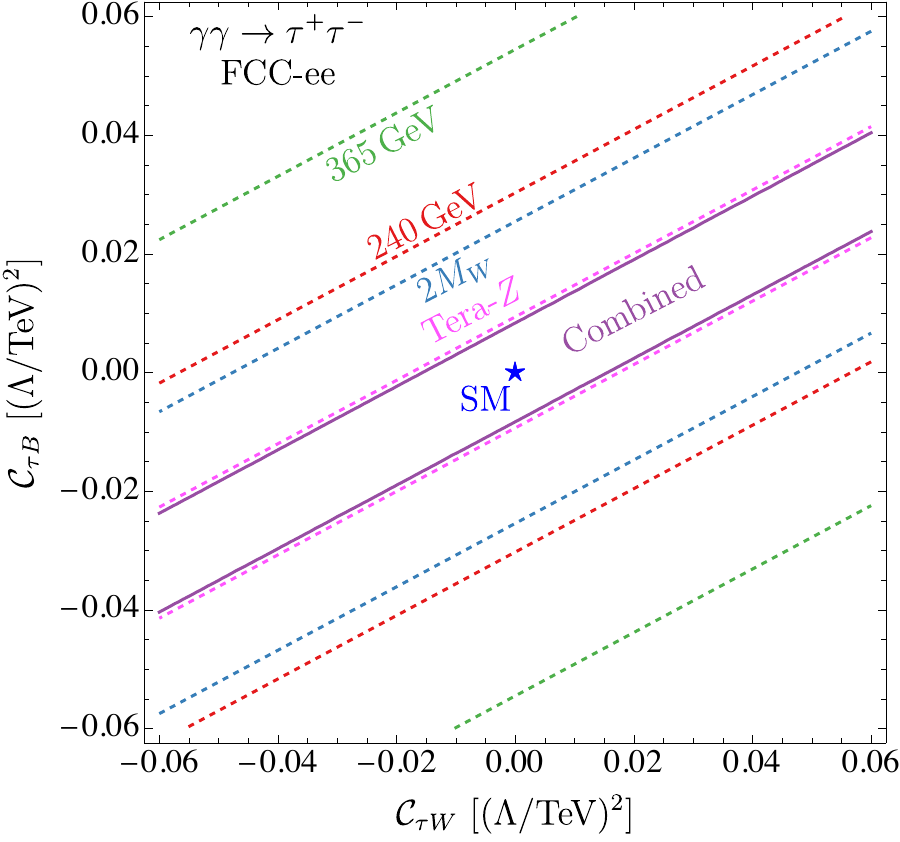}
\includegraphics[width=0.42\textwidth]{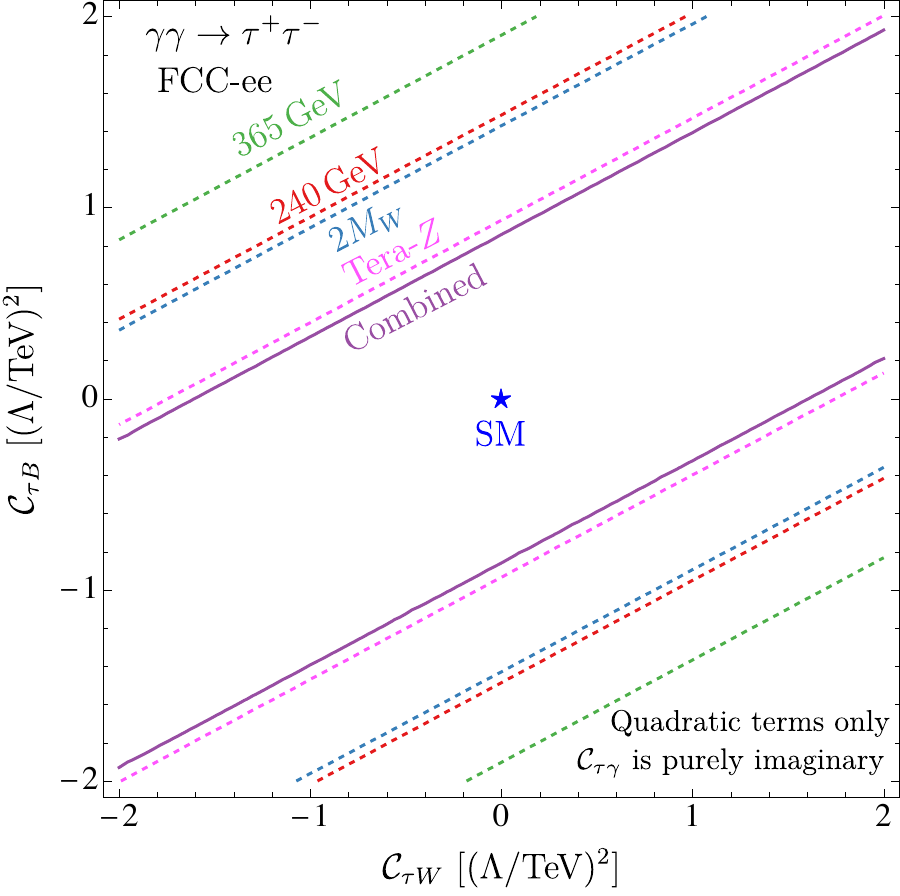}
\caption{
95\% CL sensitivity contours in the
$(\mathcal C_{\tau W},\mathcal C_{\tau B})$ plane from photon-induced
$\tau$-pair production at FCC-ee. The left panel shows the result from
the full cross section for real Wilson coefficients, while the right
panel shows the quadratic-only sensitivity for purely imaginary Wilson
coefficients. Dashed curves denote individual operating stages, the
solid purple curve denotes their combination, and the blue star marks
the SM point.
}
\label{fig:contours_FCC_aa2tt}
\end{figure}

\paragraph*{Wilson-coefficient interpretation.}
The same information can be represented in the $(\mathcal C_{\tau W},\mathcal C_{\tau B})$ plane, as shown in Fig.~\ref{fig:contours_FCC_aa2tt}. The photon-induced process is mainly sensitive to the electromagnetic combination $\mathcal C_{\tau\gamma}=c_W\mathcal C_{\tau B}-s_W\mathcal C_{\tau W}$. Consequently, the contours exhibit approximate flat directions along the orthogonal weak-dipole combination. We recall that the Wilson coefficients are here evaluated at the EW scale, and we are not including any running up to a higher scale where they might be generated.
The comparison between the two panels makes the role of the interference transparent. The full result for real coefficients is significantly more constraining than the quadratic-only case, confirming that the $\gamma\gamma$ channel is driven by the linear interference term and is therefore especially powerful for $\Delta a_\tau$. Conversely, the EDM sensitivity follows from the quadratic contribution alone.
The projected 95\% CL bounds are summarized in Table~\ref{tab:constraints_gmgm}. The Tera-$Z$ stage gives the strongest individual sensitivity because it provides the largest integrated luminosity, while the higher-energy stages add statistically independent information to the combined result.
\begin{table}[htb]
\centering
\renewcommand{\arraystretch}{1.18}
\setlength{\tabcolsep}{5pt}
\begin{tabular}{c|ccccc}
\hline
$\sqrt{s}$
& $M_Z$
& $2M_W$
& $240$ GeV
& $365$ GeV
& Combined
\\
\hline
$|\Delta a_\tau|$          & $3.4\times10^{-5}$  & $9.2\times10^{-5}$  & $1.1\times10^{-4}$  & $2.0\times10^{-4}$  & $3.0\times10^{-5}$  \\
$|d_\tau|$ [$e\,{\rm cm}$] & $1.9\times10^{-17}$ & $2.9\times10^{-17}$ & $3.0\times10^{-17}$ & $3.8\times10^{-17}$ & $1.7\times10^{-17}$ \\
\hline
\end{tabular}
\caption{
Projected $95\%$ CL constraints on $|\Delta a_\tau|$ and $|d_\tau|$ from
$\gamma\gamma\to\tau^+\tau^-$ at FCC-ee. Bounds on the magnetic dipole
moment assume a real $\mathcal{C}_{\tau\gamma}$, while those on the EDM
assume a purely imaginary coefficient.}
\label{tab:constraints_gmgm}
\end{table}

\paragraph*{Luminosity scaling and comparison with existing bounds.}
Since the photon-induced cross section is dominated by low $m(\tau^+\tau^-)$ events, increasing the nominal collider energy does not generate a sizeable energy enhancement. The reach is therefore driven mainly by luminosity. In the linear regime relevant for $\Delta a_\tau$, the sensitivity scales approximately as $\mathcal L^{-1/2}$, while for the EDM, which enters quadratically, it scales as $\mathcal L^{-1/4}$:
\begin{subequations}
\label{eq:scale_adtau}
\begin{align}
|\Delta a_\tau^{\mathcal A}|
&\simeq
|\Delta a_\tau^{\mathcal B}|
\left(\frac{\mathcal L_{\mathcal B}}{\mathcal L_{\mathcal A}}\right)^{1/2},
\label{eq:scale_atau}
\\[1mm]
|d_\tau^{\mathcal A}|
&\simeq
|d_\tau^{\mathcal B}|
\left(\frac{\mathcal L_{\mathcal B}}{\mathcal L_{\mathcal A}}\right)^{1/4}.
\label{eq:scale_dtau}
\end{align}
\end{subequations}
These scaling relations reproduce the full numerical results to within about ten percent between neighbouring FCC-ee stages, where the photon-photon partonic luminosities are similar. Extrapolating from the 240 GeV run to the LEP-II luminosity gives $|\Delta a_\tau|\lesssim 0.017$ and $|d_\tau|\lesssim 3.5\times10^{-16}\,e\,\mathrm{cm}$, in good agreement with the DELPHI constraints. This agreement provides an additional validation of the projection strategy.

The same scaling also clarifies why current LHC constraints, although very powerful, do not automatically improve dramatically with luminosity. They are obtained in high-energy regions with large $m(\tau^+\tau^-)$ and large $p_T(\tau^\pm)$, where the helicity-suppressed interference is small and the BSM contribution is effectively dominated by the quadratic term. A luminosity extrapolation to HL-LHC therefore leads only to an improvement of order a factor of two.

The photon-induced channel leads to three main conclusions. First, it provides the best FCC-ee sensitivity to the anomalous magnetic moment $\Delta a_\tau$, because the photon luminosity selects the threshold region where the linear interference term is dominant. Second, it is almost exclusively sensitive to the electromagnetic dipole combination $C_{\tau\gamma}$, leaving a flat direction along the weak-dipole combination. Finally, the same threshold dominance that makes the channel powerful for $\Delta a_\tau$ limits its competitiveness for $d_\tau$, for which direct $e^+e^-\to\tau^+\tau^-$ production remains the most effective probe.

\subsection{\boldmath{$e^+e^-\to\tau^+\tau^-\gamma$}}
\label{sec:ee2tta}

Radiative $\tau$-pair production provides a complementary probe of the dipole operators. Compared with the $e^+e^-\to\tau^+\tau^-$ channel, the additional hard photon gives access to different kinematic configurations and makes it possible to test the electromagnetic and weak dipole interactions in a more exclusive final state. As will be shown below, this channel is less competitive than $\tau$-pair production, but it provides an important consistency check of the SMEFT interpretation and of the relative role of the photon and $Z$ dipole operators.

\subsubsection{SMEFT description}
\begin{figure}[htb]
\centering
\includegraphics[width=0.22\textwidth]{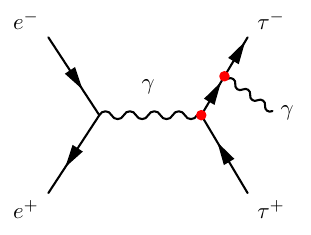}
\includegraphics[width=0.22\textwidth]{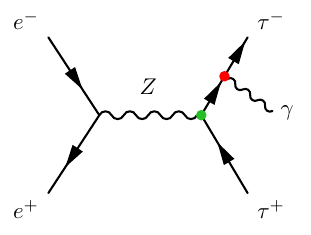}
\includegraphics[width=0.22\textwidth]{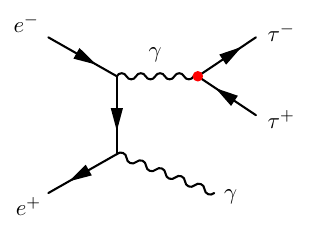}
\includegraphics[width=0.22\textwidth]{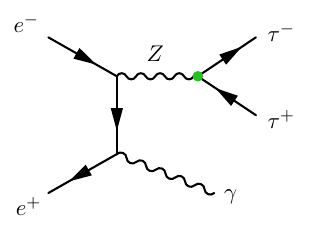}
\caption{
Same as Fig.~\ref{fig:FD_ee2tt} but for $e^+e^- \to \tau^+\tau^-\gamma$. 
}
\label{fig:FD_ee2tta}
\end{figure}

Representative leading-order diagrams are shown in Fig.~\ref{fig:FD_ee2tta}. The red and green nodes denote insertions of the electromagnetic and weak dipole operators, $\mathcal O_{\tau\gamma}$ and $\mathcal O_{\tau Z}$, respectively. The same topologies without these insertions give the SM amplitudes. The process therefore probes both Wilson-coefficient combinations introduced in Eq.~\eqref{eq:CgammaCz}, although with a different kinematic weighting than in the two-body channel.

Near the $Z$ pole, the dominant contribution is the radiative decay of a nearly on-shell $Z$ boson into the $\tau^+\tau^-\gamma$ final state. Photon-exchange diagrams are not resonantly enhanced, while configurations corresponding to associated $\gamma Z$ production can be reduced by requiring the observed photon to be sufficiently hard. In practice, in addition to the common angular and isolation requirements defined in Sec.~\ref{sec:pheno}, we impose
\begin{equation}
\label{eq:ee2tta_cuts}
m(\tau^+\tau^-\gamma)>0.8\sqrt{s},
\qquad
p_T(\tau^\pm,\gamma)>0.1\sqrt{s}.
\end{equation}
These cuts select events in the hard scattering regime and suppress soft or radiative-return configurations that do not efficiently probe the dipole vertices.

For the numerical analysis we organize the cross section as
\begin{equation}
\label{eq:ee2tta_xs}
\sigma_{e^+e^-\to\tau^+\tau^-\gamma} =
\sigma_{\tau\tau\gamma}^{\rm SM} + \sigma_{\tau\tau\gamma}^{\rm Lin} + \sigma_{\tau\tau\gamma}^{\rm Quad},
\end{equation}
where the second term denotes the interference between the SM and the dimension-six dipole amplitudes, and the last term denotes the quadratic dimension-six contribution. As in the $e^+e^-\to\tau^+\tau^-$ process, the CP-even dipole moments are constrained using the full expression, while the CP-odd dipole moments are extracted from the quadratic contribution.

\paragraph*{Structure of the amplitudes.}
The radiative final state receives contributions from both final-state-radiation topologies, in which the photon is emitted from the $\tau$ line, and initial-state-radiation topologies, in which it is emitted from the incoming electron or positron. At the $Z$ pole, the former are directly sensitive to the dipole-modified $\tau\tau\gamma$ and $\tau\tau Z$ vertices. Away from the pole, initial-state radiation can instead bring the invariant mass of the $\tau^+\tau^-$ pair close to $M_Z$, producing an on-shell $Z$ boson accompanied by a hard photon. This feature will be used below to separate regions dominantly sensitive to the weak and electromagnetic dipole operators.

The radiative channel was already considered in LEP analyses of the $\tau$ electromagnetic dipole moments~\cite{Grifols:1990ha,L3:1998gov,OPAL:1998dsa}. Here we extend this strategy by treating the electromagnetic and weak dipole interactions simultaneously within the SMEFT framework.

\subsubsection{Phenomenological analysis}
The phenomenological analysis follows the same statistical procedure and ISR treatment described in Sec.~\ref{sec:pheno} and applied to  the $e^+e^-\to\tau^+\tau^-$ channel. Since the radiative process contains an additional photon, its total rate is smaller than the non-radiative one. Its main role is therefore not to provide the leading bounds, but to test the same dipole operators in a different final state and with different mixtures of photon- and $Z$-mediated amplitudes.

\begin{figure}[htb]
\centering
\includegraphics[width=0.42\textwidth]{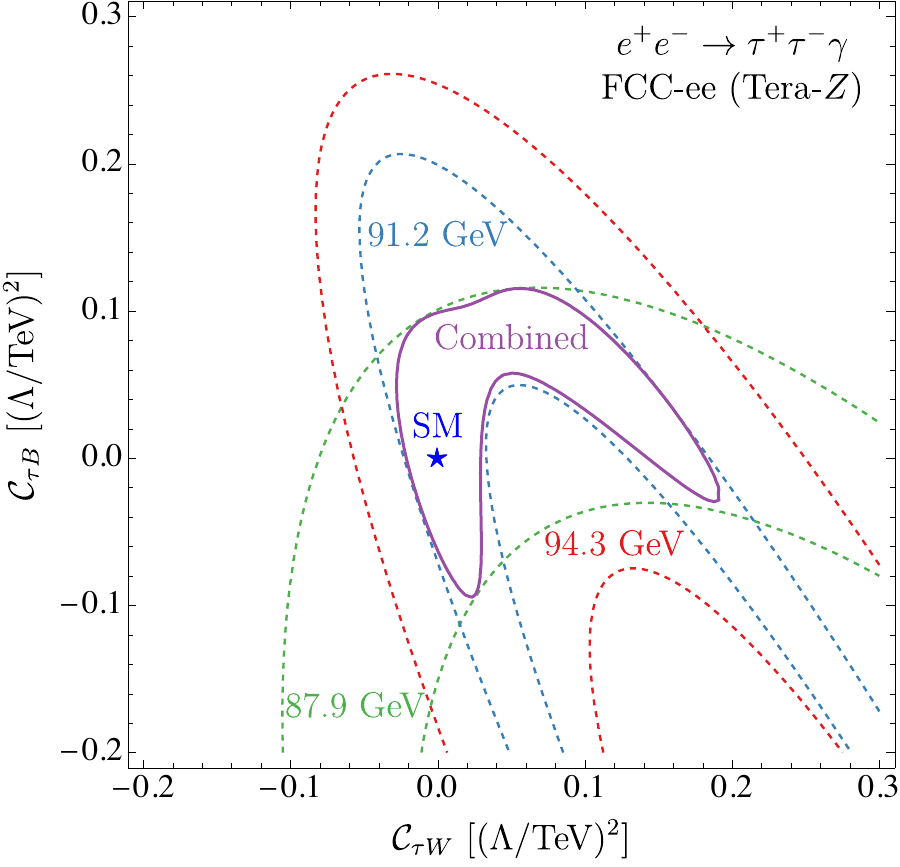}
\includegraphics[width=0.42\textwidth]{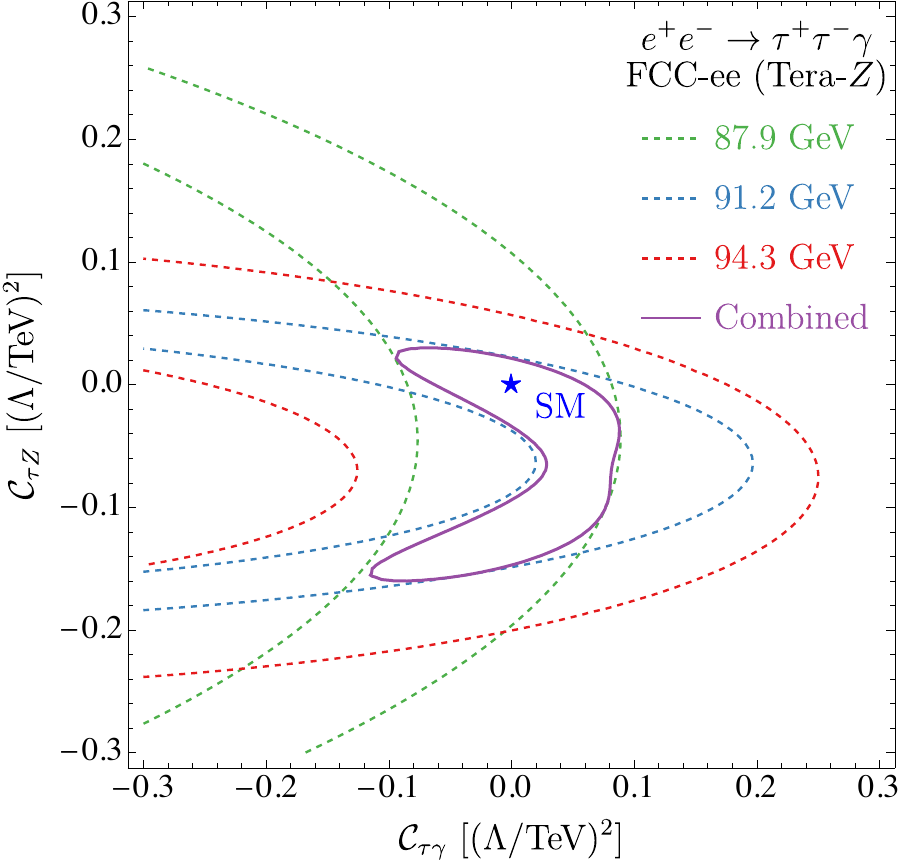}
\includegraphics[width=0.42\textwidth]{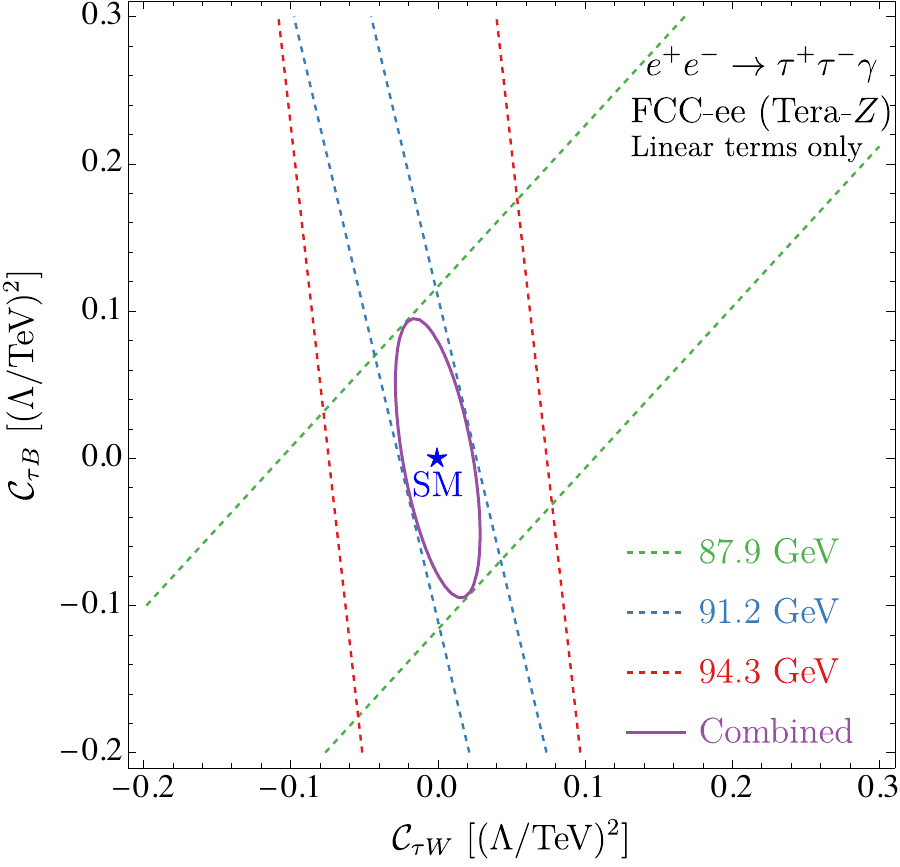}
\includegraphics[width=0.42\textwidth]{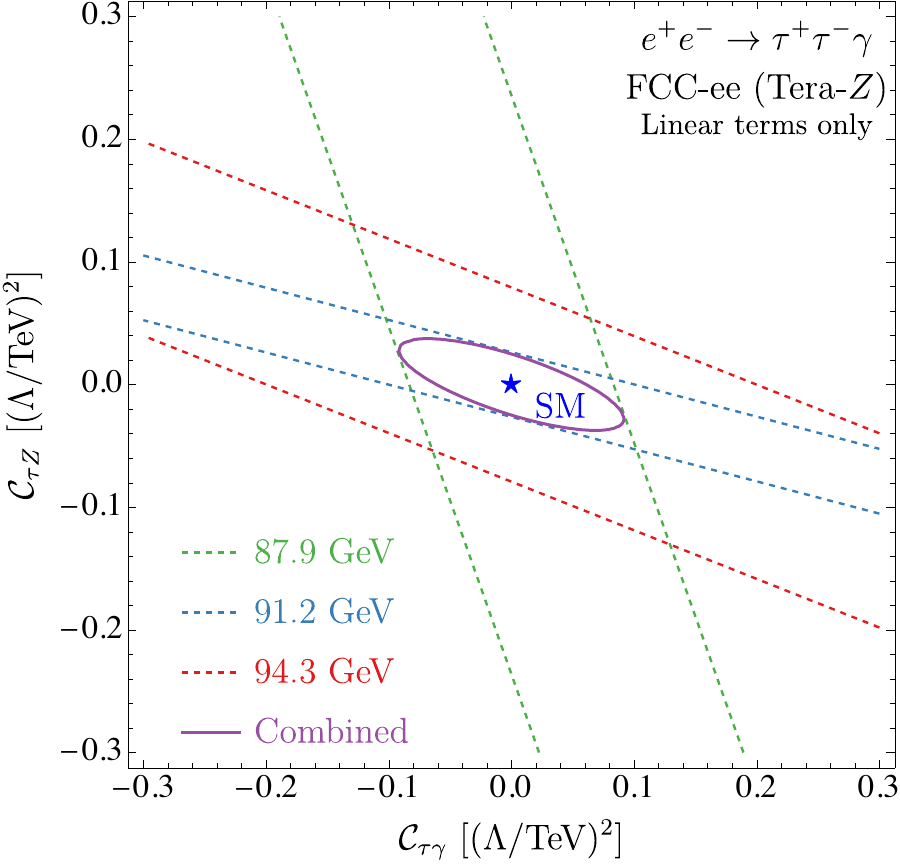}
\includegraphics[width=0.42\textwidth]{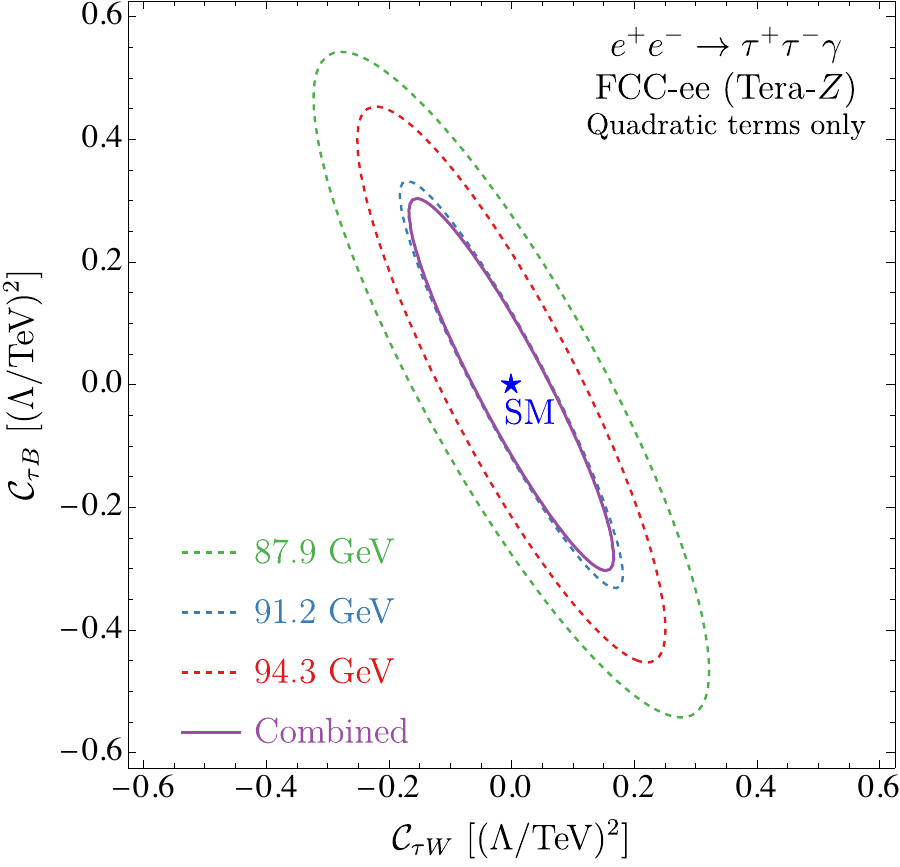}
\includegraphics[width=0.42\textwidth]{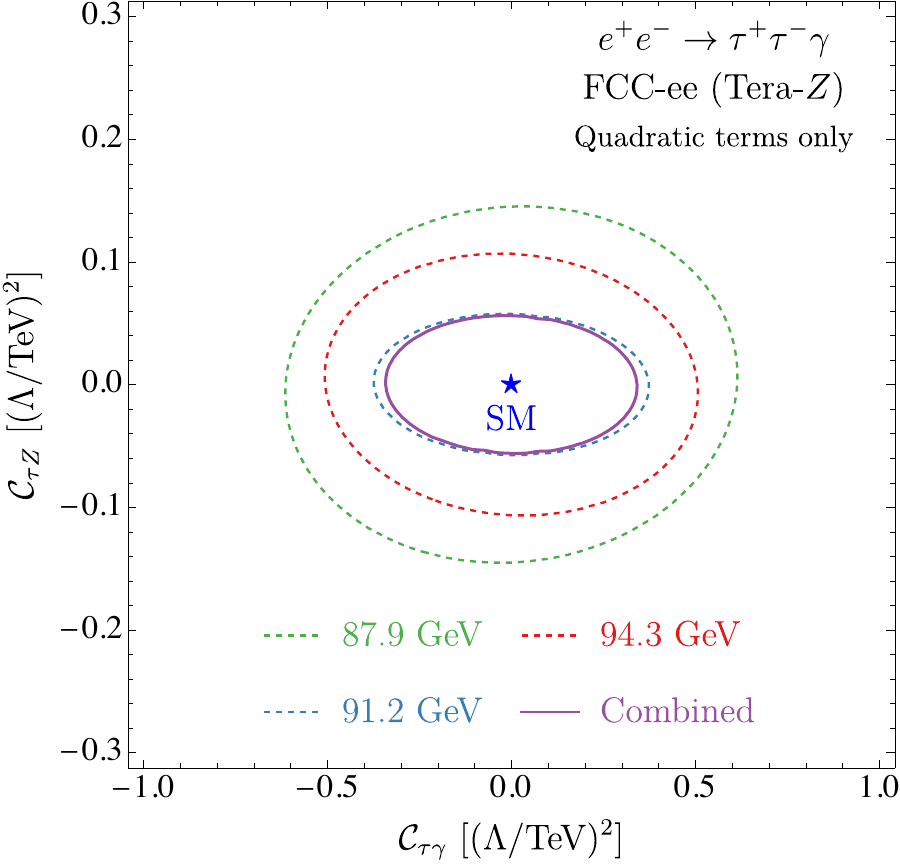}
\caption{
Expected $95\%$ CL sensitivity to the $\tau$ dipole Wilson coefficients from $e^+e^-\to \tau^+\tau^-\gamma$ at the FCC-ee Tera-$Z$ stage. The layout follows Fig.~\ref{fig:ee_3TeraZ}: from top to bottom the columns show the full dimension-six prediction, the linear interference contribution only, and the quadratic BSM contribution only. Dashed contours denote individual center-of-mass energies, while solid contours show their combination.}
\label{fig:contours_ZF_ee2tta}
\end{figure}

\paragraph*{Tera-$Z$ sensitivity.}
At the Tera-$Z$ stage, ISR distorts the resonance in the same way as for the $e^+e^-\to\tau^+\tau^-$ channel. This effect was illustrated in Fig.~\ref{fig:ee_ISR},  which also displays the $e^+e^-\to\tau^+\tau^-\gamma$ cross section around the $Z$ pole. We therefore use the ISR-improved cross sections and the kinematic selection in Eq.~\eqref{eq:ee2tta_cuts} to derive the projected sensitivity.
Fig.~\ref{fig:contours_ZF_ee2tta} shows the resulting $95\%$ CL contours at the three Tera-$Z$ operating points and their combination. The left column is displayed in the gauge basis $(\mathcal C_{\tau W},\mathcal C_{\tau B})$, while the right column shows the corresponding constraints in the physical $(\mathcal C_{\tau\gamma},\mathcal C_{\tau Z})$ basis. As in Fig.~\ref{fig:ee_3TeraZ}, the three rows correspond to the complete SMEFT prediction, the isolated linear interference term, and the pure quadratic contribution.

\paragraph*{Complete SMEFT prediction.}
The full-expression contours exhibit the same qualitative structure encountered in the $e^+e^-\to\tau^+\tau^-$ channel, but with more pronounced non-linear features. At a fixed energy, the interference and quadratic terms can partially cancel, producing ring-like or banana-shaped allowed regions. Combining the three Tera-$Z$ points reduces the degeneracy because the relative photon--$Z$ contribution changes across the resonance. A residual secondary branch remains away from the SM point, reflecting the reduced constraining power of the radiative channel compared with $e^+e^-\to\tau^+\tau^-$. A more differential analysis of the photon and $\tau$ kinematics could further reduce this degeneracy.

\paragraph*{Linear and quadratic contributions.}
The linear contribution remains important at the $Z$ pole, but it does not dominate as cleanly as in the $e^+e^-\to\tau^+\tau^-$ channel. This is visible from the difference between the full contours and the linear-only contours in Fig.~\ref{fig:contours_ZF_ee2tta}. The quadratic contribution controls the sensitivity to the CP-odd dipole moments and also determines the approximate size of the residual flat directions in the CP-even case.

The marginalized constraints obtained from the three Tera-$Z$ energies are
\begin{equation}
\label{eq:constraints_eetta_all}
\begin{aligned}
\Delta a_\tau
&\in [-4.7,\,3.6]\times10^{-4},
&
|d_\tau|
&<7.8\times10^{-18}\ e\,{\rm cm},
\\
\Delta a_\tau^Z
&\in [-6.4,\,1.2]\times10^{-4},
&
|d_\tau^Z|
&<1.3\times10^{-18}\ e\,{\rm cm}.
\end{aligned}
\end{equation}
The corresponding one-operator-at-a-time limits are reported in Table~\ref{tab:constraints_eetta_ZF}. As expected, the radiative channel is less sensitive than the $\tau^+\tau^-$ mode, but it probes the same SMEFT coefficients through a distinct set of kinematic configurations.

\begin{table}[htb]
\centering
\renewcommand{\arraystretch}{1.18}
\setlength{\tabcolsep}{5pt}
\begin{tabular}{c|cccccc}
\hline
$\sqrt{s}$ [GeV]
& Max$(\Delta a_\tau)$
& Min$(\Delta a_\tau)$
& Max$(\Delta a^Z_\tau)$
& Min$(\Delta a^Z_\tau)$
& $|d_\tau|$ [$e\,{\rm cm}$]
& $|d^Z_\tau|$ [$e\,{\rm cm}$]
\\
\hline
$87.9$   & $3.3\times10^{-4}$ & $-3.5\times10^{-4}$ & $4.3\times10^{-4}$ & $-7.9\times10^{-4}$ & $1.4\times10^{-17}$ & $3.3\times10^{-18}$ \\
$91.2$   & $3.8\times10^{-4}$ & $-4.5\times10^{-4}$ & $9.0\times10^{-5}$ & $-1.5\times10^{-4}$ & $8.5\times10^{-18}$ & $1.3\times10^{-18}$ \\
$94.3$   & $7.2\times10^{-4}$ & $-1.0\times10^{-3}$ & $2.3\times10^{-4}$ & $-8.1\times10^{-4}$ & $1.2\times10^{-17}$ & $2.4\times10^{-18}$ \\
Combined & $2.4\times10^{-4}$ & $-2.6\times10^{-4}$ & $8.6\times10^{-5}$ & $-1.4\times10^{-4}$ & $7.8\times10^{-18}$ & $1.3\times10^{-18}$ \\
\hline
\end{tabular}
\caption{
Projected $95\%$ CL one-operator-at-a-time constraints on the $\tau$
dipole moments from $e^+e^-\to\tau^+\tau^-\gamma$ at the FCC-ee
Tera-$Z$ stage. Bounds on the magnetic dipole moments assume real
Wilson coefficients, while those on the electric dipole moments assume
purely imaginary coefficients.}
\label{tab:constraints_eetta_ZF}
\end{table}

\paragraph*{Higher-energy operation.}
At the higher-energy FCC-ee stages, radiative $\tau$-pair production contains two qualitatively different event topologies. If the photon is emitted from the initial state, the invariant mass of the $\tau^+\tau^-$ system can return to the $Z$ pole, corresponding to $e^+e^-\to\gamma Z$ followed by $Z\to\tau^+\tau^-$. If instead the event is away from the $Z$ peak, the sensitivity is more directly associated with the electromagnetic dipole interaction. We therefore separate the two regions through
\begin{equation}
\label{eq:ee2tta_mtt_cut}
|m(\tau^+\tau^-)-M_Z|<10~{\rm GeV},
\qquad
|m(\tau^+\tau^-)-M_Z|>10~{\rm GeV},
\end{equation}
which select, respectively, $Z$-enhanced and off-shell configurations. The remaining kinematic requirements are kept as in Eq.~\eqref{eq:ee2tta_cuts}.

\begin{figure}[htb]
\includegraphics[width=0.4\textwidth]{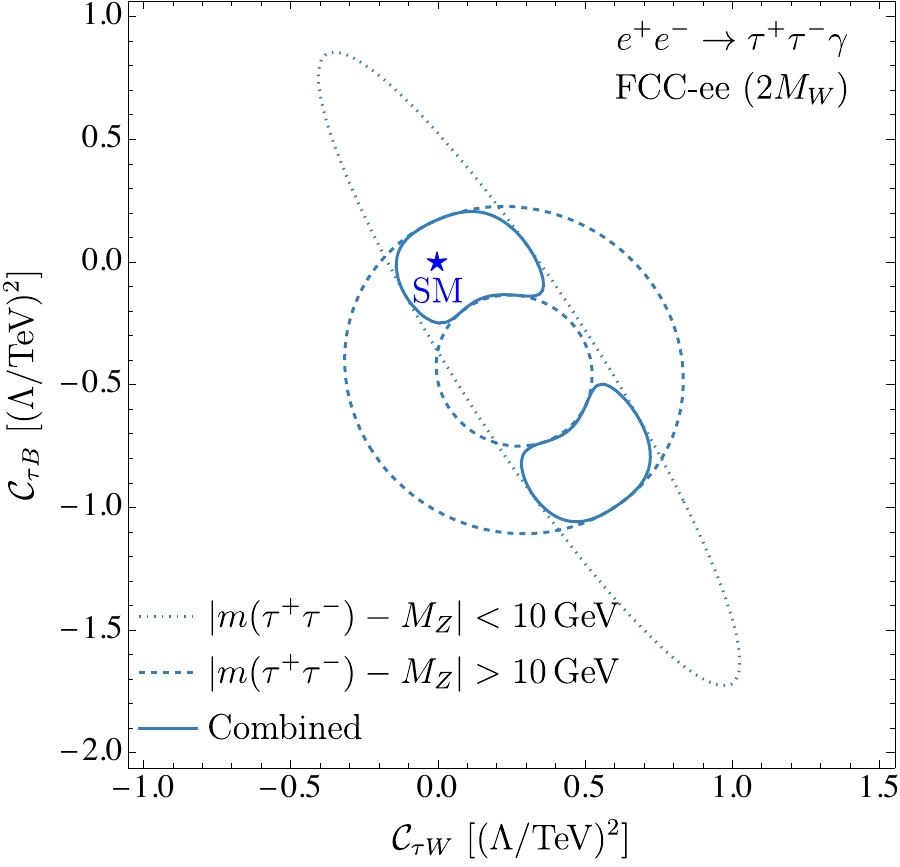}
\includegraphics[width=0.4\textwidth]{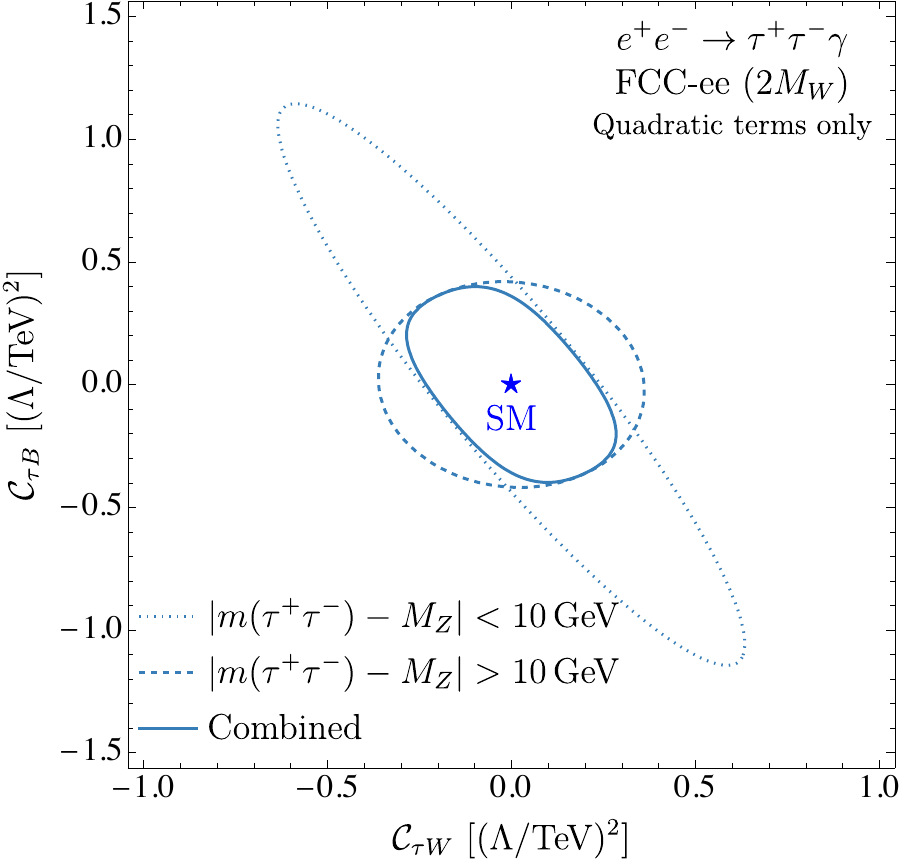}
\includegraphics[width=0.4\textwidth]{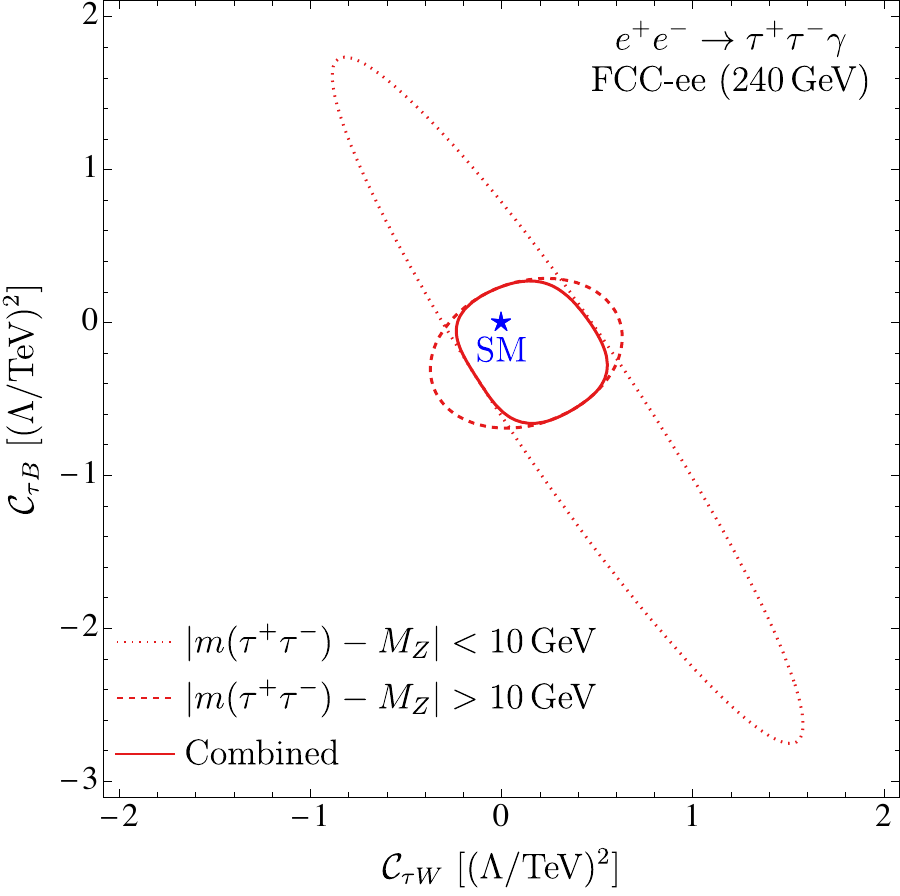}
\includegraphics[width=0.4\textwidth]{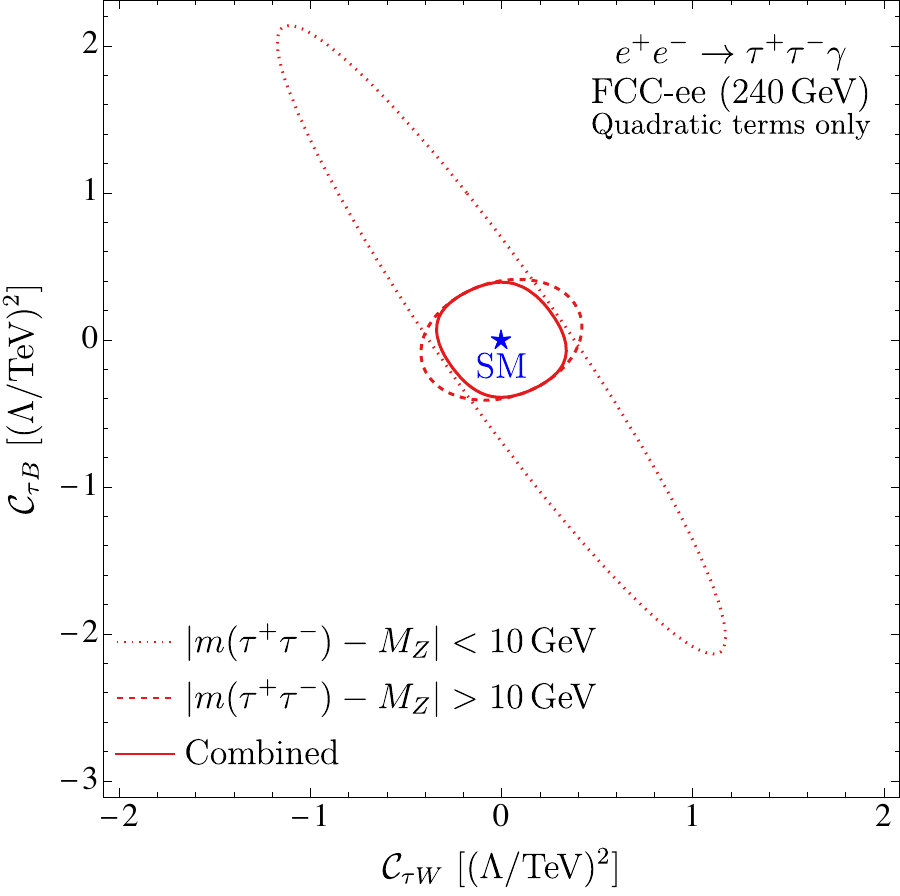}
\includegraphics[width=0.4\textwidth]{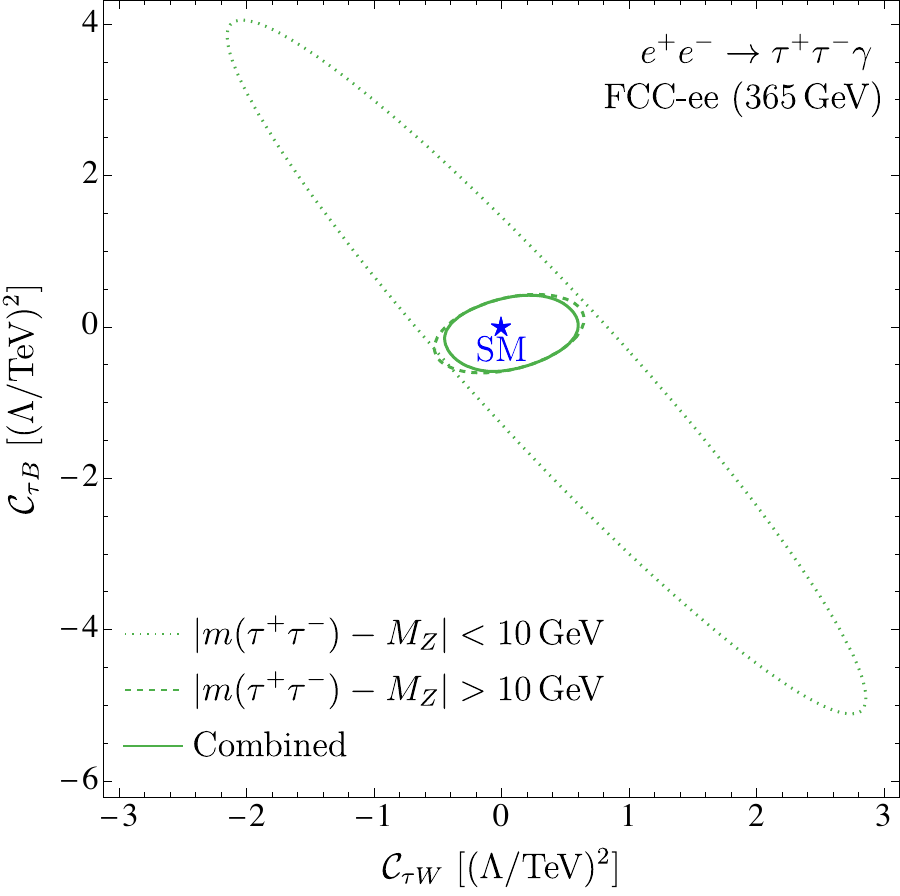}
\includegraphics[width=0.4\textwidth]{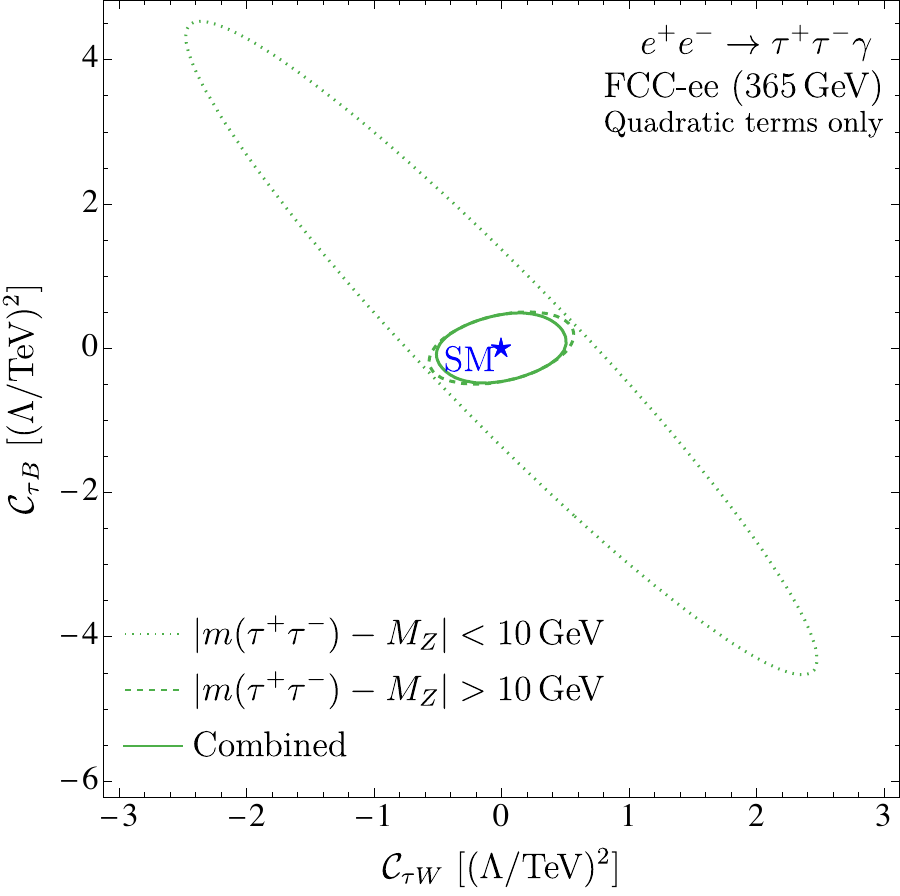}
\caption{
Expected 95\% CL sensitivity to the $\tau$ dipole Wilson coefficients in the $(\mathcal{C}_{\tau W},\mathcal{C}_{\tau B})$ plane from $e^+e^-\to\tau^+\tau^-\gamma$ at the higher-energy FCC-ee operating stages ($\sqrt{s}=2M_W$, 240 GeV, and 365 GeV). Dotted contours correspond to the ``on-$Z$-pole'' selection, $|m(\tau^+\tau^-)-M_Z|<10$ GeV, while dashed contours represent the complementary ``off-pole'' region, $|m(\tau^+\tau^-)-M_Z|>10$ GeV. The solid contours show the statistical combination of the two selections. The upper panels display the full dimension-six prediction, whereas the lower panels show only the quadratic dimension-six contribution.
} 
\label{fig:contours_FCC_ee2tta1}
\end{figure}

\begin{figure}[htb]
\includegraphics[width=0.43\textwidth]{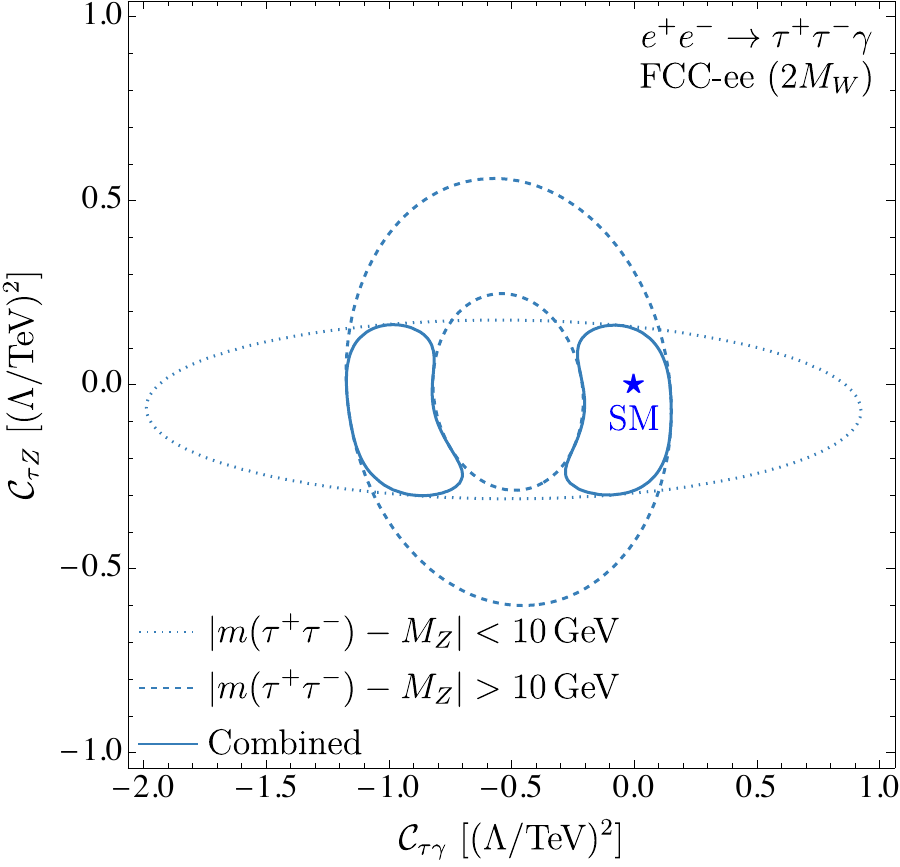}
\includegraphics[width=0.43\textwidth]{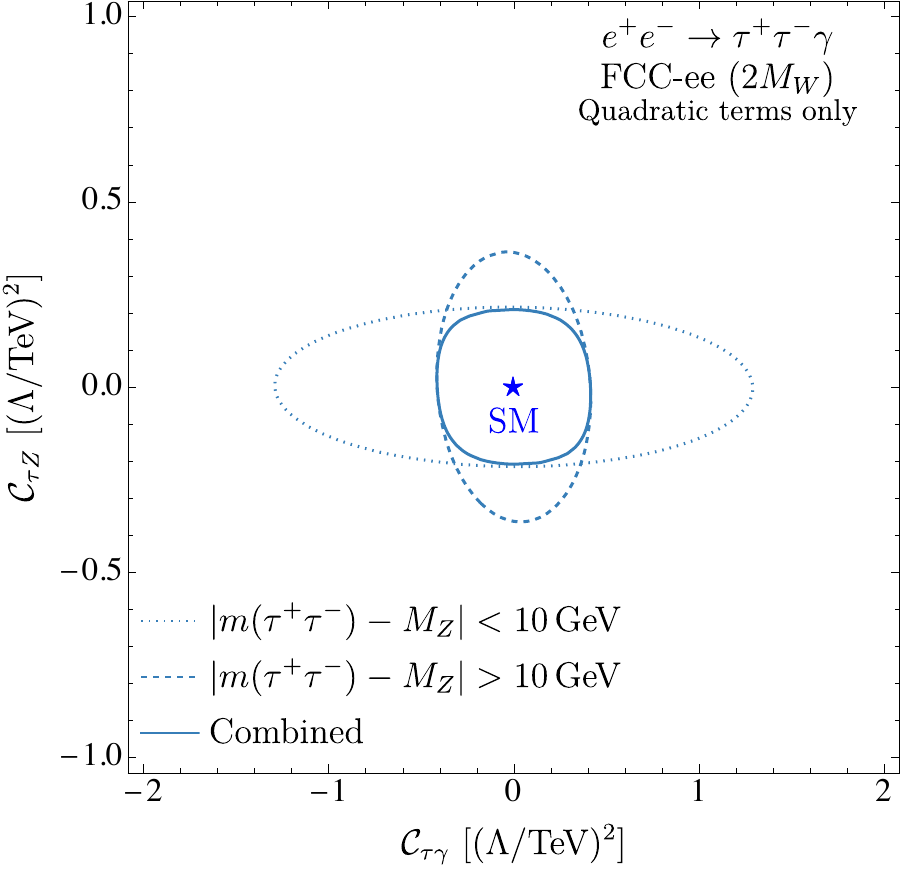}
\includegraphics[width=0.43\textwidth]{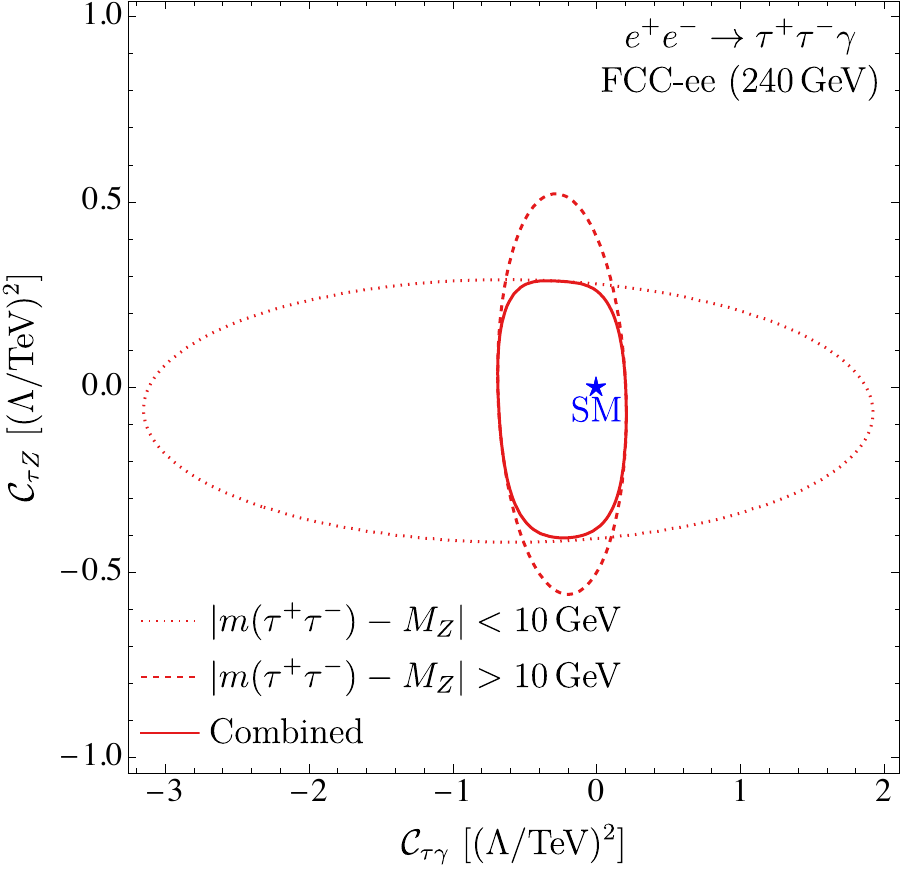}
\includegraphics[width=0.43\textwidth]{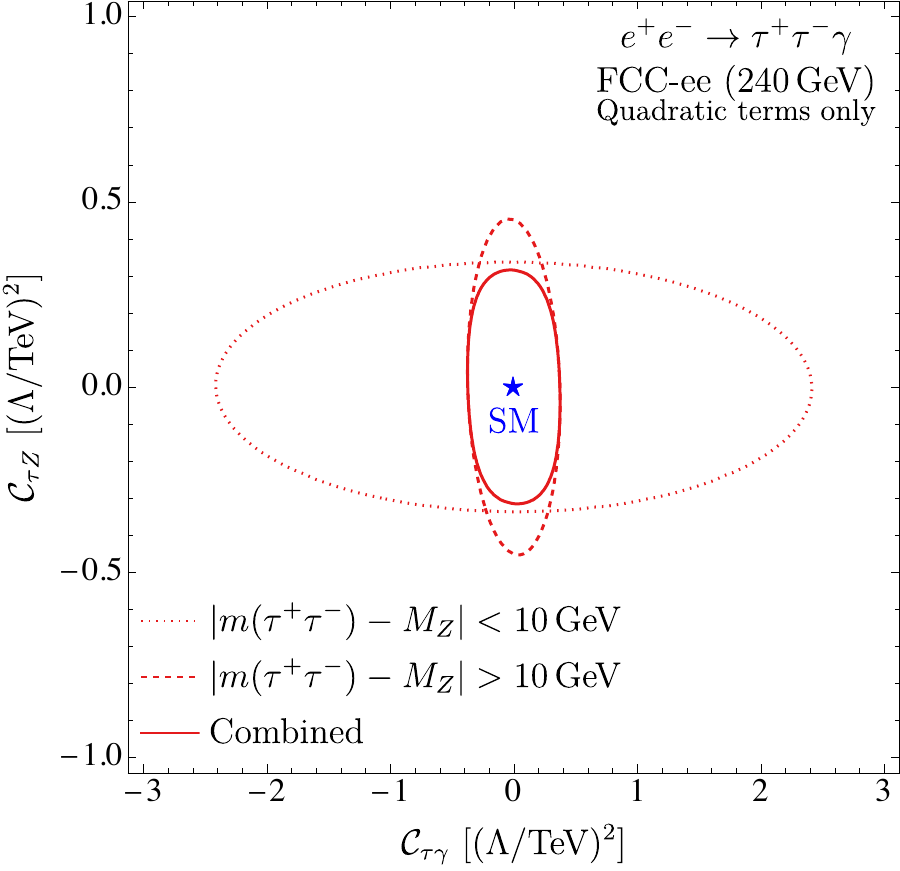}
\includegraphics[width=0.43\textwidth]{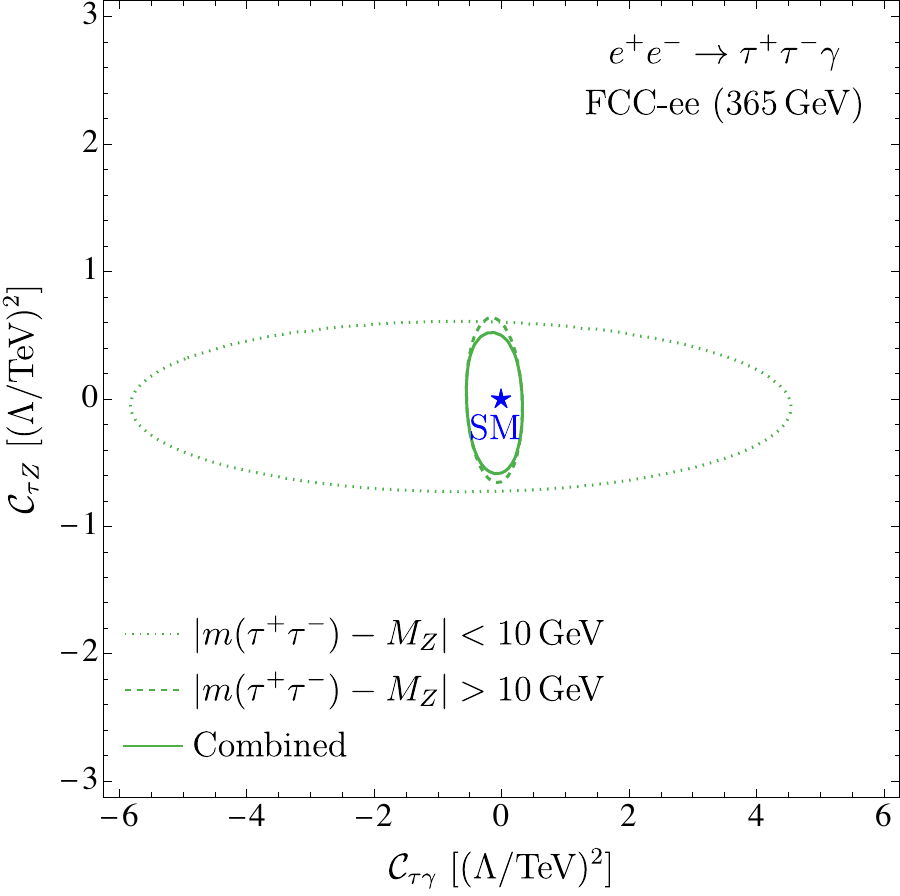}
\includegraphics[width=0.43\textwidth]{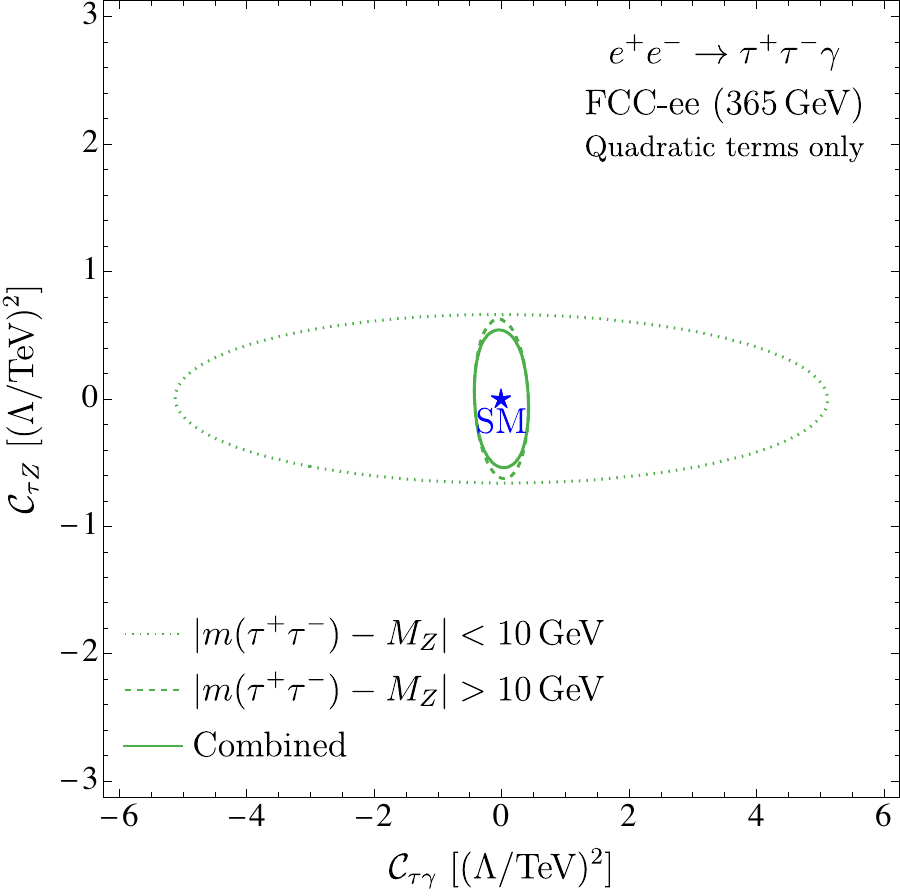}
\caption{Same as Fig.~\ref{fig:contours_FCC_ee2tta1} but in the $(\mathcal{C}_{\tau \gamma}, \mathcal{C}_{\tau Z})$ plane.} 
\label{fig:contours_FCC_ee2tta2}
\end{figure}

The detailed sensitivity contours for the individual higher-energy FCC-ee operating stages are shown in Figs.~\ref{fig:contours_FCC_ee2tta1} and \ref{fig:contours_FCC_ee2tta2}. The projected 95\% CL sensitivity contours are shown in the $(\mathcal{C}_{\tau W},\mathcal{C}_{\tau B})$ plane in Fig.~\ref{fig:contours_FCC_ee2tta1} and in the $(\mathcal{C}_{\tau\gamma},\mathcal{C}_{\tau Z})$ plane in Fig.~\ref{fig:contours_FCC_ee2tta2}. The left column displays the full dimension-six prediction, which determines the sensitivity to the CP-even dipole moments $\Delta a_\tau^{(Z)}$, while the right column shows the quadratic dimension-six contribution, from which the sensitivities to the CP-odd dipole moments $d_\tau^{(Z)}$ are obtained. The dotted and dashed contours correspond to the on- and off-pole running points, respectively, whereas the solid contours show their statistical combination. As expected, the off-pole measurements provide stronger sensitivity to $\mathcal{O}_{\tau\gamma}$, while the on-pole data are primarily sensitive to $\mathcal{O}_{\tau Z}$.


\begin{figure}[tb]
\centering
\includegraphics[width=0.44\textwidth]{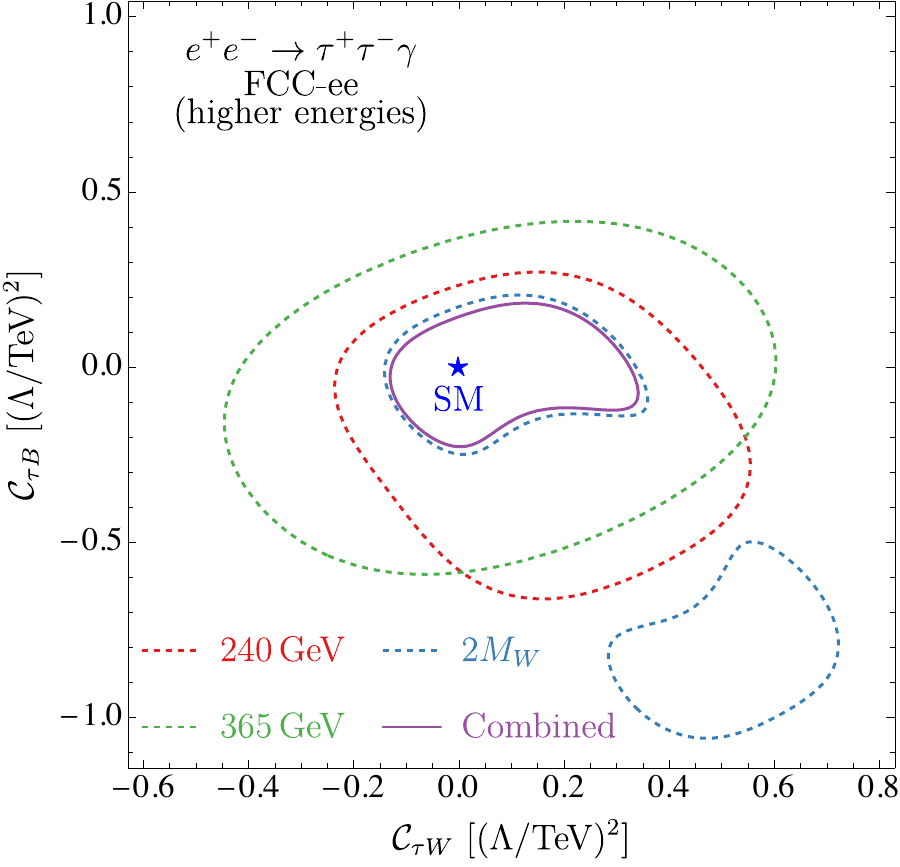}
\includegraphics[width=0.44\textwidth]{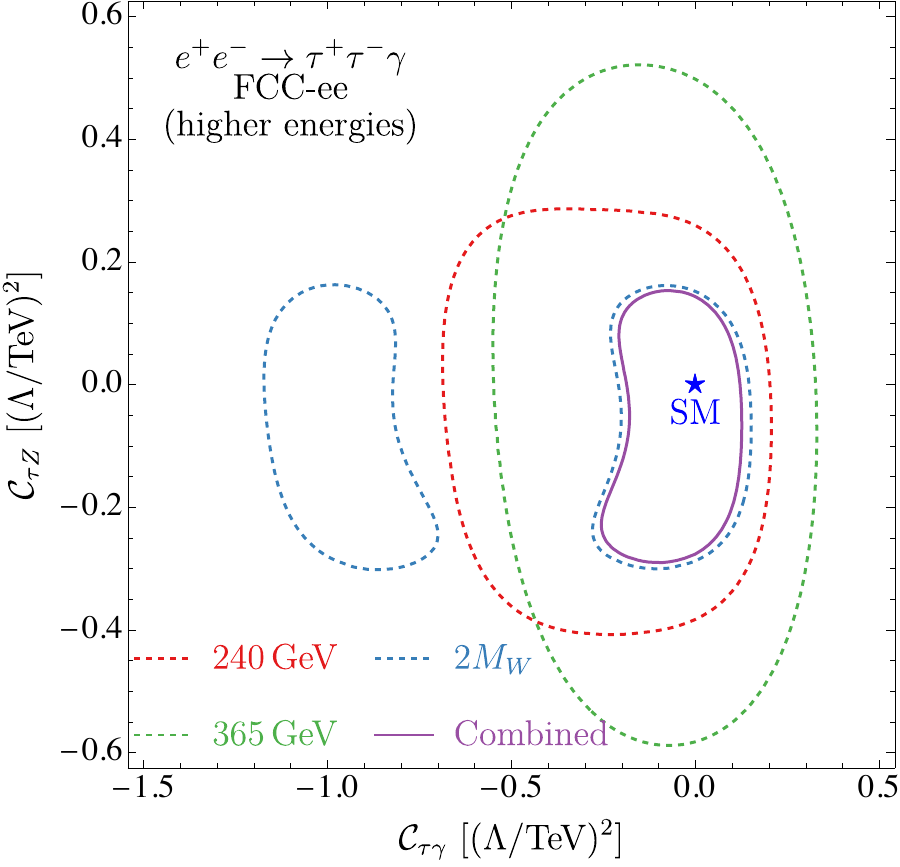}\\
\includegraphics[width=0.44\textwidth]{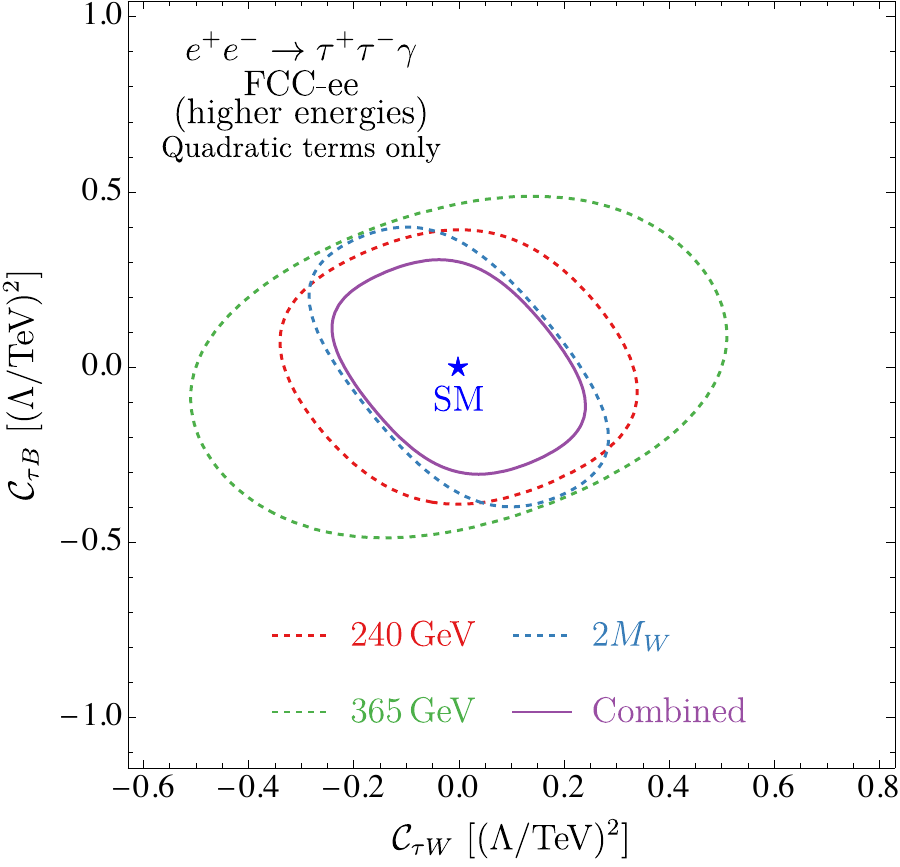}
\includegraphics[width=0.44\textwidth]{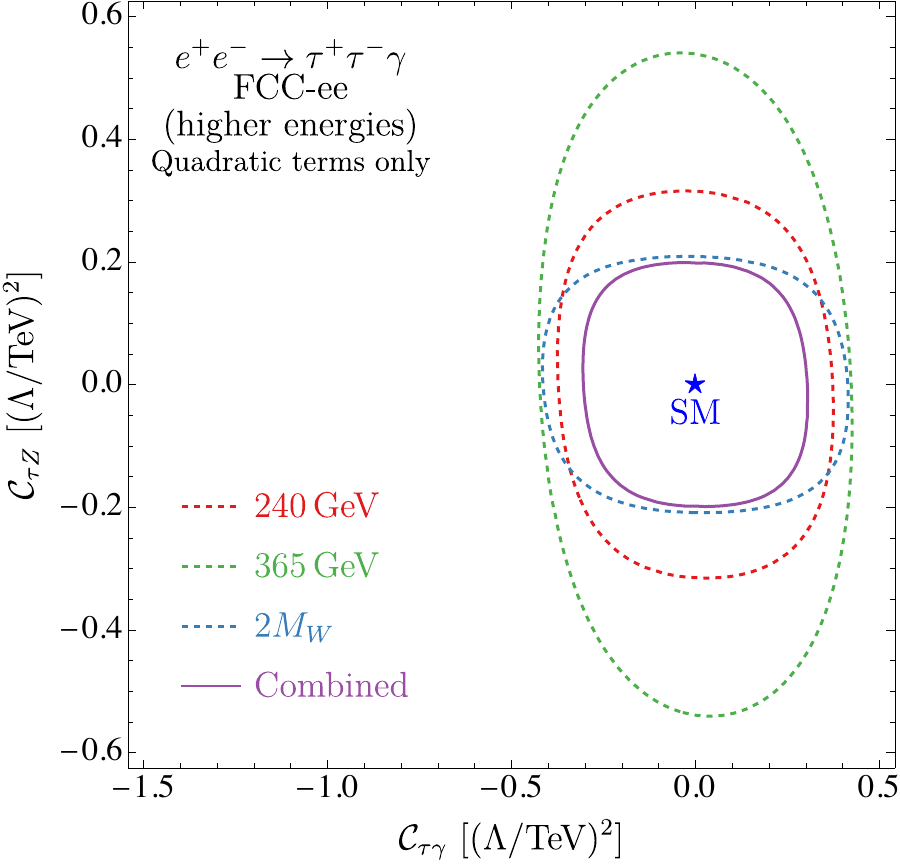}
\caption{
Expected $95\%$ CL sensitivity to the $\tau$ dipole Wilson coefficients from $e^+e^- \to \tau^+\tau^-\gamma$ at the higher-energy FCC-ee stages. The upper panels show the full dimension-six prediction, while the lower panels show the quadratic contribution only. Dashed contours denote individual energy runs, and solid purple contours show their combination.}
\label{fig:contours_FCC_ee2tta3}
\end{figure}

Their combination is displayed in Fig.~\ref{fig:contours_FCC_ee2tta3}. The same qualitative transition seen in the $e^+e^-\to\tau^+\tau^-$ channel is visible here:  the linear interference is still relevant at the lower-energy stages, while the quadratic contribution becomes increasingly important at larger $\sqrt{s}$. At $240$ GeV the cancellation between linear and quadratic terms shifts the full contour away from the origin, whereas by $365$ GeV the quadratic contribution largely controls the sensitivity.
The resulting one-operator-at-a-time bounds, including the combination with the Tera-$Z$ stage, are summarized in Table~\ref{tab:constraints_eetta}. The Tera-$Z$ stage remains the dominant source of sensitivity, while the $\sqrt{s}\simeq2M_W$ run provides the most competitive information among the higher-energy points because of its larger luminosity and residual sensitivity to the linear interference term.

\begin{table}[htb]
\centering
\renewcommand{\arraystretch}{1.18}
\setlength{\tabcolsep}{5pt}
\begin{tabular}{c|cccccc}
\hline
$\sqrt{s}$
& Max$(\Delta a_\tau)$
& Min$(\Delta a_\tau)$
& Max$(\Delta a^Z_\tau)$
& Min$(\Delta a^Z_\tau)$
& $|d_\tau|$ [$e\,{\rm cm}$]
& $|d^Z_\tau|$ [$e\,{\rm cm}$]
\\
\hline
$M_Z$     & $2.4\times10^{-4}$ & $-2.6\times10^{-4}$ & $8.6\times10^{-5}$ & $-1.4\times10^{-4}$ & $7.8\times10^{-18}$ & $1.3\times10^{-18}$ \\
$2M_W$    & $6.0\times10^{-4}$ & $-8.6\times10^{-4}$ & $6.1\times10^{-4}$ & $-1.2\times10^{-3}$ & $9.4\times10^{-18}$ & $4.8\times10^{-18}$ \\
$240$ GeV & $8.3\times10^{-4}$ & $-2.8\times10^{-3}$ & $1.0\times10^{-3}$ & $-1.5\times10^{-3}$ & $8.5\times10^{-18}$ & $7.2\times10^{-18}$ \\
$365$ GeV & $1.3\times10^{-3}$ & $-2.2\times10^{-3}$ & $2.0\times10^{-3}$ & $-2.3\times10^{-3}$ & $9.6\times10^{-18}$ & $1.2\times10^{-17}$ \\
Combined  & $2.2\times10^{-4}$ & $-2.4\times10^{-4}$ & $8.5\times10^{-5}$ & $-1.4\times10^{-4}$ & $6.1\times10^{-18}$ & $1.3\times10^{-18}$ \\
\hline
\end{tabular}
\caption{
Projected $95\%$ CL one-operator-at-a-time constraints on the $\tau$
dipole moments from $e^+e^-\to\tau^+\tau^-\gamma$ at FCC-ee stages.}
\label{tab:constraints_eetta}
\end{table}

Radiative $\tau$-pair production is not the leading channel for any of the dipole moments considered in this work. Its importance is instead diagnostic. It probes the same electromagnetic and weak dipole operators as $\tau$-pair production, but with a hard photon in the final state and with different mixtures of final-state and initial-state radiation. The Tera-$Z$ stage provides the strongest constraints in this channel, while the higher-energy stages test a complementary regime in which the quadratic contribution becomes more important. The resulting bounds are therefore useful mainly as a cross-check of the global SMEFT interpretation rather than as the dominant driver of the final combined sensitivity.

\subsection{\boldmath{$H\to\tau^+\tau^-\gamma$}}
\label{sec:h2ttg}

The radiative Higgs decay $H\to\tau^+\tau^-\gamma$ provides an independent and complementary probe of the electromagnetic dipole interaction of the $\tau$ lepton~\cite{Galon:2016ngp,Rajaraman:2018uyb,Buttazzo:2020ibd}. 
Unlike the production channels discussed in the previous sections, this process probes the dipole operator directly in a Higgs decay, thereby testing the same gauge-invariant SMEFT interaction in a different physical realization. 
Although the Higgs sample expected at the FCC-ee,  $\mathcal{O}(2\times 10^6)$ events, 
is several orders of magnitude smaller than the Higgs yield at the LHC, the exceptionally clean experimental environment and the model-independent identification of Higgs events through the recoil-mass technique make rare Higgs decays experimentally accessible. 
It therefore provides an independent determination of the $\tau$ electromagnetic dipole couplings and an important cross-check of the constraints extracted from $\tau$-pair production.


\subsubsection{SMEFT description}
\label{sec:h2ttg_smeft}
\begin{figure}[htb]
\centering
\includegraphics[width=0.25\textwidth]{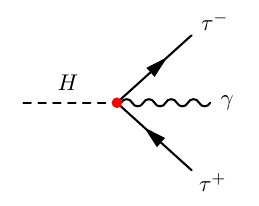}
\includegraphics[width=0.25\textwidth]{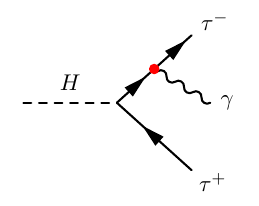}
\includegraphics[width=0.25\textwidth]{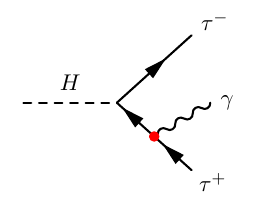}
\caption{
Leading-order Feynman diagrams for the radiative Higgs decay $H\to\tau^+\tau^-\gamma$ in the SMEFT. The red dots indicate vertices involving the electromagnetic dipole operator $\mathcal O_{\tau\gamma}$. The leftmost diagram arises exclusively from this dimension-6 operator, whereas in the two rightmost diagrams the marked vertices can receive contributions from $\mathcal O_{\tau\gamma}$ in addition to their SM contributions.
}
\label{fig:FD_h2tta}
\end{figure}

The leading-order diagrams contributing to $H\to\tau^+\tau^-\gamma$ are shown in Fig.~\ref{fig:FD_h2tta}. The dipole operator $\mathcal O_{\tau\gamma}$ contributes both through the contact $H\tau\tau\gamma$ interaction and through photon emission from the final-state $\tau$ leptons. 

Unlike the processes $e^+e^-\to\tau^+\tau^-$, $\gamma\gamma\to\tau^+\tau^-$, and $e^+e^-\to\tau^+\tau^-\gamma$, the dipole amplitude does not suffer from the helicity suppression associated with the interference between the SM vector current and the dimension-six dipole interaction. Consequently, the linear SMEFT contribution is expected to dominate the sensitivity over most of the parameter space relevant for future Higgs factories.

The BSM contribution to the radiative Higgs width can be written as
\begin{equation}
\Gamma^{(\rm NP)}(H\to\tau^+\tau^-\gamma)
=
\frac{e\,y_\tau M_H^3}
{64\pi^3}
\,
\frac{{\rm Re}(\mathcal C_{\tau\gamma})}
{\Lambda^2}
+
\frac{M_H^5}
{768\pi^3}
\,
\frac{|\mathcal C_{\tau\gamma}|^2}
{\Lambda^4},
\label{eq:higgs_width_decomposition}
\end{equation}
where the first term is the interference between SM and NP amplitudes, and the second term is the square of the NP amplitude. The sensitivity to the anomalous magnetic moment originates primarily from the linear dependence on the Wilson coefficient, whereas the electric dipole moment, which is proportional to the imaginary part of $\mathcal{C}_{\tau\gamma}$, contributes only through the quadratic term. The dominance of the interference term  in the former case constitutes one of the main differences with respect to the $\tau$-pair production channels at high energies.

The SM contribution to $H\to\tau^+\tau^-\gamma$ comes from the radiative corrections to $H\to \tau^+\tau^-$. It contains the familiar infrared enhancement associated with soft or collinear photon emission, and strongly depends on the kinematical cuts that we will impose. In addition, electroweak loop corrections become sizeable in the kinematic region where the invariant mass of the $\tau$ pair approaches the $Z$ resonance, $m(\tau^+\tau^-)\simeq M_Z$~\cite{Han:2017yhy}. Both effects play an important role in the phenomenological analysis presented below. The weak dipole operator $\mathcal O_{\tau Z}$ also induces the decay $H\to\tau^+\tau^-Z$. However, owing to the much smaller branching fraction, this channel is not expected to provide competitive constraints at future Higgs factories and will therefore not be considered further.


\subsubsection{Phenomenological analysis}
\begin{figure}[htb]
\centering
\includegraphics[width=0.23\textwidth]{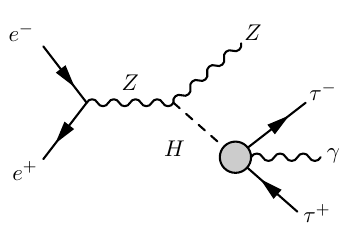}
\includegraphics[width=0.23\textwidth]{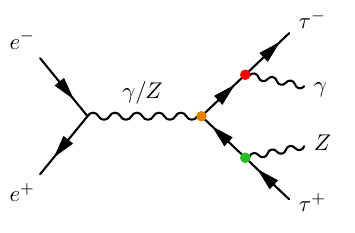}
\includegraphics[width=0.23\textwidth]{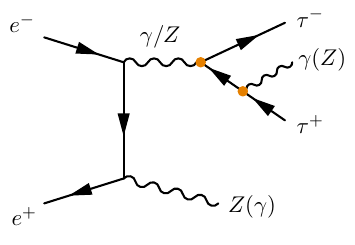}
\includegraphics[width=0.23\textwidth]{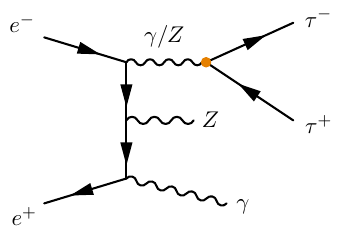}
\caption{
Representative leading-order Feynman diagrams for $e^+e^-\to Z\tau^+\tau^-\gamma$ in the SMEFT. The gray circle represents the radiative Higgs decay $H\to\tau^+\tau^-\gamma$, whose contributing topologies are shown in Fig.~\ref{fig:FD_h2tta}. Red and green dots denote insertions of the electromagnetic and weak dipole operators, respectively, while orange dots mark vertices that can receive contributions from both operators. In the absence of dimension-6 contributions, the orange-marked vertices reduce to their SM counterparts.
}
\label{fig:FD_ee2ttaz}
\end{figure}

A realistic assessment of the sensitivity to the radiative Higgs decay requires the study of the complete production process $e^+e^-\to Z\tau^+\tau^-\gamma$, rather than the Higgs decay in isolation. Representative leading-order diagrams are shown in Fig.~\ref{fig:FD_ee2ttaz}, where the gray circle denotes the $H\to\tau^+\tau^-\gamma$ decay sub-process discussed above, while the colored vertices indicate possible insertions of the dipole operators.

\paragraph*{Signal definition and event selection.}
The dominant irreducible background originates from non-resonant $e^+e^-\to Z\tau^+\tau^-\gamma$ production, together with events where the $\tau$ pair is produced through an on-shell $Z$ boson. To isolate the signal of radiative Higgs decay we require $|m(\tau^+\tau^-\gamma)-M_H|<10~{\rm GeV}$, which selects the narrow Higgs resonance while efficiently suppressing the continuum background. To reject the dominant resonant contribution from $Z\to\tau^+\tau^-$ we further impose $|m(\tau^+\tau^-)-M_Z|>10~{\rm GeV}$~\footnote{The chosen mass window is conservative. Tightening the cut to $|m(\tau^+\tau^-\gamma)-M_H|<1~\mathrm{GeV}$ would reduce the SM background by approximately a factor of two, leading to projected constraints on $\Delta a_\tau$ ($d_\tau$) that are stronger by factors of $1/\sqrt{2}$ ($1/\sqrt[4]{2}$), respectively.}. Besides removing the largest SM background, this veto also suppresses contributions from the kinematic region where electroweak loop corrections to the SM decay become strongly enhanced. Finally, to suppress the infrared enhancement associated with soft-photon emission while retaining most of the signal, we require $p_T(\tau^\pm,\gamma)>M_H/10$. These cuts are applied in addition to the common acceptance criteria introduced in Sec.~\ref{sec:pheno}.

\begin{figure}[htb]
\centering
\includegraphics[width=0.42\textwidth]{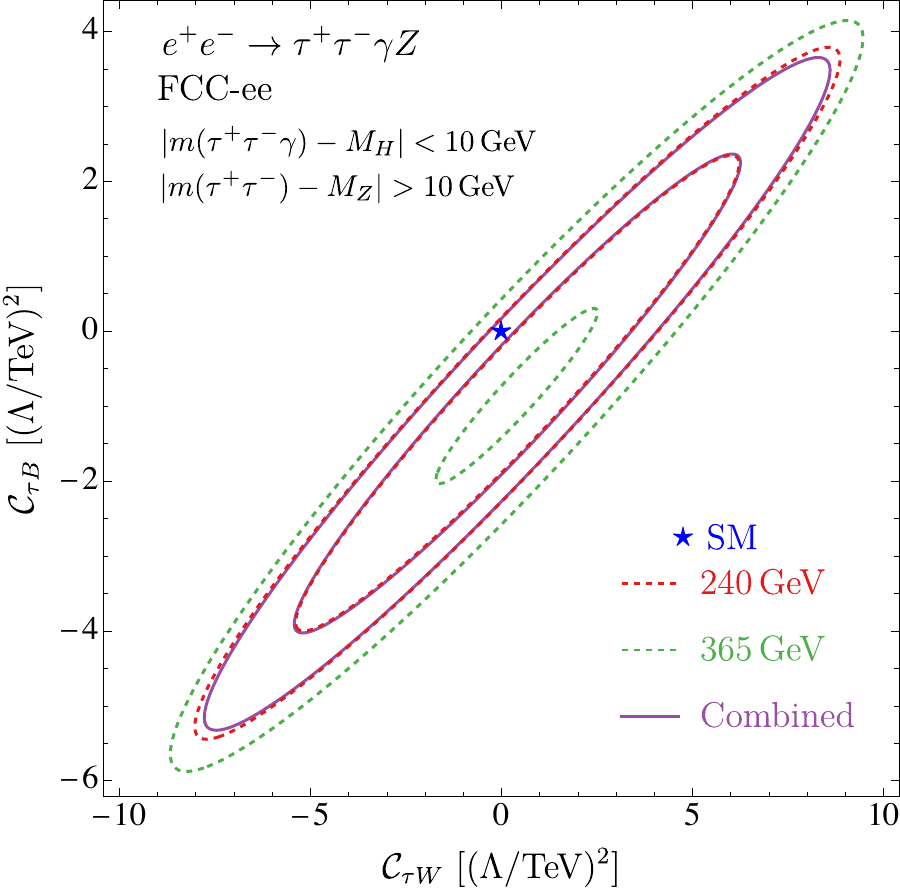}
\includegraphics[width=0.43\textwidth]{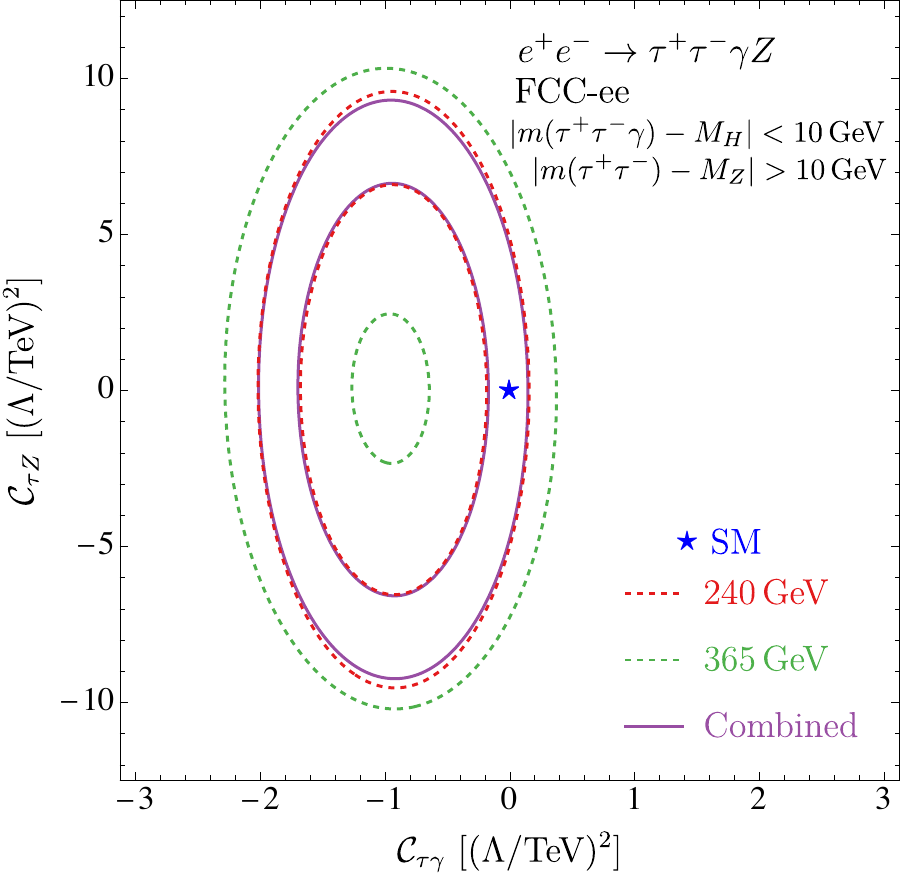}
\includegraphics[width=0.42\textwidth]{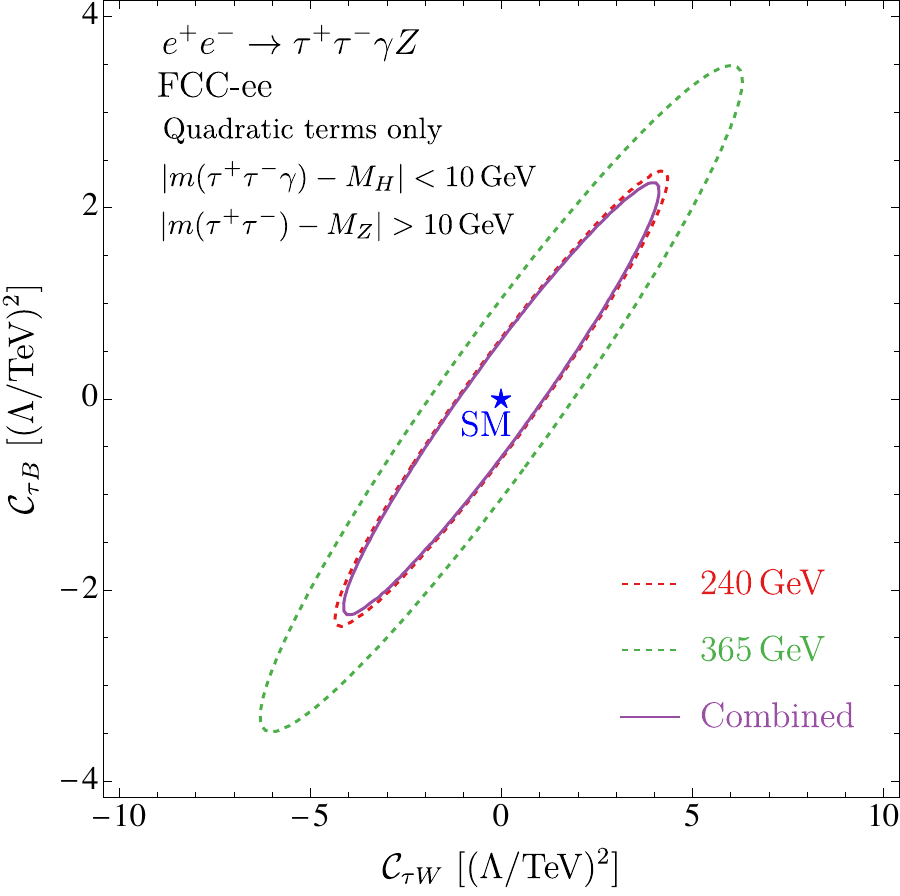}
\includegraphics[width=0.43\textwidth]{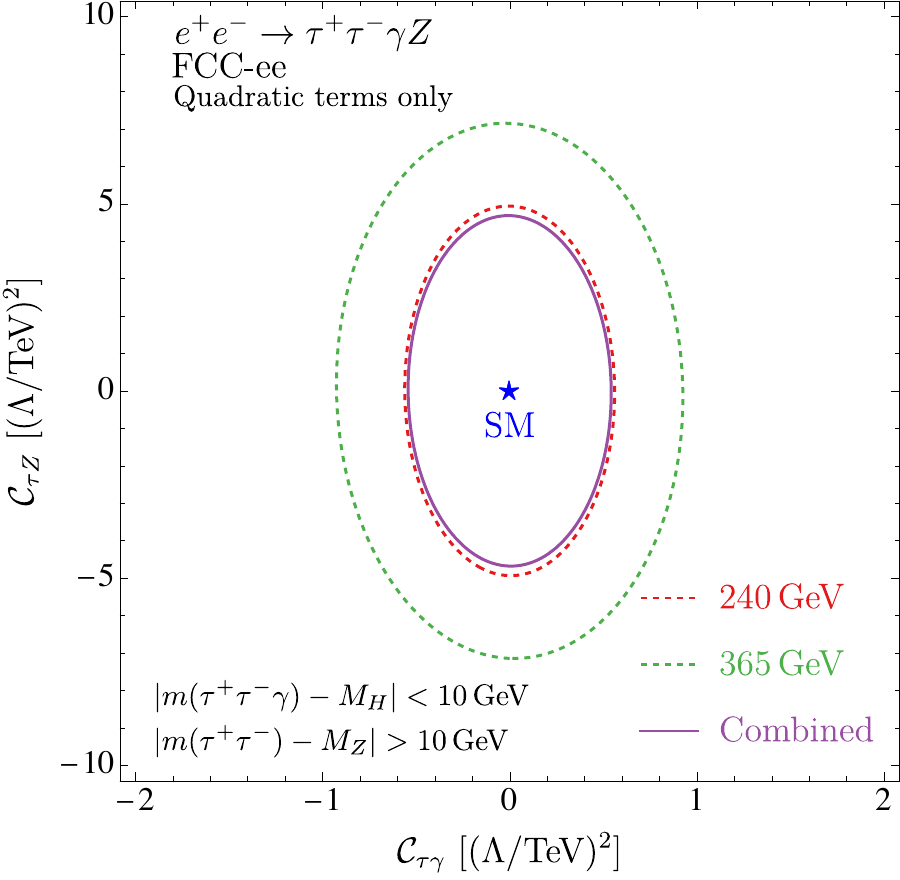}
\caption{
Same as Fig.~\ref{fig:contours_FCC_ee2tta3}, but for $H\to\tau^+\tau^-\gamma$.
}
\label{fig:contours_FCC_ee2ttaz}
\end{figure}

\paragraph*{Sensitivity at the Higgs factory.}
The resulting $95\%$ CL sensitivity contours at the $\sqrt{s}=240$ GeV and $\sqrt{s}=365$ GeV FCC-ee operating stages are shown in Fig.~\ref{fig:contours_FCC_ee2ttaz}. As in the previous sections, the upper panels display the complete SMEFT prediction, whereas the lower panels show the contribution from the quadratic dimension-six terms alone. 

Unlike the $\tau$-pair production channels discussed previously, the
full SMEFT prediction is dominated by the linear interference between
the SM and dipole amplitudes, in agreement with the
analytical discussion of Sec.~\ref{sec:h2ttg_smeft}. Consequently, 
the allowed regions obtained from the complete calculation are
substantially smaller than those obtained from the quadratic
contribution alone.


\paragraph*{Physical interpretation.}
The radiative Higgs decay is primarily sensitive to the electromagnetic dipole operator $\mathcal O_{\tau\gamma}$. As a consequence, the projected constraints on $\Delta a_\tau$ and $d_\tau$ are considerably stronger than those on the weak dipole moments. The latter receive contributions only through non-resonant production diagrams and are further suppressed by the $m(\tau^+\tau^-)$ veto, which intentionally removes the kinematic region most sensitive to $\mathcal O_{\tau Z}$.
The combined constraints obtained from the
240 GeV and 365 GeV operating stages improve upon the expected
HL-LHC sensitivity, while remaining weaker than those derived from
$e^+e^-\to\tau^+\tau^-$ and $\gamma\gamma\to\tau^+\tau^-$.
This hierarchy directly reflects the much smaller Higgs production rate
compared with the enormous $\tau$ samples available through resonant
$Z$ production and photon fusion.
The resulting $95\%$ CL one-operator-at-a-time bounds on the $\tau$
dipole moments from $H\to\tau^+\tau^-\gamma$ at FCC-ee are summarized in Table~\ref{tab:constraints_eettaz}.

The radiative Higgs decay probes the same gauge-invariant dipole
operator discussed throughout this work in a qualitatively different
production mechanism. Its sensitivity is driven primarily by the
unsuppressed interference between the SM and dimension-six
dipole amplitudes, making it particularly sensitive to the anomalous
magnetic moment.
Although this channel cannot match the exceptional precision achievable through $e^+e^-\to\tau^+\tau^-$ or photon-photon scattering, it provides an experimentally clean and theoretically independent probe of the electromagnetic dipole interaction.
Its inclusion therefore strengthens the overall SMEFT program by
providing an important consistency test of the dipole interpretation
obtained from the other production channels.

\begin{table}[htb]
\centering
\renewcommand{\arraystretch}{1.18}
\setlength{\tabcolsep}{5pt}
\begin{tabular}{c|cccccc}
\hline
$\sqrt{s}$
& Max$(\Delta a_\tau)$
& Min$(\Delta a_\tau)$
& Max$(\Delta a^Z_\tau)$
& Min$(\Delta a^Z_\tau)$
& $|d_\tau|$ [$e\,{\rm cm}$]
& $|d^Z_\tau|$ [$e\,{\rm cm}$]
\\
\hline
$240$ GeV & $6.4\times10^{-4}$ & $-7.5\times10^{-4}$ & $2.0\times10^{-2}$ & $-2.0\times10^{-2}$ & $1.3\times10^{-17}$ & $1.1\times10^{-16}$ \\
$365$ GeV & $1.5\times10^{-3}$ & $-2.6\times10^{-3}$ & $2.8\times10^{-2}$ & $-2.9\times10^{-2}$ & $2.1\times10^{-17}$ & $1.2\times10^{-16}$ \\
Combined  & $6.0\times10^{-4}$ & $-7.0\times10^{-4}$ & $1.9\times10^{-2}$ & $-1.9\times10^{-2}$ & $1.2\times10^{-17}$ & $1.1\times10^{-16}$ \\
\hline
\end{tabular}
\caption{
Projected $95\%$ CL one-operator-at-a-time constraints on the $\tau$
dipole moments from $H\to\tau^+\tau^-\gamma$ at FCC-ee.}
\label{tab:constraints_eettaz}
\end{table}

\subsection{Combined sensitivity and global picture}
\label{sec:global_picture}
We conclude our phenomenological analysis by combining all production
channels and FCC-ee operating stages discussed in the previous sections.
This provides the ultimate sensitivity achievable at future
electron--positron colliders within the SMEFT framework and allows us
to identify the complementary role played by the different production
mechanisms in probing the electromagnetic and weak dipole moments of
the $\tau$ lepton.

\begin{figure}[htb]
\centering
\includegraphics[width=0.435\textwidth]{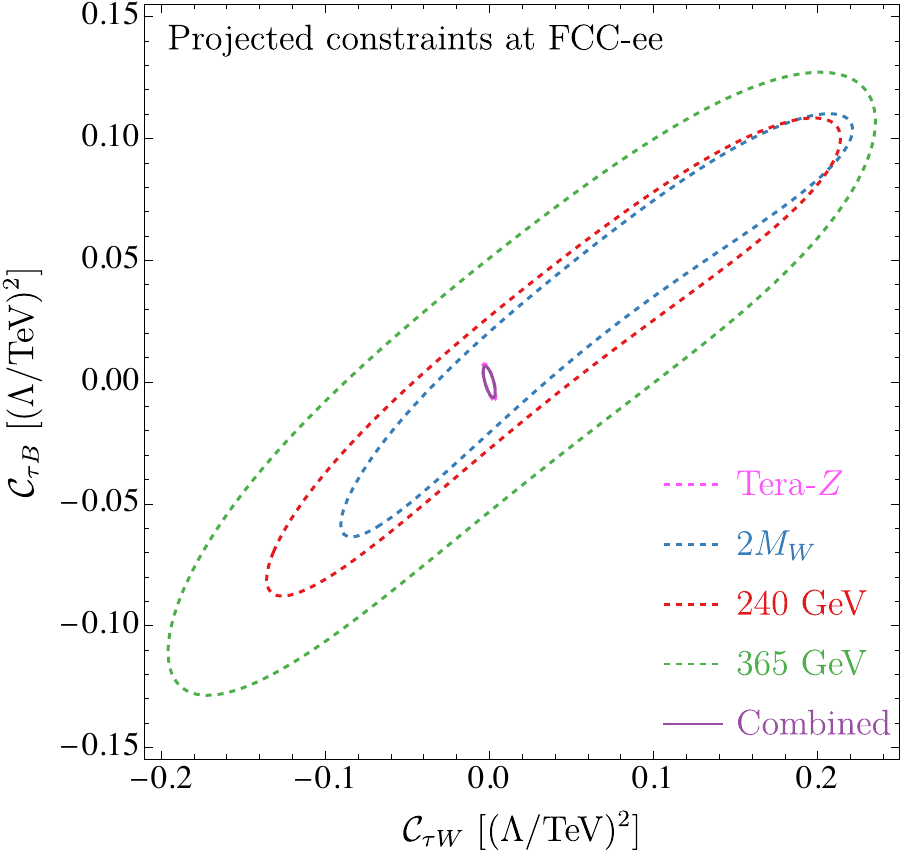}
\includegraphics[width=0.44\textwidth]{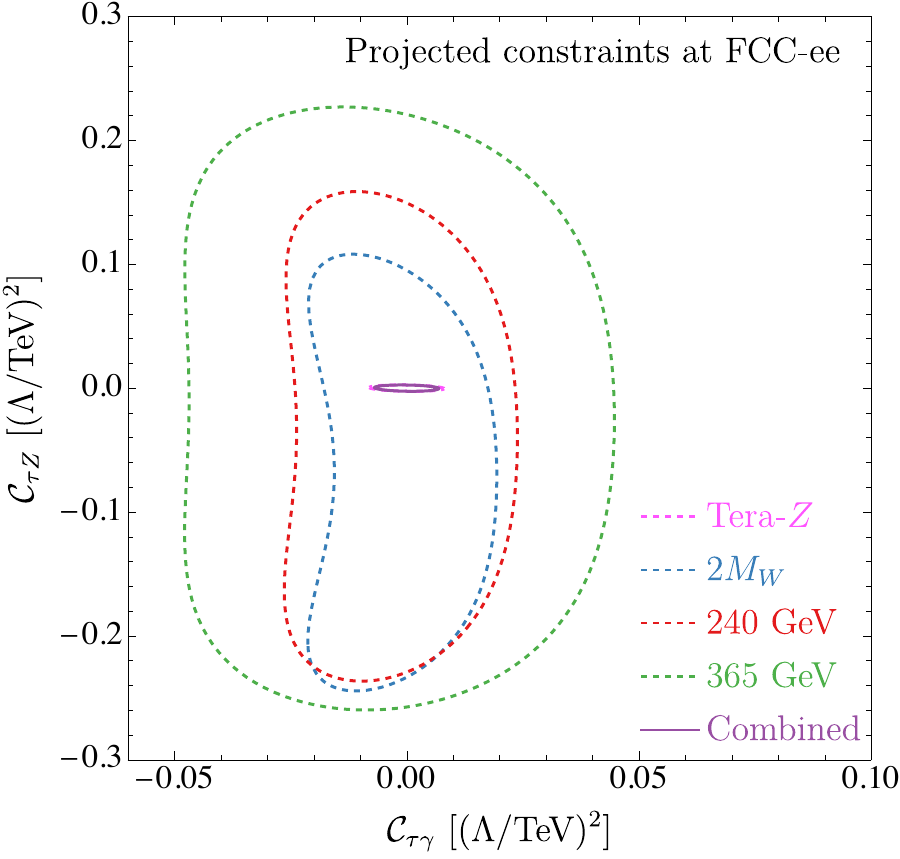}
\includegraphics[width=0.44\textwidth]{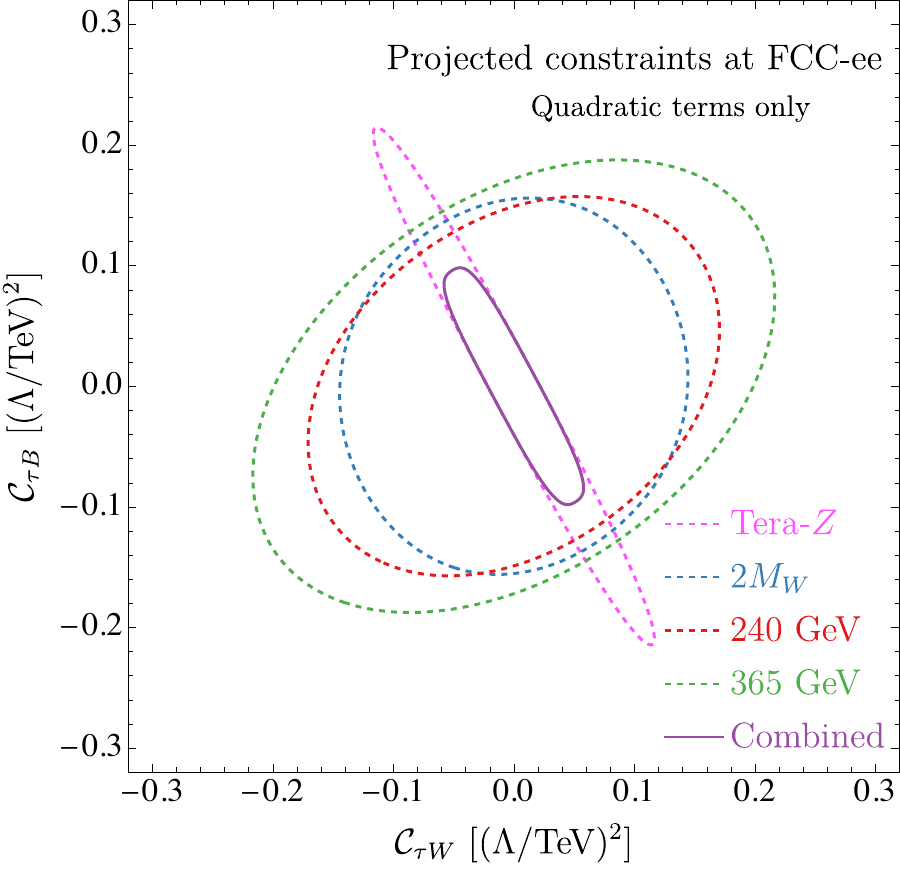}
\includegraphics[width=0.44\textwidth]{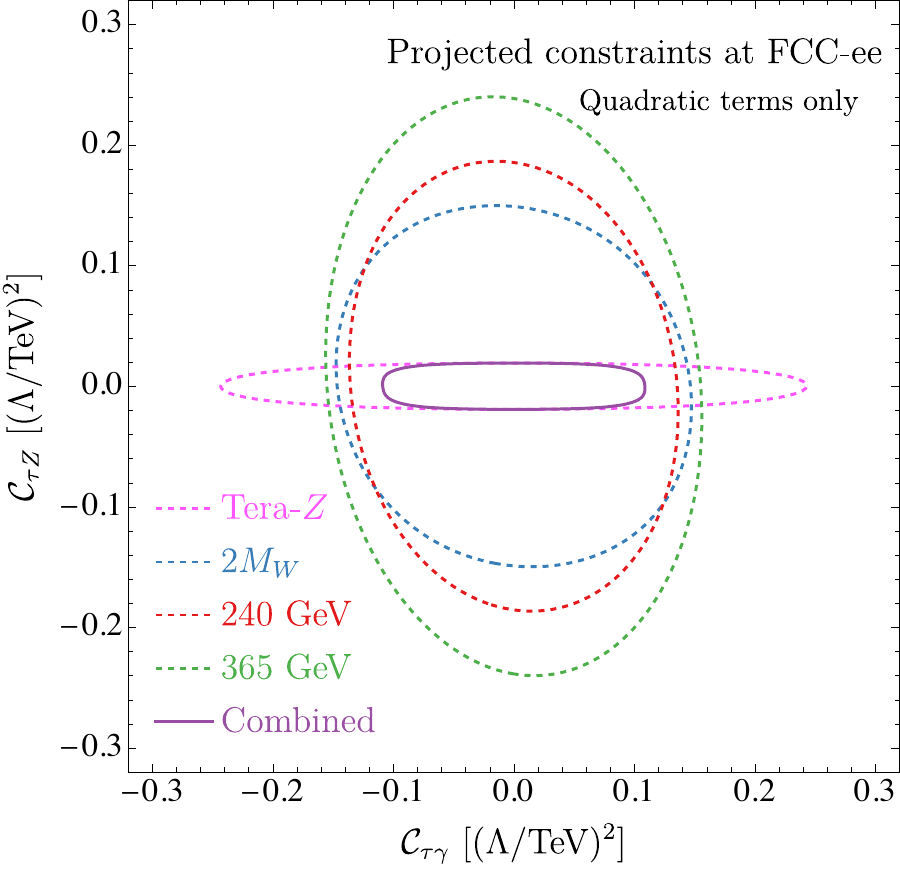}
\caption{
Global 95\% CL sensitivity to the $\tau$ dipole Wilson coefficients
obtained by combining all channels and FCC-ee operating stages
considered in this work. The left panels show the constraints in the
gauge basis $(\mathcal C_{\tau W},\mathcal C_{\tau B})$, while the
right panels display the corresponding reach in the physical basis
$(\mathcal C_{\tau\gamma},\mathcal C_{\tau Z})$. The upper panels are
obtained from the complete dimension-six prediction, whereas the lower
panels isolate the quadratic BSM contribution. Dashed contours denote
individual FCC-ee operating stages, while solid contours show their
statistical combination.
}
\label{fig:contours_FCC_all}
\end{figure}

The resulting 95\% CL sensitivity contours are shown in
Fig.~\ref{fig:contours_FCC_all} in both the
$(\mathcal C_{\tau W},\mathcal C_{\tau B})$ basis and the
$(\mathcal C_{\tau\gamma},\mathcal C_{\tau Z})$ basis.
The upper panels are obtained from the complete dimension-six
prediction and determine the sensitivity to the anomalous magnetic
moments, whereas the lower panels isolate the quadratic contribution,
which provides benchmark projections for the electric dipole moments.
The dashed contours correspond to the individual FCC-ee operating
stages, while the solid contours show their statistical combination.
The corresponding one-operator-at-a-time constraints are summarized in
Table~\ref{tab:constraints_FCCee}.
\begin{table}[htb]
\centering
\renewcommand{\arraystretch}{1.18}
\setlength{\tabcolsep}{5pt}
\begin{tabular}{c|cccccc}
\hline
$\sqrt{s}$
& Max$(\Delta a_\tau)$
& Min$(\Delta a_\tau)$
& Max$(\Delta a^Z_\tau)$
& Min$(\Delta a^Z_\tau)$
& $|d_\tau|$ [$e\,{\rm cm}$]
& $|d^Z_\tau|$ [$e\,{\rm cm}$]
\\
\hline
$M_Z$     & $3.2\times10^{-5}$ & $-3.2\times10^{-5}$ & $9.9\times10^{-6}$ & $-1.0\times10^{-5}$ & $5.5\times10^{-18}$ & $4.4\times10^{-19}$ \\
$2M_W$    & $7.2\times10^{-5}$ & $-7.3\times10^{-5}$ & $3.8\times10^{-4}$ & $-9.4\times10^{-4}$ & $3.3\times10^{-18}$ & $3.4\times10^{-18}$ \\
$240$ GeV & $9.6\times10^{-5}$ & $-9.8\times10^{-5}$ & $6.0\times10^{-4}$ & $-9.2\times10^{-4}$ & $3.1\times10^{-18}$ & $4.2\times10^{-18}$ \\
$365$ GeV & $1.8\times10^{-4}$ & $-1.9\times10^{-4}$ & $8.9\times10^{-4}$ & $-1.0\times10^{-3}$ & $3.5\times10^{-18}$ & $5.4\times10^{-18}$ \\
Combined  & $2.8\times10^{-5}$ & $-2.8\times10^{-5}$ & $9.9\times10^{-6}$ & $-1.0\times10^{-5}$ & $2.5\times10^{-18}$ & $4.4\times10^{-19}$ \\
\hline
\end{tabular}
\caption{
Projected 95\% CL one-operator-at-a-time constraints on the electromagnetic and weak dipole moments of the $\tau$ lepton from the combination of all channels considered in this work, shown for the individual FCC-ee operating stages and for their global combination.}
\label{tab:constraints_FCCee}
\end{table}
Several lessons emerge from the combined analysis:
\begin{itemize}
\item {\it Complementarity of production mechanisms.}
The four channels considered in this work probe different aspects of
the same underlying dipole operators. Resonant $e^+e^-\to\tau^+\tau^-$ production provides the strongest sensitivity to the weak dipole moments through $Z$ exchange and energy scans across the resonance. 
It provides the strongest bound on the $\tau$ EDM at the higher-energy FCC-ee stages.
Photon-induced $\gamma\gamma\to\tau^+\tau^-$ production instead dominates the sensitivity to the electromagnetic dipole operator, owing to the enhancement of linear interference effects in the near-threshold
region. The radiative processes $e^+e^-\to\tau^+\tau^-\gamma$ and
$H\to\tau^+\tau^-\gamma$ are less competitive numerically 
at FCC-ee energies, but provide independent consistency tests of the same SMEFT operators in different kinematic and production environments. 
This complementarity substantially improves the robustness of the SMEFT interpretation compared with any individual measurement and is the central advantage of the future
high-luminosity $e^+e^-$ program.

\item {\it Role of the Tera-$Z$ program.}
The Tera-$Z$ stage plays a unique role throughout the analysis. Its
enormous integrated luminosity compensates the helicity suppression of
the interference terms in direct $\tau$-pair production, making the
sensitivity to the CP-even weak dipole moments predominantly linear in
the Wilson coefficients. Moreover, the energy scan across the three
$Z$-pole operating points efficiently removes the approximate flat
directions between $\mathcal C_{\tau\gamma}$ and
$\mathcal C_{\tau Z}$. This explains why the global sensitivity to
$\Delta a_\tau^Z$ and $d_\tau^Z$ is essentially set by the
Tera-$Z$ program.

\item{\it Linear versus quadratic SMEFT regimes.}
An important outcome of this analysis is that future
electron--positron colliders probe different regions of the SMEFT
expansion depending on the process and energy stage. The Tera-$Z$
measurement of $e^+e^-\to\tau^+\tau^-$ and the threshold-dominated
$\gamma\gamma\to\tau^+\tau^-$ channel are controlled by linear
interference effects for CP-even dipoles. By contrast, higher-energy
production becomes progressively more sensitive to the quadratic
dimension-six contribution, which is particularly relevant for the
electric dipole moments. The global FCC-ee program therefore probes
both the interference- and quadratic-dominated regimes of the same
SMEFT operators.

\item{\it Electromagnetic versus weak dipoles.}
The final combined sensitivities reflect this complementarity. The
strongest constraint on $\Delta a_\tau$ originates from photon-photon scattering, where the photon luminosity enhances the near-threshold region and the
linear interference with $\mathcal C_{\tau\gamma}$ dominates. The
constraint on $\Delta a_\tau^Z$ is instead determined almost entirely
by resonant $e^+e^-\to\tau^+\tau^-$ production at the $Z$ pole.
For the electric dipole moments, the best sensitivity to $d_\tau$ is
obtained from $e^+e^-\to\tau^+\tau^-$ at the higher-energy stages, where the SM background in
direct $\tau$-pair production is reduced and the quadratic dipole
contribution becomes more important. The weak electric dipole moment
$d_\tau^Z$ remains dominated by the Tera-$Z$ measurement, owing to the
resonant enhancement of the $Z$-mediated amplitude.
\end{itemize}

\section{Summary}
\label{sec:summary}

The short lifetime of the $\tau$ lepton prevents direct measurements of
its dipole moments and makes collider processes the primary probes of
these observables. In this work we have studied the sensitivity of
future high-luminosity $e^+e^-$ colliders to the electromagnetic and
weak dipole moments of the $\tau$ lepton within the SMEFT. We have
focused on the FCC-ee running program as a benchmark and analyzed the
complementary channels
\[
e^+e^- \to \tau^+\tau^-,
\qquad
\gamma\gamma\to\tau^+\tau^-,
\qquad
e^+e^- \to \tau^+\tau^-\gamma,
\qquad
H\to\tau^+\tau^-\gamma .
\]
Together, these processes probe different combinations of the
gauge-invariant dipole operators and different regimes of the SMEFT
expansion.

The dipole interactions were parametrized in terms of the dimension-six
operators $\mathcal O_{\tau B}$ and $\mathcal O_{\tau W}$, or
equivalently, in the physical basis,
$\mathcal O_{\tau\gamma}$ and $\mathcal O_{\tau Z}$.
This basis makes explicit the connection between the Wilson
coefficients and the electromagnetic and weak dipole moments
$\Delta a_\tau$, $d_\tau$, $\Delta a_\tau^Z$ and $d_\tau^Z$.
We have also verified that the projected FCC-ee sensitivities lie well
inside the perturbative-unitarity region, so that the SMEFT description
is self-consistent over the energy range considered.

A central outcome of our analysis is that the different production
mechanisms probe genuinely complementary dynamical regimes.
The leading sensitivity to the weak dipole moments is provided by
$e^+e^-\to\tau^+\tau^-$ at the Tera-$Z$ stage of the FCC-ee program. The
combination of resonant $Z$-boson production and the enormous integrated
luminosity makes the interference terms dominant despite their helicity
suppression. In addition, the scan across the $Z$ resonance lifts the
approximate flat direction between $\mathcal C_{\tau\gamma}$ and
$\mathcal C_{\tau Z}$. This results in the projected bounds 
\[
|\Delta a_\tau^Z| < 1.0\times10^{-5},
\qquad
|d_\tau^Z| < 4.4\times10^{-19}\,e\,\mathrm{cm}.
\]

Photon-induced production, $e^+e^-\to e^+e^-\tau^+\tau^-$, dominated by
$\gamma\gamma\to\tau^+\tau^-$, provides the strongest probe of the
anomalous magnetic moment. The convolution with the photon luminosity
strongly enhances the near-threshold region, $m_{\tau\tau}\gtrsim2m_\tau$, where the usual high-energy helicity suppression of the interference between the SM and the CP-even electromagnetic dipole amplitude is no longer effective. As a result,
the sensitivity is dominated by the linear contribution in
$\mathcal C_{\tau\gamma}$. Because essentially the same near-threshold
partonic region controls the signal at all FCC-ee operating stages,
the improvement obtained by increasing the collider energy is modest.
The outstanding sensitivity of the Tera-$Z$ program is therefore
driven primarily by its unprecedented integrated luminosity, leading to:
\[
|\Delta a_\tau| < 3.4\times 10^{-5}.
\]
The best sensitivity to the electric dipole moment is obtained from
the process $e^+e^-\to \tau^+\tau^-$ at the higher-energy stages, 
where the SM background in direct $\tau$-pair production is reduced and 
the quadratic dipole contribution becomes more relevant:
\[
|d_\tau| < 2.5\times 10^{-18}\, e\,{\rm cm}.
\]

The radiative channels $e^+e^-\to\tau^+\tau^-\gamma$ and 
$H\to\tau^+\tau^-\gamma$ are less competitive numerically, but provide
important independent probes of the same SMEFT operators.
The former tests the dipole interactions in a hard-photon final state,
while the latter probes the electromagnetic dipole operator directly
through the Higgs--dipole contact interaction. These channels therefore
serve as valuable consistency checks of any possible deviation observed
in the leading $\tau$-pair production modes.

The projected bounds obtained in this work improve the current experimental limits by several orders of magnitude and establish future electron--positron colliders as unparalleled precision probes of dipole-induced new physics in the $\tau$ sector.
Although our numerical projections are based on the FCC-ee program,
the main conclusions apply broadly to other future high-luminosity
$e^+e^-$ facilities with comparable energy and luminosity profiles,
such as CEPC. The results at the 240 and 365 GeV stages also provide a
useful benchmark for linear-collider scenarios. More generally, the
combination of resonant $Z$ production, photon-induced $\tau$-pair
production, radiative final states and Higgs decays offers a coherent
and overconstrained strategy to test the electromagnetic and weak
dipole interactions of the $\tau$ lepton at the next generation of
lepton colliders.

Looking further ahead to the high-energy frontier, multi-TeV muon colliders offer a distinct and highly complementary avenue to probe $\tau$ dipole interactions~\cite{Buttazzo:2026amk}. A comprehensive investigation of $\tau$ dipole moments at multi-TeV muon colliders and a detailed comparison with FCC-ee sensitivities are presented in our
companion study~\cite{Buttazzo:2026jfu}.

The projected sensitivities presented here are not intended as definitive
forecasts of the limits that will ultimately be obtained at the FCC-ee.
Rather, they quantify the potential reach of the different channels and
identify the processes and kinematic regimes that warrant dedicated
experimental and theoretical investigation. A robust determination of the
achievable limits will require detector-level simulations, realistic $\tau$
reconstruction and background modeling, a systematic-uncertainty assessment,
and higher-order theoretical predictions.


\begin{acknowledgments}
We thank B. Batell, I. Brivio, C. Degrande, G. Durieux, T. Han, M. Hoferichter, F. Jaffredo, J. Liu, O. Mattelaer, M. Passera, S. Renner, M. Selvaggi, X. Wang, and Z. Yu for useful discussions. 
This research is partially supported by the IISN convention 4.4517.08, ``Theory of fundamental interactions.'' Y. Ma acknowledges the support as a Postdoctoral Researcher of the Fonds de la Recherche Scientifique de Belgique (F.R.S.-FNRS), Belgium.
Computational resources have been provided by the supercomputing facilities of the Universit\'e catholique de Louvain (CISM/UCL) and the Consortium des \'Equipements de Calcul Intensif en 
F\'ed\'eration Wallonie Bruxelles (C\'ECI) funded by the Fond de la Recherche Scientifique de Belgique (F.R.S.-FNRS) under convention 2.5020.11 and by the Walloon Region.
G. Levati gratefully acknowledges financial support by the Swiss National Science Foundation (Project No.\ 
TMCG-2\_213690). Z.Q. Wang acknowledges support from the China Scholarship Council (CSC)- UCLouvain Co‑funding PhD Fellowship (Grant No.\ 202206040017).
P. Paradisi received funding by the European Union’s Horizon 2020 research and innovation program under the Marie Sklodowska-Curie grant agreements n.~860881~-~HIDDeN, n.~101086085~-~ASYMMETRY, by the Italian MUR Departments of Excellence grant 2023-2027 ``Quantum Frontiers'' and by the European Union - Next Generation EU and by the Italian Ministry of University and Research (MUR) via the PRIN 2022 project n. 2022K4B58X - AxionOrigins. D. Buttazzo is partially funded by the European Union - Next Generation EU and MUR through the grant PRIN 202289JEW4. We also acknowledge support from the COMETA COST Action CA22130.
\end{acknowledgments}



\bibliographystyle{utphys}

\bibliography{ref.bib}

\providecommand{\href}[2]{#2}\begingroup\raggedright\begin{thebibliography}{100}

\bibitem{Fan:2022eto}
X.~Fan, T.~G. Myers, B.~A.~D. Sukra, and G.~Gabrielse, ``{Measurement of the
  Electron Magnetic Moment},''
  \href{http://dx.doi.org/10.1103/PhysRevLett.130.071801}{{\em Phys. Rev.
  Lett.} {\bfseries 130} no.~7, (2023) 071801},
  \href{http://arxiv.org/abs/2209.13084}{{\ttfamily arXiv:2209.13084
  [physics.atom-ph]}}.

\bibitem{Muong-2:2023cdq}
{\bfseries Muon g-2} Collaboration, D.~P. Aguillard {\em et~al.},
  ``{Measurement of the Positive Muon Anomalous Magnetic Moment to 0.20~ppm},''
  \href{http://dx.doi.org/10.1103/PhysRevLett.131.161802}{{\em Phys. Rev.
  Lett.} {\bfseries 131} no.~16, (2023) 161802},
  \href{http://arxiv.org/abs/2308.06230}{{\ttfamily arXiv:2308.06230
  [hep-ex]}}.

\bibitem{Muong-2:2025xyk}
{\bfseries Muon g-2} Collaboration, D.~P. Aguillard {\em et~al.},
  ``{Measurement of the Positive Muon Anomalous Magnetic Moment to 127~ppb},''
  \href{http://dx.doi.org/10.1103/7clf-sm2v}{{\em Phys. Rev. Lett.} {\bfseries
  135} no.~10, (2025) 101802},
  \href{http://arxiv.org/abs/2506.03069}{{\ttfamily arXiv:2506.03069
  [hep-ex]}}.

\bibitem{Roussy:2022cmp}
T.~S. Roussy {\em et~al.}, ``{An improved bound on the
  electron\textquoteright{}s electric dipole moment},''
  \href{http://dx.doi.org/10.1126/science.adg4084}{{\em Science} {\bfseries
  381} no.~6653, (2023) adg4084},
  \href{http://arxiv.org/abs/2212.11841}{{\ttfamily arXiv:2212.11841
  [physics.atom-ph]}}.

\bibitem{Muong-2:2008ebm}
{\bfseries Muon $g-2$} Collaboration, G.~W. Bennett {\em et~al.}, ``{An
  Improved Limit on the Muon Electric Dipole Moment},''
  \href{http://dx.doi.org/10.1103/PhysRevD.80.052008}{{\em Phys. Rev. D}
  {\bfseries 80} (2009) 052008},
  \href{http://arxiv.org/abs/0811.1207}{{\ttfamily arXiv:0811.1207 [hep-ex]}}.

\bibitem{Morel:2020dww}
L.~Morel, Z.~Yao, P.~Clad{\'e}, and S.~Guellati-Kh{\'e}lifa, ``{Determination
  of the fine-structure constant with an accuracy of 81 parts per trillion},''
  \href{http://dx.doi.org/10.1038/s41586-020-2964-7}{{\em Nature} {\bfseries
  588} no.~7836, (2020) 61--65}.

\bibitem{Parker:2018vye}
R.~H. Parker, C.~Yu, W.~Zhong, B.~Estey, and H.~M{\"u}ller, ``{Measurement of
  the fine-structure constant as a test of the Standard Model},''
  \href{http://dx.doi.org/10.1126/science.aap7706}{{\em Science} {\bfseries
  360} (2018) 191}, \href{http://arxiv.org/abs/1812.04130}{{\ttfamily
  arXiv:1812.04130 [physics.atom-ph]}}.

\bibitem{Aoyama:2019ryr}
T.~Aoyama, T.~Kinoshita, and M.~Nio, ``{Theory of the Anomalous Magnetic Moment
  of the Electron},'' \href{http://dx.doi.org/10.3390/atoms7010028}{{\em Atoms}
  {\bfseries 7} no.~1, (2019) 28}.

\bibitem{Volkov:2019phy}
S.~Volkov, ``{Calculating the five-loop QED contribution to the electron
  anomalous magnetic moment: Graphs without lepton loops},''
  \href{http://dx.doi.org/10.1103/PhysRevD.100.096004}{{\em Phys. Rev. D}
  {\bfseries 100} no.~9, (2019) 096004},
  \href{http://arxiv.org/abs/1909.08015}{{\ttfamily arXiv:1909.08015
  [hep-ph]}}.

\bibitem{Volkov:2024yzc}
S.~Volkov, ``{Calculation of the total 10th order QED contribution to the
  electron magnetic moment},''
  \href{http://dx.doi.org/10.1103/PhysRevD.110.036001}{{\em Phys. Rev. D}
  {\bfseries 110} no.~3, (2024) 036001},
  \href{http://arxiv.org/abs/2404.00649}{{\ttfamily arXiv:2404.00649
  [hep-ph]}}.

\bibitem{Aoyama:2024aly}
T.~Aoyama, M.~Hayakawa, A.~Hirayama, and M.~Nio, ``{Verification of the
  tenth-order QED contribution to the anomalous magnetic moment of the electron
  from diagrams without fermion loops},''
  \href{http://dx.doi.org/10.1103/PhysRevD.111.L031902}{{\em Phys. Rev. D}
  {\bfseries 111} no.~3, (2025) L031902},
  \href{http://arxiv.org/abs/2412.06473}{{\ttfamily arXiv:2412.06473
  [hep-ph]}}.

\bibitem{DiLuzio:2024sps}
L.~Di~Luzio, A.~Keshavarzi, A.~Masiero, and P.~Paradisi, ``{Model-Independent
  Tests of the Hadronic Vacuum Polarization Contribution to the Muon g-2},''
  \href{http://dx.doi.org/10.1103/PhysRevLett.134.011902}{{\em Phys. Rev.
  Lett.} {\bfseries 134} no.~1, (2025) 011902},
  \href{http://arxiv.org/abs/2408.01123}{{\ttfamily arXiv:2408.01123
  [hep-ph]}}.

\bibitem{Hoferichter:2025fea}
M.~Hoferichter, P.~Stoffer, and M.~Zillinger, ``{Hadronic light-by-light
  scattering in the anomalous magnetic moments of electron and
  {\ensuremath{\tau}}},''
  \href{http://dx.doi.org/10.1016/j.physletb.2025.139565}{{\em Phys. Lett. B}
  {\bfseries 866} (2025) 139565},
  \href{http://arxiv.org/abs/2504.10582}{{\ttfamily arXiv:2504.10582
  [hep-ph]}}.

\bibitem{Aliberti:2025beg}
R.~Aliberti {\em et~al.}, ``{The anomalous magnetic moment of the muon in the
  Standard Model: an update},''
  \href{http://dx.doi.org/10.1016/j.physrep.2025.08.002}{{\em Phys. Rept.}
  {\bfseries 1143} (2025) 1--158},
  \href{http://arxiv.org/abs/2505.21476}{{\ttfamily arXiv:2505.21476
  [hep-ph]}}.

\bibitem{Aoyama:2012wk}
T.~Aoyama, M.~Hayakawa, T.~Kinoshita, and M.~Nio, ``{Complete Tenth-Order QED
  Contribution to the Muon g-2},''
  \href{http://dx.doi.org/10.1103/PhysRevLett.109.111808}{{\em Phys. Rev.
  Lett.} {\bfseries 109} (2012) 111808},
  \href{http://arxiv.org/abs/1205.5370}{{\ttfamily arXiv:1205.5370 [hep-ph]}}.

\bibitem{Czarnecki:2002nt}
A.~Czarnecki, W.~J. Marciano, and A.~Vainshtein, ``{Refinements in electroweak
  contributions to the muon anomalous magnetic moment},''
  \href{http://dx.doi.org/10.1103/PhysRevD.67.073006}{{\em Phys. Rev. D}
  {\bfseries 67} (2003) 073006},
  \href{http://arxiv.org/abs/hep-ph/0212229}{{\ttfamily arXiv:hep-ph/0212229}}.
  [Erratum: Phys. Rev. D {\bf 73}, 119901 (2006)].

\bibitem{Gnendiger:2013pva}
C.~Gnendiger, D.~St{\"o}ckinger, and H.~St{\"o}ckinger-Kim, ``{The electroweak
  contributions to $(g-2)_\mu$ after the Higgs boson mass measurement},''
  \href{http://dx.doi.org/10.1103/PhysRevD.88.053005}{{\em Phys. Rev. D}
  {\bfseries 88} (2013) 053005},
  \href{http://arxiv.org/abs/1306.5546}{{\ttfamily arXiv:1306.5546 [hep-ph]}}.

\bibitem{Ludtke:2024ase}
J.~L{\"u}dtke, M.~Procura, and P.~Stoffer, ``{Dispersion relations for the
  hadronic VVA correlator},''
  \href{http://dx.doi.org/10.1007/JHEP04(2025)130}{{\em JHEP} {\bfseries 04}
  (2025) 130}, \href{http://arxiv.org/abs/2410.11946}{{\ttfamily
  arXiv:2410.11946 [hep-ph]}}.

\bibitem{Hoferichter:2025yih}
M.~Hoferichter, J.~L{\"u}dtke, L.~Naterop, M.~Procura, and P.~Stoffer,
  ``{Improved Evaluation of the Electroweak Contribution to Muon g-2},''
  \href{http://dx.doi.org/10.1103/PhysRevLett.134.201801}{{\em Phys. Rev.
  Lett.} {\bfseries 134} no.~20, (2025) 201801},
  \href{http://arxiv.org/abs/2503.04883}{{\ttfamily arXiv:2503.04883
  [hep-ph]}}.

\bibitem{RBC:2018dos}
{\bfseries RBC, UKQCD} Collaboration, T.~Blum, P.~A. Boyle, V.~G{\"u}lpers,
  T.~Izubuchi, L.~Jin, C.~Jung, A.~J{\"u}ttner, C.~Lehner, A.~Portelli, and
  J.~T. Tsang, ``{Calculation of the hadronic vacuum polarization contribution
  to the muon anomalous magnetic moment},''
  \href{http://dx.doi.org/10.1103/PhysRevLett.121.022003}{{\em Phys. Rev.
  Lett.} {\bfseries 121} no.~2, (2018) 022003},
  \href{http://arxiv.org/abs/1801.07224}{{\ttfamily arXiv:1801.07224
  [hep-lat]}}.

\bibitem{Giusti:2019xct}
D.~Giusti, V.~Lubicz, G.~Martinelli, F.~Sanfilippo, and S.~Simula,
  ``{Electromagnetic and strong isospin-breaking corrections to the muon $g -
  2$ from Lattice QCD+QED},''
  \href{http://dx.doi.org/10.1103/PhysRevD.99.114502}{{\em Phys. Rev. D}
  {\bfseries 99} no.~11, (2019) 114502},
  \href{http://arxiv.org/abs/1901.10462}{{\ttfamily arXiv:1901.10462
  [hep-lat]}}.

\bibitem{Borsanyi:2020mff}
S.~Borsanyi {\em et~al.}, ``{Leading hadronic contribution to the muon magnetic
  moment from lattice QCD},''
  \href{http://dx.doi.org/10.1038/s41586-021-03418-1}{{\em Nature} {\bfseries
  593} no.~7857, (2021) 51--55},
  \href{http://arxiv.org/abs/2002.12347}{{\ttfamily arXiv:2002.12347
  [hep-lat]}}.

\bibitem{Lehner:2020crt}
C.~Lehner and A.~S. Meyer, ``{Consistency of hadronic vacuum polarization
  between lattice QCD and the R-ratio},''
  \href{http://dx.doi.org/10.1103/PhysRevD.101.074515}{{\em Phys. Rev. D}
  {\bfseries 101} (2020) 074515},
  \href{http://arxiv.org/abs/2003.04177}{{\ttfamily arXiv:2003.04177
  [hep-lat]}}.

\bibitem{Wang:2022lkq}
{\bfseries $\chi$QCD} Collaboration, G.~Wang, T.~Draper, K.-F. Liu, and Y.-B.
  Yang, ``{Muon g-2 with overlap valence fermions},''
  \href{http://dx.doi.org/10.1103/PhysRevD.107.034513}{{\em Phys. Rev. D}
  {\bfseries 107} no.~3, (2023) 034513},
  \href{http://arxiv.org/abs/2204.01280}{{\ttfamily arXiv:2204.01280
  [hep-lat]}}.

\bibitem{Aubin:2022hgm}
C.~Aubin, T.~Blum, M.~Golterman, and S.~Peris, ``{Muon anomalous magnetic
  moment with staggered fermions: Is the lattice spacing small enough?},''
  \href{http://dx.doi.org/10.1103/PhysRevD.106.054503}{{\em Phys. Rev. D}
  {\bfseries 106} no.~5, (2022) 054503},
  \href{http://arxiv.org/abs/2204.12256}{{\ttfamily arXiv:2204.12256
  [hep-lat]}}.

\bibitem{Ce:2022kxy}
M.~C{\`e} {\em et~al.}, ``{Window observable for the hadronic vacuum
  polarization contribution to the muon g-2 from lattice QCD},''
  \href{http://dx.doi.org/10.1103/PhysRevD.106.114502}{{\em Phys. Rev. D}
  {\bfseries 106} no.~11, (2022) 114502},
  \href{http://arxiv.org/abs/2206.06582}{{\ttfamily arXiv:2206.06582
  [hep-lat]}}.

\bibitem{ExtendedTwistedMass:2022jpw}
{\bfseries ETM} Collaboration, C.~Alexandrou {\em et~al.}, ``{Lattice
  calculation of the short and intermediate time-distance hadronic vacuum
  polarization contributions to the muon magnetic moment using twisted-mass
  fermions},'' \href{http://dx.doi.org/10.1103/PhysRevD.107.074506}{{\em Phys.
  Rev. D} {\bfseries 107} no.~7, (2023) 074506},
  \href{http://arxiv.org/abs/2206.15084}{{\ttfamily arXiv:2206.15084
  [hep-lat]}}.

\bibitem{RBC:2023pvn}
{\bfseries RBC, UKQCD} Collaboration, T.~Blum {\em et~al.}, ``{Update of
  Euclidean windows of the hadronic vacuum polarization},''
  \href{http://dx.doi.org/10.1103/PhysRevD.108.054507}{{\em Phys. Rev. D}
  {\bfseries 108} no.~5, (2023) 054507},
  \href{http://arxiv.org/abs/2301.08696}{{\ttfamily arXiv:2301.08696
  [hep-lat]}}.

\bibitem{Kuberski:2024bcj}
S.~Kuberski, M.~C{\`e}, G.~von Hippel, H.~B. Meyer, K.~Ottnad, A.~Risch, and
  H.~Wittig, ``{Hadronic vacuum polarization in the muon g {\ensuremath{-}} 2:
  the short-distance contribution from lattice QCD},''
  \href{http://dx.doi.org/10.1007/JHEP03(2024)172}{{\em JHEP} {\bfseries 03}
  (2024) 172}, \href{http://arxiv.org/abs/2401.11895}{{\ttfamily
  arXiv:2401.11895 [hep-lat]}}.

\bibitem{Boccaletti:2024guq}
A.~Boccaletti {\em et~al.}, ``{High precision calculation of the hadronic
  vacuum polarisation contribution to the muon anomaly},'' 7, 2024.

\bibitem{Spiegel:2024dec}
S.~Spiegel and C.~Lehner, ``{High-precision continuum limit study of the HVP
  short-distance window},'' \href{http://dx.doi.org/10.1103/mj3d-yq87}{{\em
  Phys. Rev. D} {\bfseries 111} no.~11, (2025) 114517},
  \href{http://arxiv.org/abs/2410.17053}{{\ttfamily arXiv:2410.17053
  [hep-lat]}}.

\bibitem{RBC:2024fic}
{\bfseries RBC, UKQCD} Collaboration, T.~Blum {\em et~al.}, ``{Long-Distance
  Window of the Hadronic Vacuum Polarization for the Muon g-2},''
  \href{http://dx.doi.org/10.1103/PhysRevLett.134.201901}{{\em Phys. Rev.
  Lett.} {\bfseries 134} no.~20, (2025) 201901},
  \href{http://arxiv.org/abs/2410.20590}{{\ttfamily arXiv:2410.20590
  [hep-lat]}}.

\bibitem{Djukanovic:2024cmq}
D.~Djukanovic, G.~von Hippel, S.~Kuberski, H.~B. Meyer, N.~Miller, K.~Ottnad,
  J.~Parrino, A.~Risch, and H.~Wittig, ``{The hadronic vacuum polarization
  contribution to the muon g {\ensuremath{-}} 2 at long distances},''
  \href{http://dx.doi.org/10.1007/JHEP04(2025)098}{{\em JHEP} {\bfseries 04}
  (2025) 098}, \href{http://arxiv.org/abs/2411.07969}{{\ttfamily
  arXiv:2411.07969 [hep-lat]}}.

\bibitem{ExtendedTwistedMass:2024nyi}
{\bfseries ETM} Collaboration, C.~Alexandrou {\em et~al.}, ``{Strange and charm
  quark contributions to the muon anomalous magnetic moment in lattice QCD with
  twisted-mass fermions},''
  \href{http://dx.doi.org/10.1103/PhysRevD.111.054502}{{\em Phys. Rev. D}
  {\bfseries 111} no.~5, (2025) 054502},
  \href{http://arxiv.org/abs/2411.08852}{{\ttfamily arXiv:2411.08852
  [hep-lat]}}.

\bibitem{MILC:2024ryz}
{\bfseries Fermilab Lattice, HPQCD, MILC} Collaboration, A.~Bazavov {\em
  et~al.}, ``{Hadronic vacuum polarization for the muon g-2 from lattice QCD:
  Complete short and intermediate windows},''
  \href{http://dx.doi.org/10.1103/PhysRevD.111.094508}{{\em Phys. Rev. D}
  {\bfseries 111} no.~9, (2025) 094508},
  \href{http://arxiv.org/abs/2411.09656}{{\ttfamily arXiv:2411.09656
  [hep-lat]}}.

\bibitem{FermilabLatticeHPQCD:2024ppc}
{\bfseries Fermilab Lattice, HPQCD, MILC} Collaboration, A.~Bazavov {\em
  et~al.}, ``{Hadronic Vacuum Polarization for the Muon g-2 from Lattice QCD:
  Long-Distance and Full Light-Quark Connected Contribution},''
  \href{http://dx.doi.org/10.1103/d583-yhfs}{{\em Phys. Rev. Lett.} {\bfseries
  135} no.~1, (2025) 011901}, \href{http://arxiv.org/abs/2412.18491}{{\ttfamily
  arXiv:2412.18491 [hep-lat]}}.

\bibitem{Keshavarzi:2019abf}
A.~Keshavarzi, D.~Nomura, and T.~Teubner, ``{$g-2$ of charged leptons, $\alpha
  (M^2_Z)$ , and the hyperfine splitting of muonium},''
  \href{http://dx.doi.org/10.1103/PhysRevD.101.014029}{{\em Phys. Rev. D}
  {\bfseries 101} no.~1, (2020) 014029},
  \href{http://arxiv.org/abs/1911.00367}{{\ttfamily arXiv:1911.00367
  [hep-ph]}}.

\bibitem{Kurz:2014wya}
A.~Kurz, T.~Liu, P.~Marquard, and M.~Steinhauser, ``{Hadronic contribution to
  the muon anomalous magnetic moment to next-to-next-to-leading order},''
  \href{http://dx.doi.org/10.1016/j.physletb.2014.05.043}{{\em Phys. Lett. B}
  {\bfseries 734} (2014) 144--147},
  \href{http://arxiv.org/abs/1403.6400}{{\ttfamily arXiv:1403.6400 [hep-ph]}}.

\bibitem{Colangelo:2015ama}
G.~Colangelo, M.~Hoferichter, M.~Procura, and P.~Stoffer, ``{Dispersion
  relation for hadronic light-by-light scattering: theoretical foundations},''
  \href{http://dx.doi.org/10.1007/JHEP09(2015)074}{{\em JHEP} {\bfseries 09}
  (2015) 074}, \href{http://arxiv.org/abs/1506.01386}{{\ttfamily
  arXiv:1506.01386 [hep-ph]}}.

\bibitem{Masjuan:2017tvw}
P.~Masjuan and P.~S{\'a}nchez-Puertas, ``{Pseudoscalar-pole contribution to the
  $(g_{\mu}-2)$: a rational approach},''
  \href{http://dx.doi.org/10.1103/PhysRevD.95.054026}{{\em Phys. Rev. D}
  {\bfseries 95} no.~5, (2017) 054026},
  \href{http://arxiv.org/abs/1701.05829}{{\ttfamily arXiv:1701.05829
  [hep-ph]}}.

\bibitem{Colangelo:2017qdm}
G.~Colangelo, M.~Hoferichter, M.~Procura, and P.~Stoffer, ``{Rescattering
  effects in the hadronic-light-by-light contribution to the anomalous magnetic
  moment of the muon},''
  \href{http://dx.doi.org/10.1103/PhysRevLett.118.232001}{{\em Phys. Rev.
  Lett.} {\bfseries 118} no.~23, (2017) 232001},
  \href{http://arxiv.org/abs/1701.06554}{{\ttfamily arXiv:1701.06554
  [hep-ph]}}.

\bibitem{Colangelo:2017fiz}
G.~Colangelo, M.~Hoferichter, M.~Procura, and P.~Stoffer, ``{Dispersion
  relation for hadronic light-by-light scattering: two-pion contributions},''
  \href{http://dx.doi.org/10.1007/JHEP04(2017)161}{{\em JHEP} {\bfseries 04}
  (2017) 161}, \href{http://arxiv.org/abs/1702.07347}{{\ttfamily
  arXiv:1702.07347 [hep-ph]}}.

\bibitem{Hoferichter:2018dmo}
M.~Hoferichter, B.-L. Hoid, B.~Kubis, S.~Leupold, and S.~P. Schneider,
  ``{Pion-pole contribution to hadronic light-by-light scattering in the
  anomalous magnetic moment of the muon},''
  \href{http://dx.doi.org/10.1103/PhysRevLett.121.112002}{{\em Phys. Rev.
  Lett.} {\bfseries 121} no.~11, (2018) 112002},
  \href{http://arxiv.org/abs/1805.01471}{{\ttfamily arXiv:1805.01471
  [hep-ph]}}.

\bibitem{Hoferichter:2018kwz}
M.~Hoferichter, B.-L. Hoid, B.~Kubis, S.~Leupold, and S.~P. Schneider,
  ``{Dispersion relation for hadronic light-by-light scattering: pion pole},''
  \href{http://dx.doi.org/10.1007/JHEP10(2018)141}{{\em JHEP} {\bfseries 10}
  (2018) 141}, \href{http://arxiv.org/abs/1808.04823}{{\ttfamily
  arXiv:1808.04823 [hep-ph]}}.

\bibitem{Eichmann:2019tjk}
G.~Eichmann, C.~S. Fischer, E.~Weil, and R.~Williams, ``{Single pseudoscalar
  meson pole and pion box contributions to the anomalous magnetic moment of the
  muon},'' \href{http://dx.doi.org/10.1016/j.physletb.2019.134855}{{\em Phys.
  Lett. B} {\bfseries 797} (2019) 134855},
  \href{http://arxiv.org/abs/1903.10844}{{\ttfamily arXiv:1903.10844
  [hep-ph]}}. [Erratum: Phys. Lett. B {\bf 799}, 135029 (2019)].

\bibitem{Bijnens:2019ghy}
J.~Bijnens, N.~Hermansson-Truedsson, and A.~Rodr{\'\i}guez-S{\'a}nchez,
  ``{Short-distance constraints for the HLbL contribution to the muon anomalous
  magnetic moment},''
  \href{http://dx.doi.org/10.1016/j.physletb.2019.134994}{{\em Phys. Lett. B}
  {\bfseries 798} (2019) 134994},
  \href{http://arxiv.org/abs/1908.03331}{{\ttfamily arXiv:1908.03331
  [hep-ph]}}.

\bibitem{Leutgeb:2019gbz}
J.~Leutgeb and A.~Rebhan, ``{Axial vector transition form factors in
  holographic QCD and their contribution to the anomalous magnetic moment of
  the muon},'' \href{http://dx.doi.org/10.1103/PhysRevD.101.114015}{{\em Phys.
  Rev. D} {\bfseries 101} no.~11, (2020) 114015},
  \href{http://arxiv.org/abs/1912.01596}{{\ttfamily arXiv:1912.01596
  [hep-ph]}}.

\bibitem{Cappiello:2019hwh}
L.~Cappiello, O.~Cat{\`a}, G.~D'Ambrosio, D.~Greynat, and A.~Iyer,
  ``{Axial-vector and pseudoscalar mesons in the hadronic light-by-light
  contribution to the muon $(g-2)$},''
  \href{http://dx.doi.org/10.1103/PhysRevD.102.016009}{{\em Phys. Rev. D}
  {\bfseries 102} no.~1, (2020) 016009},
  \href{http://arxiv.org/abs/1912.02779}{{\ttfamily arXiv:1912.02779
  [hep-ph]}}.

\bibitem{Masjuan:2020jsf}
P.~Masjuan, P.~Roig, and P.~S{\'a}nchez-Puertas, ``{The interplay of transverse
  degrees of freedom and axial-vector mesons with short-distance constraints in
  g {\ensuremath{-}} 2},''
  \href{http://dx.doi.org/10.1088/1361-6471/ac3892}{{\em J. Phys. G} {\bfseries
  49} no.~1, (2022) 015002}, \href{http://arxiv.org/abs/2005.11761}{{\ttfamily
  arXiv:2005.11761 [hep-ph]}}.

\bibitem{Bijnens:2020xnl}
J.~Bijnens, N.~Hermansson-Truedsson, L.~Laub, and
  A.~Rodr{\'\i}guez-S{\'a}nchez, ``{Short-distance HLbL contributions to the
  muon anomalous magnetic moment beyond perturbation theory},''
  \href{http://dx.doi.org/10.1007/JHEP10(2020)203}{{\em JHEP} {\bfseries 10}
  (2020) 203}, \href{http://arxiv.org/abs/2008.13487}{{\ttfamily
  arXiv:2008.13487 [hep-ph]}}.

\bibitem{Bijnens:2021jqo}
J.~Bijnens, N.~Hermansson-Truedsson, L.~Laub, and
  A.~Rodr{\'\i}guez-S{\'a}nchez, ``{The two-loop perturbative correction to the
  $(g - 2)_\mu$ HLbL at short distances},''
  \href{http://dx.doi.org/10.1007/JHEP04(2021)240}{{\em JHEP} {\bfseries 04}
  (2021) 240}, \href{http://arxiv.org/abs/2101.09169}{{\ttfamily
  arXiv:2101.09169 [hep-ph]}}.

\bibitem{Danilkin:2021icn}
I.~Danilkin, M.~Hoferichter, and P.~Stoffer, ``{A dispersive estimate of scalar
  contributions to hadronic light-by-light scattering},''
  \href{http://dx.doi.org/10.1016/j.physletb.2021.136502}{{\em Phys. Lett. B}
  {\bfseries 820} (2021) 136502},
  \href{http://arxiv.org/abs/2105.01666}{{\ttfamily arXiv:2105.01666
  [hep-ph]}}.

\bibitem{Stamen:2022uqh}
D.~Stamen, D.~Hariharan, M.~Hoferichter, B.~Kubis, and P.~Stoffer, ``{Kaon
  electromagnetic form factors in dispersion theory},''
  \href{http://dx.doi.org/10.1140/epjc/s10052-022-10348-3}{{\em Eur. Phys. J.
  C} {\bfseries 82} no.~5, (2022) 432},
  \href{http://arxiv.org/abs/2202.11106}{{\ttfamily arXiv:2202.11106
  [hep-ph]}}.

\bibitem{Leutgeb:2022lqw}
J.~Leutgeb, J.~Mager, and A.~Rebhan, ``{Hadronic light-by-light contribution to
  the muon g-2 from holographic QCD with solved U(1)A problem},''
  \href{http://dx.doi.org/10.1103/PhysRevD.107.054021}{{\em Phys. Rev. D}
  {\bfseries 107} no.~5, (2023) 054021},
  \href{http://arxiv.org/abs/2211.16562}{{\ttfamily arXiv:2211.16562
  [hep-ph]}}.

\bibitem{Hoferichter:2023tgp}
M.~Hoferichter, B.~Kubis, and M.~Zanke, ``{Axial-vector transition form factors
  and $e^+ e^- \to f_1 \pi^+ \pi^-$},''
  \href{http://dx.doi.org/10.1007/JHEP08(2023)209}{{\em JHEP} {\bfseries 08}
  (2023) 209}, \href{http://arxiv.org/abs/2307.14413}{{\ttfamily
  arXiv:2307.14413 [hep-ph]}}.

\bibitem{Hoferichter:2024fsj}
M.~Hoferichter, P.~Stoffer, and M.~Zillinger, ``{An optimized basis for
  hadronic light-by-light scattering},''
  \href{http://dx.doi.org/10.1007/JHEP04(2024)092}{{\em JHEP} {\bfseries 04}
  (2024) 092}, \href{http://arxiv.org/abs/2402.14060}{{\ttfamily
  arXiv:2402.14060 [hep-ph]}}.

\bibitem{Estrada:2024cfy}
E.~J. Estrada, S.~Gonz{\`a}lez-Sol{\'\i}s, A.~Guevara, and P.~Roig, ``{Improved
  {\ensuremath{\pi}}$^{0}$, {\ensuremath{\eta}}, {\ensuremath{\eta}}'
  transition form factors in resonance chiral theory and their $
  {a}_{\mu}^{\textrm{HLbL}} $ contribution},''
  \href{http://dx.doi.org/10.1007/JHEP12(2024)203}{{\em JHEP} {\bfseries 12}
  (2024) 203}, \href{http://arxiv.org/abs/2409.10503}{{\ttfamily
  arXiv:2409.10503 [hep-ph]}}.

\bibitem{Deineka:2024mzt}
O.~Deineka, I.~Danilkin, and M.~Vanderhaeghen, ``{Dispersive estimate of the
  a0(980) contribution to (g-2){\ensuremath{\mu}}},''
  \href{http://dx.doi.org/10.1103/PhysRevD.111.034009}{{\em Phys. Rev. D}
  {\bfseries 111} no.~3, (2025) 034009},
  \href{http://arxiv.org/abs/2410.12894}{{\ttfamily arXiv:2410.12894
  [hep-ph]}}.

\bibitem{Eichmann:2024glq}
G.~Eichmann, C.~S. Fischer, T.~Haeuser, and O.~Regenfelder, ``{Axial-vector and
  scalar contributions to hadronic light-by-light scattering},''
  \href{http://dx.doi.org/10.1140/epjc/s10052-025-14055-7}{{\em Eur. Phys. J.
  C} {\bfseries 85} no.~4, (2025) 445},
  \href{http://arxiv.org/abs/2411.05652}{{\ttfamily arXiv:2411.05652
  [hep-ph]}}.

\bibitem{Bijnens:2024jgh}
J.~Bijnens, N.~Hermansson-Truedsson, and A.~Rodr{\'\i}guez-S{\'a}nchez,
  ``{Constraints on the hadronic light-by-light tensor in corner kinematics for
  the muon g {\ensuremath{-}} 2},''
  \href{http://dx.doi.org/10.1007/JHEP03(2025)094}{{\em JHEP} {\bfseries 03}
  (2025) 094}, \href{http://arxiv.org/abs/2411.09578}{{\ttfamily
  arXiv:2411.09578 [hep-ph]}}.

\bibitem{Hoferichter:2024vbu}
M.~Hoferichter, P.~Stoffer, and M.~Zillinger, ``{Complete Dispersive Evaluation
  of the Hadronic Light-by-Light Contribution to Muon g-2},''
  \href{http://dx.doi.org/10.1103/PhysRevLett.134.061902}{{\em Phys. Rev.
  Lett.} {\bfseries 134} no.~6, (2025) 061902},
  \href{http://arxiv.org/abs/2412.00190}{{\ttfamily arXiv:2412.00190
  [hep-ph]}}.

\bibitem{Hoferichter:2024bae}
M.~Hoferichter, P.~Stoffer, and M.~Zillinger, ``{Dispersion relation for
  hadronic light-by-light scattering: subleading contributions},''
  \href{http://dx.doi.org/10.1007/JHEP02(2025)121}{{\em JHEP} {\bfseries 02}
  (2025) 121}, \href{http://arxiv.org/abs/2412.00178}{{\ttfamily
  arXiv:2412.00178 [hep-ph]}}.

\bibitem{Holz:2024lom}
S.~Holz, M.~Hoferichter, B.-L. Hoid, and B.~Kubis, ``{Precision Evaluation of
  the {\ensuremath{\eta}}- and {\ensuremath{\eta}}'-Pole Contributions to
  Hadronic Light-by-Light Scattering in the Anomalous Magnetic Moment of the
  Muon},'' \href{http://dx.doi.org/10.1103/PhysRevLett.134.171902}{{\em Phys.
  Rev. Lett.} {\bfseries 134} no.~17, (2025) 171902},
  \href{http://arxiv.org/abs/2411.08098}{{\ttfamily arXiv:2411.08098
  [hep-ph]}}.

\bibitem{Holz:2024diw}
S.~Holz, M.~Hoferichter, B.-L. Hoid, and B.~Kubis, ``{Dispersion relation for
  hadronic light-by-light scattering: $\eta$ and $\eta'$ poles},''
  \href{http://dx.doi.org/10.1007/JHEP04(2025)147}{{\em JHEP} {\bfseries 04}
  (2025) 147}, \href{http://arxiv.org/abs/2412.16281}{{\ttfamily
  arXiv:2412.16281 [hep-ph]}}.

\bibitem{Cappiello:2025fyf}
L.~Cappiello, J.~Leutgeb, J.~Mager, and A.~Rebhan, ``{Tensor meson transition
  form factors in holographic QCD and the muon g {\ensuremath{-}} 2},''
  \href{http://dx.doi.org/10.1007/JHEP07(2025)033}{{\em JHEP} {\bfseries 07}
  (2025) 033}, \href{http://arxiv.org/abs/2501.09699}{{\ttfamily
  arXiv:2501.09699 [hep-ph]}}.

\bibitem{Colangelo:2014qya}
G.~Colangelo, M.~Hoferichter, A.~Nyffeler, M.~Passera, and P.~Stoffer,
  ``{Remarks on higher-order hadronic corrections to the muon
  g{\ensuremath{-}}2},''
  \href{http://dx.doi.org/10.1016/j.physletb.2014.06.012}{{\em Phys. Lett. B}
  {\bfseries 735} (2014) 90--91},
  \href{http://arxiv.org/abs/1403.7512}{{\ttfamily arXiv:1403.7512 [hep-ph]}}.

\bibitem{Blum:2019ugy}
T.~Blum, N.~Christ, M.~Hayakawa, T.~Izubuchi, L.~Jin, C.~Jung, and C.~Lehner,
  ``{Hadronic Light-by-Light Scattering Contribution to the Muon Anomalous
  Magnetic Moment from Lattice QCD},''
  \href{http://dx.doi.org/10.1103/PhysRevLett.124.132002}{{\em Phys. Rev.
  Lett.} {\bfseries 124} no.~13, (2020) 132002},
  \href{http://arxiv.org/abs/1911.08123}{{\ttfamily arXiv:1911.08123
  [hep-lat]}}.

\bibitem{Chao:2021tvp}
E.-H. Chao, R.~J. Hudspith, A.~G{\'e}rardin, J.~R. Green, H.~B. Meyer, and
  K.~Ottnad, ``{Hadronic light-by-light contribution to $(g-2)_\mu $ from
  lattice QCD: a complete calculation},''
  \href{http://dx.doi.org/10.1140/epjc/s10052-021-09455-4}{{\em Eur. Phys. J.
  C} {\bfseries 81} no.~7, (2021) 651},
  \href{http://arxiv.org/abs/2104.02632}{{\ttfamily arXiv:2104.02632
  [hep-lat]}}.

\bibitem{Chao:2022xzg}
E.-H. Chao, R.~J. Hudspith, A.~G{\'e}rardin, J.~R. Green, and H.~B. Meyer,
  ``{The charm-quark contribution to light-by-light scattering in the muon
  $(g-2)$ from lattice QCD},''
  \href{http://dx.doi.org/10.1140/epjc/s10052-022-10589-2}{{\em Eur. Phys. J.
  C} {\bfseries 82} no.~8, (2022) 664},
  \href{http://arxiv.org/abs/2204.08844}{{\ttfamily arXiv:2204.08844
  [hep-lat]}}.

\bibitem{Blum:2023vlm}
{\bfseries RBC, UKQCD} Collaboration, T.~Blum, N.~Christ, M.~Hayakawa,
  T.~Izubuchi, L.~Jin, C.~Jung, C.~Lehner, and C.~Tu, ``{Hadronic
  light-by-light contribution to the muon anomaly from lattice QCD with
  infinite volume QED at physical pion mass},''
  \href{http://dx.doi.org/10.1103/PhysRevD.111.014501}{{\em Phys. Rev. D}
  {\bfseries 111} no.~1, (2025) 014501},
  \href{http://arxiv.org/abs/2304.04423}{{\ttfamily arXiv:2304.04423
  [hep-lat]}}.

\bibitem{Fodor:2024jyn}
Z.~Fodor, A.~Gerardin, L.~Lellouch, K.~K. Szabo, B.~C. Toth, and C.~Zimmermann,
  ``{Hadronic light-by-light scattering contribution to the anomalous magnetic
  moment of the muon at the physical pion mass},''
  \href{http://dx.doi.org/10.1103/wdrk-7nrt}{{\em Phys. Rev. D} {\bfseries 111}
  no.~11, (2025) 114509}, \href{http://arxiv.org/abs/2411.11719}{{\ttfamily
  arXiv:2411.11719 [hep-lat]}}.

\bibitem{Hertzog:2025ssc}
D.~W. Hertzog and M.~Hoferichter, ``{The anomalous magnetic moment of the muon:
  status and perspectives},''
  \href{http://dx.doi.org/10.1146/annurev-nucl-102422-040841}{{\em Ann. Rev.
  Nucl. Part. Sci.} {\bfseries 76} (2026) 343},
  \href{http://arxiv.org/abs/2512.16980}{{\ttfamily arXiv:2512.16980
  [hep-ph]}}.

\bibitem{Giudice:2012ms}
G.~F. Giudice, P.~Paradisi, and M.~Passera, ``{Testing new physics with the
  electron g-2},'' \href{http://dx.doi.org/10.1007/JHEP11(2012)113}{{\em JHEP}
  {\bfseries 11} (2012) 113}, \href{http://arxiv.org/abs/1208.6583}{{\ttfamily
  arXiv:1208.6583 [hep-ph]}}.

\bibitem{Crivellin:2021spu}
A.~Crivellin, M.~Hoferichter, and J.~M. Roney, ``{Toward testing the magnetic
  moment of the tau at one part per million},''
  \href{http://dx.doi.org/10.1103/PhysRevD.106.093007}{{\em Phys. Rev. D}
  {\bfseries 106} no.~9, (2022) 093007},
  \href{http://arxiv.org/abs/2111.10378}{{\ttfamily arXiv:2111.10378
  [hep-ph]}}.

\bibitem{Athron:2025ets}
P.~Athron, K.~M{\"o}hling, D.~St{\"o}ckinger, and H.~St{\"o}ckinger-Kim, ``{The
  muon magnetic moment and physics beyond the standard model},''
  \href{http://dx.doi.org/10.1016/j.ppnp.2025.104225}{{\em Prog. Part. Nucl.
  Phys.} {\bfseries 148} (2026) 104225},
  \href{http://arxiv.org/abs/2507.09289}{{\ttfamily arXiv:2507.09289
  [hep-ph]}}.

\bibitem{Barbieri:1995uv}
R.~Barbieri, G.~R. Dvali, and L.~J. Hall, ``{Predictions from a U(2) flavor
  symmetry in supersymmetric theories},''
  \href{http://dx.doi.org/10.1016/0370-2693(96)00318-8}{{\em Phys. Lett. B}
  {\bfseries 377} (1996) 76--82},
  \href{http://arxiv.org/abs/hep-ph/9512388}{{\ttfamily arXiv:hep-ph/9512388}}.

\bibitem{Barbieri:1996ae}
R.~Barbieri and L.~J. Hall, ``{A Grand unified supersymmetric theory of
  flavor},'' {\em Nuovo Cim. A} {\bfseries 110} (1997) 1--30,
  \href{http://arxiv.org/abs/hep-ph/9605224}{{\ttfamily arXiv:hep-ph/9605224}}.

\bibitem{Barbieri:2012uh}
R.~Barbieri, D.~Buttazzo, F.~Sala, and D.~M. Straub, ``{Flavour physics from an
  approximate $U(2)^3$ symmetry},''
  \href{http://dx.doi.org/10.1007/JHEP07(2012)181}{{\em JHEP} {\bfseries 07}
  (2012) 181}, \href{http://arxiv.org/abs/1203.4218}{{\ttfamily arXiv:1203.4218
  [hep-ph]}}.

\bibitem{Matsedonskyi:2014iha}
O.~Matsedonskyi, ``{On Flavour and Naturalness of Composite Higgs Models},''
  \href{http://dx.doi.org/10.1007/JHEP02(2015)154}{{\em JHEP} {\bfseries 02}
  (2015) 154}, \href{http://arxiv.org/abs/1411.4638}{{\ttfamily arXiv:1411.4638
  [hep-ph]}}.

\bibitem{Panico:2016ull}
G.~Panico and A.~Pomarol, ``{Flavor hierarchies from dynamical scales},''
  \href{http://dx.doi.org/10.1007/JHEP07(2016)097}{{\em JHEP} {\bfseries 07}
  (2016) 097}, \href{http://arxiv.org/abs/1603.06609}{{\ttfamily
  arXiv:1603.06609 [hep-ph]}}.

\bibitem{Glioti:2024hye}
A.~Glioti, R.~Rattazzi, L.~Ricci, and L.~Vecchi, ``{Exploring the flavor
  symmetry landscape},''
  \href{http://dx.doi.org/10.21468/SciPostPhys.18.6.201}{{\em SciPost Phys.}
  {\bfseries 18} no.~6, (2025) 201},
  \href{http://arxiv.org/abs/2402.09503}{{\ttfamily arXiv:2402.09503
  [hep-ph]}}.

\bibitem{Froggatt:1978nt}
C.~D. Froggatt and H.~B. Nielsen, ``{Hierarchy of Quark Masses, Cabibbo Angles
  and CP Violation},''
  \href{http://dx.doi.org/10.1016/0550-3213(79)90316-X}{{\em Nucl. Phys. B}
  {\bfseries 147} (1979) 277--298}.

\bibitem{Berezhiani:1983hm}
Z.~G. Berezhiani, ``{The Weak Mixing Angles in Gauge Models with Horizontal
  Symmetry: A New Approach to Quark and Lepton Masses},''
  \href{http://dx.doi.org/10.1016/0370-2693(83)90737-2}{{\em Phys. Lett. B}
  {\bfseries 129} (1983) 99--102}.

\bibitem{Bordone:2017bld}
M.~Bordone, C.~Cornella, J.~Fuentes-Martin, and G.~Isidori, ``{A three-site
  gauge model for flavor hierarchies and flavor anomalies},''
  \href{http://dx.doi.org/10.1016/j.physletb.2018.02.011}{{\em Phys. Lett. B}
  {\bfseries 779} (2018) 317--323},
  \href{http://arxiv.org/abs/1712.01368}{{\ttfamily arXiv:1712.01368
  [hep-ph]}}.

\bibitem{Fuentes-Martin:2022xnb}
J.~Fuentes-Martin, G.~Isidori, J.~M. Lizana, N.~Selimovic, and B.~A. Stefanek,
  ``{Flavor hierarchies, flavor anomalies, and Higgs mass from a warped extra
  dimension},'' \href{http://dx.doi.org/10.1016/j.physletb.2022.137382}{{\em
  Phys. Lett. B} {\bfseries 834} (2022) 137382},
  \href{http://arxiv.org/abs/2203.01952}{{\ttfamily arXiv:2203.01952
  [hep-ph]}}.

\bibitem{Davighi:2023iks}
J.~Davighi and G.~Isidori, ``{Non-universal gauge interactions addressing the
  inescapable link between Higgs and flavour},''
  \href{http://dx.doi.org/10.1007/JHEP07(2023)147}{{\em JHEP} {\bfseries 07}
  (2023) 147}, \href{http://arxiv.org/abs/2303.01520}{{\ttfamily
  arXiv:2303.01520 [hep-ph]}}.

\bibitem{Nakai:2025dmp}
Y.~Nakai, Y.~Shigekami, P.~Sun, and Z.~Zhang, ``{Models for the electric dipole
  moment and anomalous magnetic moment of the tau lepton},''
  \href{http://dx.doi.org/10.1103/rnrt-9gjy}{{\em Phys. Rev. D} {\bfseries 113}
  no.~3, (2026) 035001}, \href{http://arxiv.org/abs/2508.05935}{{\ttfamily
  arXiv:2508.05935 [hep-ph]}}.

\bibitem{Eidelman:2007sb}
S.~Eidelman and M.~Passera, ``{Theory of the tau lepton anomalous magnetic
  moment},'' \href{http://dx.doi.org/10.1142/S0217732307022694}{{\em Mod. Phys.
  Lett. A} {\bfseries 22} (2007) 159--179},
  \href{http://arxiv.org/abs/hep-ph/0701260}{{\ttfamily arXiv:hep-ph/0701260}}.

\bibitem{Eidelman:2016aih}
S.~Eidelman, D.~Epifanov, M.~Fael, L.~Mercolli, and M.~Passera, ``{$\tau$
  dipole moments via radiative leptonic $\tau$ decays},''
  \href{http://dx.doi.org/10.1007/JHEP03(2016)140}{{\em JHEP} {\bfseries 03}
  (2016) 140}, \href{http://arxiv.org/abs/1601.07987}{{\ttfamily
  arXiv:1601.07987 [hep-ph]}}.

\bibitem{Wittig:2025azm}
H.~Wittig, ``{Lepton anomalous magnetic moments: Theory},''
  \href{http://arxiv.org/abs/2512.18382}{{\ttfamily arXiv:2512.18382
  [hep-ph]}}.

\bibitem{Buchmuller:1985jz}
W.~Buchmuller and D.~Wyler, ``{Effective Lagrangian Analysis of New
  Interactions and Flavor Conservation},''
  \href{http://dx.doi.org/10.1016/0550-3213(86)90262-2}{{\em Nucl. Phys. B}
  {\bfseries 268} (1986) 621--653}.

\bibitem{CMS:2024qjo}
{\bfseries CMS} Collaboration, A.~Hayrapetyan {\em et~al.}, ``{Observation of
  $\gamma\gamma\to\tau\tau$ in proton-proton collisions and limits on the
  anomalous electromagnetic moments of the $\tau$ lepton},''
  \href{http://dx.doi.org/10.1088/1361-6633/ad6fcb}{{\em Rept. Prog. Phys.}
  {\bfseries 87} no.~10, (2024) 107801},
  \href{http://arxiv.org/abs/2406.03975}{{\ttfamily arXiv:2406.03975
  [hep-ex]}}.

\bibitem{ATLAS:2025oiy}
{\bfseries ATLAS} Collaboration, G.~Aad {\em et~al.}, ``{A measurement of the
  high-mass $ \tau \overline{\tau} $ production cross-section at $ \sqrt{s}=13
  $ TeV with the ATLAS detector and constraints on new particles and
  couplings},'' \href{http://dx.doi.org/10.1007/JHEP10(2025)054}{{\em JHEP}
  {\bfseries 10} (2025) 054}, \href{http://arxiv.org/abs/2503.19836}{{\ttfamily
  arXiv:2503.19836 [hep-ex]}}.

\bibitem{Belle:2021ybo}
{\bfseries Belle} Collaboration, K.~Inami {\em et~al.}, ``{An improved search
  for the electric dipole moment of the $\tau$ lepton},''
  \href{http://dx.doi.org/10.1007/JHEP04(2022)110}{{\em JHEP} {\bfseries 04}
  (2022) 110}, \href{http://arxiv.org/abs/2108.11543}{{\ttfamily
  arXiv:2108.11543 [hep-ex]}}.

\bibitem{Bernabeu:2006wf}
J.~Bernabeu, G.~A. Gonzalez-Sprinberg, and J.~Vidal, ``{CP violation and
  electric-dipole-moment at low energy tau production with polarized
  electrons},'' \href{http://dx.doi.org/10.1016/j.nuclphysb.2006.11.023}{{\em
  Nucl. Phys. B} {\bfseries 763} (2007) 283--292},
  \href{http://arxiv.org/abs/hep-ph/0610135}{{\ttfamily arXiv:hep-ph/0610135}}.

\bibitem{Bernabeu:2007rr}
J.~Bernabeu, G.~A. Gonzalez-Sprinberg, J.~Papavassiliou, and J.~Vidal, ``{Tau
  anomalous magnetic moment form-factor at super B/flavor factories},''
  \href{http://dx.doi.org/10.1016/j.nuclphysb.2007.09.001}{{\em Nucl. Phys. B}
  {\bfseries 790} (2008) 160--174},
  \href{http://arxiv.org/abs/0707.2496}{{\ttfamily arXiv:0707.2496 [hep-ph]}}.

\bibitem{Bernabeu:2008ii}
J.~Bernabeu, G.~A. Gonzalez-Sprinberg, and J.~Vidal, ``{Tau spin correlations
  and the anomalous magnetic moment},''
  \href{http://dx.doi.org/10.1088/1126-6708/2009/01/062}{{\em JHEP} {\bfseries
  01} (2009) 062}, \href{http://arxiv.org/abs/0807.2366}{{\ttfamily
  arXiv:0807.2366 [hep-ph]}}.

\bibitem{USBelleIIGroup:2022qro}
{\bfseries US Belle II Group, Belle II/SuperKEKB e- Polarization Upgrade
  Working Group} Collaboration, D.~M. Asner {\em et~al.}, ``{Snowmass 2021
  White Paper on Upgrading SuperKEKB with a Polarized Electron Beam: Discovery
  Potential and Proposed Implementation},'' in {\em {Snowmass 2021}}.
\newblock 5, 2022.
\newblock \href{http://arxiv.org/abs/2205.12847}{{\ttfamily arXiv:2205.12847
  [physics.acc-ph]}}.

\bibitem{Aihara:2024zds}
H.~Aihara {\em et~al.}, ``{The Belle II Detector Upgrades Framework Conceptual
  Design Report},'' 6, 2024.

\bibitem{Gogniat:2025eom}
J.~Gogniat, M.~Hoferichter, and Y.~Ulrich, ``{Towards testing (g
  {\ensuremath{-}} 2)$_{\tau}$ in e$^{+}$e$^{-}$ {\textrightarrow}
  {\ensuremath{\tau}}$^{+}${\ensuremath{\tau}}$^{-}$: radiative corrections and
  projections for Belle II},''
  \href{http://dx.doi.org/10.1007/JHEP07(2025)172}{{\em JHEP} {\bfseries 07}
  (2025) 172}, \href{http://arxiv.org/abs/2505.09678}{{\ttfamily
  arXiv:2505.09678 [hep-ph]}}.

\bibitem{Hoferichter:2025zjp}
M.~Hoferichter and G.~Levati, ``{Light new physics and the {\ensuremath{\tau}}
  lepton dipole moments},'' \href{http://dx.doi.org/10.1103/7rsy-974z}{{\em
  Phys. Rev. D} {\bfseries 113} no.~9, (2026) 095002},
  \href{http://arxiv.org/abs/2511.03786}{{\ttfamily arXiv:2511.03786
  [hep-ph]}}.

\bibitem{Hoferichter:2025ijh}
M.~Hoferichter and G.~Levati, ``{Light New Physics and the {\ensuremath{\tau}}
  Lepton Dipole Moments: Prospects at Belle II},''
  \href{http://dx.doi.org/10.1103/d3rt-69rd}{{\em Phys. Rev. Lett.} {\bfseries
  136} no.~18, (2026) 181804},
  \href{http://arxiv.org/abs/2510.13966}{{\ttfamily arXiv:2510.13966
  [hep-ph]}}.

\bibitem{Belle-II:2018jsg}
{\bfseries Belle II} Collaboration, W.~Altmannshofer {\em et~al.}, ``{The Belle
  II Physics Book},'' \href{http://dx.doi.org/10.1093/ptep/ptz106}{{\em PTEP}
  {\bfseries 2019} no.~12, (2019) 123C01},
  \href{http://arxiv.org/abs/1808.10567}{{\ttfamily arXiv:1808.10567
  [hep-ex]}}. [Erratum: PTEP {\bf 2020}, 029201 (2020)].

\bibitem{Gogniat:2026zvf}
J.~Gogniat, M.~Hoferichter, and G.~Levati, ``{Four-fermion operators, Z-boson
  exchange, and {\ensuremath{\tau}} lepton dipole moments},''
  \href{http://dx.doi.org/10.1016/j.physletb.2026.140786}{{\em Phys. Lett. B}
  {\bfseries 880} (2026) 140786},
  \href{http://arxiv.org/abs/2604.16598}{{\ttfamily arXiv:2604.16598
  [hep-ph]}}.

\bibitem{Buttazzo:2026amk}
D.~Buttazzo, G.~Levati, Y.~Ma, F.~Maltoni, P.~Paradisi, and Z.~Wang, ``{Probing
  $\tau$ lepton dipole moments at future lepton colliders},''
  \href{http://arxiv.org/abs/2604.14281}{{\ttfamily arXiv:2604.14281
  [hep-ph]}}.

\bibitem{Schwinger:1948iu}
J.~S. Schwinger, ``{On Quantum electrodynamics and the magnetic moment of the
  electron},'' \href{http://dx.doi.org/10.1103/PhysRev.73.416}{{\em Phys. Rev.}
  {\bfseries 73} (1948) 416--417}.

\bibitem{Bernabeu:1994wh}
J.~Bernabeu, G.~A. Gonzalez-Sprinberg, M.~Tung, and J.~Vidal, ``{The Tau weak
  magnetic dipole moment},''
  \href{http://dx.doi.org/10.1016/0550-3213(94)00525-J}{{\em Nucl. Phys. B}
  {\bfseries 436} (1995) 474--486},
  \href{http://arxiv.org/abs/hep-ph/9411289}{{\ttfamily arXiv:hep-ph/9411289}}.

\bibitem{Kosnik:2026cfq}
N.~Ko{\v{s}}nik, Z.~Polonsky, and A.~Smolkovi{\v{c}}, ``{Current and future
  constraints on heavy New Physics from $\tau$ weak dipole moments},''
  \href{http://arxiv.org/abs/2606.07232}{{\ttfamily arXiv:2606.07232
  [hep-ph]}}.

\bibitem{Hollik:1998wk}
W.~Hollik, J.~I. Illana, C.~Schappacher, D.~Stockinger, and S.~Rigolin,
  ``{Dipole form-factors and loop induced CP violation in supersymmetry},''
  \href{http://arxiv.org/abs/hep-ph/9808408}{{\ttfamily arXiv:hep-ph/9808408}}.

\bibitem{Yamaguchi:2020eub}
Y.~Yamaguchi and N.~Yamanaka, ``{Large long-distance contributions to the
  electric dipole moments of charged leptons in the standard model},''
  \href{http://dx.doi.org/10.1103/PhysRevLett.125.241802}{{\em Phys. Rev.
  Lett.} {\bfseries 125} (2020) 241802},
  \href{http://arxiv.org/abs/2003.08195}{{\ttfamily arXiv:2003.08195
  [hep-ph]}}.

\bibitem{Yamaguchi:2020dsy}
Y.~Yamaguchi and N.~Yamanaka, ``{Quark level and hadronic contributions to the
  electric dipole moment of charged leptons in the standard model},''
  \href{http://dx.doi.org/10.1103/PhysRevD.103.013001}{{\em Phys. Rev. D}
  {\bfseries 103} no.~1, (2021) 013001},
  \href{http://arxiv.org/abs/2006.00281}{{\ttfamily arXiv:2006.00281
  [hep-ph]}}.

\bibitem{L3:1998gov}
{\bfseries L3} Collaboration, M.~Acciarri {\em et~al.}, ``{Measurement of the
  anomalous magnetic and electric dipole moments of the tau lepton},''
  \href{http://dx.doi.org/10.1016/S0370-2693(98)00736-9}{{\em Phys. Lett. B}
  {\bfseries 434} (1998) 169--179}.

\bibitem{OPAL:1998dsa}
{\bfseries OPAL} Collaboration, K.~Ackerstaff {\em et~al.}, ``{An Upper limit
  on the anomalous magnetic moment of the tau lepton},''
  \href{http://dx.doi.org/10.1016/S0370-2693(98)00520-6}{{\em Phys. Lett. B}
  {\bfseries 431} (1998) 188--198},
  \href{http://arxiv.org/abs/hep-ex/9803020}{{\ttfamily arXiv:hep-ex/9803020}}.

\bibitem{DELPHI:2003nah}
{\bfseries DELPHI} Collaboration, J.~Abdallah {\em et~al.}, ``{Study of
  tau-pair production in photon-photon collisions at LEP and limits on the
  anomalous electromagnetic moments of the tau lepton},''
  \href{http://dx.doi.org/10.1140/epjc/s2004-01852-y}{{\em Eur. Phys. J. C}
  {\bfseries 35} (2004) 159--170},
  \href{http://arxiv.org/abs/hep-ex/0406010}{{\ttfamily arXiv:hep-ex/0406010}}.

\bibitem{ATLAS:2022ryk}
{\bfseries ATLAS} Collaboration, G.~Aad {\em et~al.}, ``{Observation of the
  {\ensuremath{\gamma}}{\ensuremath{\gamma}}{\textrightarrow}{\ensuremath{\tau}}{\ensuremath{\tau}}
  Process in Pb+Pb Collisions and Constraints on the {\ensuremath{\tau}}-Lepton
  Anomalous Magnetic Moment with the ATLAS Detector},''
  \href{http://dx.doi.org/10.1103/PhysRevLett.131.151802}{{\em Phys. Rev.
  Lett.} {\bfseries 131} no.~15, (2023) 151802},
  \href{http://arxiv.org/abs/2204.13478}{{\ttfamily arXiv:2204.13478
  [hep-ex]}}.

\bibitem{CMS:2022arf}
{\bfseries CMS} Collaboration, A.~Tumasyan {\em et~al.}, ``{Observation of
  $\tau$ lepton pair production in ultraperipheral lead-lead collisions at
  $\sqrt{s_\mathrm{NN}}$ = 5.02 TeV},''
  \href{http://dx.doi.org/10.1103/PhysRevLett.131.151803}{{\em Phys. Rev.
  Lett.} {\bfseries 131} (2023) 151803},
  \href{http://arxiv.org/abs/2206.05192}{{\ttfamily arXiv:2206.05192
  [nucl-ex]}}.

\bibitem{Grifols:1990ha}
J.~A. Grifols and A.~Mendez, ``{Electromagnetic properties of the tau lepton
  from Z0 decay},'' \href{http://dx.doi.org/10.1016/0370-2693(91)90276-V}{{\em
  Phys. Lett. B} {\bfseries 255} (1991) 611--612}. [Erratum: Phys.Lett.B 259,
  512 (1991)].

\bibitem{Allwicher:2021jkr}
L.~Allwicher, L.~Di~Luzio, M.~Fedele, F.~Mescia, and M.~Nardecchia, ``{What is
  the scale of new physics behind the muon g-2?},''
  \href{http://dx.doi.org/10.1103/PhysRevD.104.055035}{{\em Phys. Rev. D}
  {\bfseries 104} no.~5, (2021) 055035},
  \href{http://arxiv.org/abs/2105.13981}{{\ttfamily arXiv:2105.13981
  [hep-ph]}}.

\bibitem{Bresciani:2026acy}
L.~C. Bresciani, P.~Paradisi, and A.~Sainaghi, ``{Unitarity Bounds and Sum
  Rules in the Standard Model Effective Field Theory},''
  \href{http://dx.doi.org/10.1103/v5xg-dmss}{{\em Phys. Rev. Lett.} {\bfseries
  137} no.~15, (2026) 151801},
  \href{http://arxiv.org/abs/2603.03423}{{\ttfamily arXiv:2603.03423
  [hep-ph]}}.

\bibitem{Degrande:2011ua}
C.~Degrande, C.~Duhr, B.~Fuks, D.~Grellscheid, O.~Mattelaer, and T.~Reiter,
  ``{UFO - The Universal FeynRules Output},''
  \href{http://dx.doi.org/10.1016/j.cpc.2012.01.022}{{\em Comput. Phys.
  Commun.} {\bfseries 183} (2012) 1201--1214},
  \href{http://arxiv.org/abs/1108.2040}{{\ttfamily arXiv:1108.2040 [hep-ph]}}.

\bibitem{Ma:2026LEPTONdipole}
Y.~Ma, ``\textsc{LEPTONdipole}: Flavor-diagonal charged-lepton dipole operators
  in the {SMEFT} ({FeynRules} / {UFO} / {FeynArts}),'' 2026.
\newblock \url{https://doi.org/10.5281/zenodo.24182752}.

\bibitem{Brivio:2020onw}
I.~Brivio, ``{SMEFTsim 3.0 {\textemdash} a practical guide},''
  \href{http://dx.doi.org/10.1007/JHEP04(2021)073}{{\em JHEP} {\bfseries 04}
  (2021) 073}, \href{http://arxiv.org/abs/2012.11343}{{\ttfamily
  arXiv:2012.11343 [hep-ph]}}.

\bibitem{Jenkins:2013zja}
E.~E. Jenkins, A.~V. Manohar, and M.~Trott, ``{Renormalization Group Evolution
  of the Standard Model Dimension Six Operators I: Formalism and lambda
  Dependence},'' \href{http://dx.doi.org/10.1007/JHEP10(2013)087}{{\em JHEP}
  {\bfseries 10} (2013) 087}, \href{http://arxiv.org/abs/1308.2627}{{\ttfamily
  arXiv:1308.2627 [hep-ph]}}.

\bibitem{Jenkins:2013wua}
E.~E. Jenkins, A.~V. Manohar, and M.~Trott, ``{Renormalization Group Evolution
  of the Standard Model Dimension Six Operators II: Yukawa Dependence},''
  \href{http://dx.doi.org/10.1007/JHEP01(2014)035}{{\em JHEP} {\bfseries 01}
  (2014) 035}, \href{http://arxiv.org/abs/1310.4838}{{\ttfamily arXiv:1310.4838
  [hep-ph]}}.

\bibitem{Alonso:2013hga}
R.~Alonso, E.~E. Jenkins, A.~V. Manohar, and M.~Trott, ``{Renormalization Group
  Evolution of the Standard Model Dimension Six Operators III: Gauge Coupling
  Dependence and Phenomenology},''
  \href{http://dx.doi.org/10.1007/JHEP04(2014)159}{{\em JHEP} {\bfseries 04}
  (2014) 159}, \href{http://arxiv.org/abs/1312.2014}{{\ttfamily arXiv:1312.2014
  [hep-ph]}}.

\bibitem{Jenkins:2017jig}
E.~E. Jenkins, A.~V. Manohar, and P.~Stoffer, ``{Low-Energy Effective Field
  Theory below the Electroweak Scale: Operators and Matching},''
  \href{http://dx.doi.org/10.1007/JHEP03(2018)016}{{\em JHEP} {\bfseries 03}
  (2018) 016}, \href{http://arxiv.org/abs/1709.04486}{{\ttfamily
  arXiv:1709.04486 [hep-ph]}}. [Erratum: JHEP 12, 043 (2023)].

\bibitem{Jenkins:2017dyc}
E.~E. Jenkins, A.~V. Manohar, and P.~Stoffer, ``{Low-Energy Effective Field
  Theory below the Electroweak Scale: Anomalous Dimensions},''
  \href{http://dx.doi.org/10.1007/JHEP01(2018)084}{{\em JHEP} {\bfseries 01}
  (2018) 084}, \href{http://arxiv.org/abs/1711.05270}{{\ttfamily
  arXiv:1711.05270 [hep-ph]}}. [Erratum: JHEP {\bf 12}, 042 (2023)].

\bibitem{Bresciani:2025toe}
L.~C. Bresciani, G.~Levati, and P.~Paradisi, ``{Amplitudes and partial wave
  unitarity bounds},'' \href{http://arxiv.org/abs/2504.12855}{{\ttfamily
  arXiv:2504.12855 [hep-ph]}}.

\bibitem{FCC:2018evy}
{\bfseries FCC} Collaboration, A.~Abada {\em et~al.}, ``{FCC-ee: The Lepton
  Collider}: {Future Circular Collider Conceptual Design Report Volume 2},''
  \href{http://dx.doi.org/10.1140/epjst/e2019-900045-4}{{\em Eur. Phys. J. ST}
  {\bfseries 228} no.~2, (2019) 261--623}.

\bibitem{Bernardi:2022hny}
G.~Bernardi {\em et~al.}, ``{The Future Circular Collider: a Summary for the US
  2021 Snowmass Process},''  (3, 2022) ,
  \href{http://arxiv.org/abs/2203.06520}{{\ttfamily arXiv:2203.06520
  [hep-ex]}}.

\bibitem{FCC:2025lpp}
{\bfseries FCC} Collaboration, M.~Benedikt {\em et~al.}, ``{Future Circular
  Collider Feasibility Study Report: Volume 1, Physics, Experiments,
  Detectors},'' \href{http://dx.doi.org/10.1140/epjc/s10052-025-15077-x}{{\em
  Eur. Phys. J. C} {\bfseries 85} no.~12, (2025) 1468},
  \href{http://arxiv.org/abs/2505.00272}{{\ttfamily arXiv:2505.00272
  [hep-ex]}}.

\bibitem{Altmann:2025feg}
J.~Altmann {\em et~al.}, \href{http://dx.doi.org/10.23731/CYRM-2025-005}{{\em
  {ECFA Higgs, electroweak, and top Factory Study}}}, vol.~5/2025 of {\em CERN
  Yellow Reports: Monographs}.
\newblock 6, 2025.
\newblock \href{http://arxiv.org/abs/2506.15390}{{\ttfamily arXiv:2506.15390
  [hep-ex]}}.

\bibitem{CEPCStudyGroup:2018ghi}
{\bfseries CEPC Study Group} Collaboration, M.~Dong {\em et~al.}, ``{CEPC
  Conceptual Design Report: Volume 2 - Physics {\&} Detector},'' 11, 2018.

\bibitem{CEPCStudyGroup:2018rmc}
{\bfseries CEPC Study Group} Collaboration, ``{CEPC Conceptual Design Report:
  Volume 1 - Accelerator},'' 9, 2018.

\bibitem{An:2018dwb}
F.~An {\em et~al.}, ``{Precision Higgs physics at the CEPC},''
  \href{http://dx.doi.org/10.1088/1674-1137/43/4/043002}{{\em Chin. Phys. C}
  {\bfseries 43} no.~4, (2019) 043002},
  \href{http://arxiv.org/abs/1810.09037}{{\ttfamily arXiv:1810.09037
  [hep-ex]}}.

\bibitem{CEPCPhysicsStudyGroup:2022uwl}
{\bfseries CEPC Physics Study Group} Collaboration, H.~Cheng {\em et~al.},
  ``{The Physics potential of the CEPC. Prepared for the US Snowmass Community
  Planning Exercise (Snowmass 2021)},'' in {\em {Snowmass 2021}}.
\newblock 5, 2022.
\newblock \href{http://arxiv.org/abs/2205.08553}{{\ttfamily arXiv:2205.08553
  [hep-ph]}}.

\bibitem{CEPCStudyGroup:2023quu}
{\bfseries CEPC Study Group} Collaboration, W.~Abdallah {\em et~al.}, ``{CEPC
  Technical Design Report: Accelerator},''
  \href{http://dx.doi.org/10.1007/s41605-024-00463-y}{{\em Radiat. Detect.
  Technol. Methods} {\bfseries 8} no.~1, (2024) 1--1105},
  \href{http://arxiv.org/abs/2312.14363}{{\ttfamily arXiv:2312.14363
  [physics.acc-ph]}}. [Erratum: Radiat.Detect.Technol.Methods 9, 184--192
  (2025)].

\bibitem{CEPCStudyGroup:2025kmw}
{\bfseries CEPC Study Group} Collaboration, S.~P. Adhya {\em et~al.}, ``{CEPC
  Technical Design Report - Reference Detector},''
  \href{http://arxiv.org/abs/2510.05260}{{\ttfamily arXiv:2510.05260
  [hep-ex]}}.

\bibitem{janot_2024_yr3v6-dgh16}
P.~Janot, C.~Grojean, F.~Zimmermann, and M.~Benedikt, {\em Integrated
  Luminosities and Sequence of Events for the FCC Feasibility Study Report},
  Dec., 2024.
\newblock \url{https://doi.org/10.17181/yr3v6-dgh16}.

\bibitem{Alwall:2014hca}
J.~Alwall, R.~Frederix, S.~Frixione, V.~Hirschi, F.~Maltoni, O.~Mattelaer,
  H.~S. Shao, T.~Stelzer, P.~Torrielli, and M.~Zaro, ``{The automated
  computation of tree-level and next-to-leading order differential cross
  sections, and their matching to parton shower simulations},''
  \href{http://dx.doi.org/10.1007/JHEP07(2014)079}{{\em JHEP} {\bfseries 07}
  (2014) 079}, \href{http://arxiv.org/abs/1405.0301}{{\ttfamily arXiv:1405.0301
  [hep-ph]}}.

\bibitem{Frederix:2018nkq}
R.~Frederix, S.~Frixione, V.~Hirschi, D.~Pagani, H.~S. Shao, and M.~Zaro,
  ``{The automation of next-to-leading order electroweak calculations},''
  \href{http://dx.doi.org/10.1007/JHEP11(2021)085}{{\em JHEP} {\bfseries 07}
  (2018) 185}, \href{http://arxiv.org/abs/1804.10017}{{\ttfamily
  arXiv:1804.10017 [hep-ph]}}. [Erratum: JHEP 11, 085 (2021)].

\bibitem{Kilian:2007gr}
W.~Kilian, T.~Ohl, and J.~Reuter, ``{WHIZARD: Simulating Multi-Particle
  Processes at LHC and ILC},''
  \href{http://dx.doi.org/10.1140/epjc/s10052-011-1742-y}{{\em Eur. Phys. J. C}
  {\bfseries 71} (2011) 1742}, \href{http://arxiv.org/abs/0708.4233}{{\ttfamily
  arXiv:0708.4233 [hep-ph]}}.

\bibitem{ParticleDataGroup:2020ssz}
{\bfseries Particle Data Group} Collaboration, P.~A. Zyla {\em et~al.},
  ``{Review of Particle Physics},''
  \href{http://dx.doi.org/10.1093/ptep/ptaa104}{{\em PTEP} {\bfseries 2020}
  no.~8, (2020) 083C01}.

\bibitem{Cowan:2010js}
G.~Cowan, K.~Cranmer, E.~Gross, and O.~Vitells, ``{Asymptotic formulae for
  likelihood-based tests of new physics},''
  \href{http://dx.doi.org/10.1140/epjc/s10052-011-1554-0}{{\em Eur. Phys. J. C}
  {\bfseries 71} (2011) 1554}, \href{http://arxiv.org/abs/1007.1727}{{\ttfamily
  arXiv:1007.1727 [physics.data-an]}}. [Erratum: Eur.Phys.J.C 73, 2501 (2013)].

\bibitem{Du:2026ynu}
X.-Y. Du, Z.-Y. Zou, X.-G. He, and K.~Xie, ``{Transverse Tau Spin Correlations
  and the Weak Electric Dipole Moment at Future $Z$ Factories},''
  \href{http://arxiv.org/abs/2609.18892}{{\ttfamily arXiv:2609.18892
  [hep-ph]}}.

\bibitem{vonWeizsacker:1934nji}
C.~F. von Weizsacker, ``{Radiation emitted in collisions of very fast
  electrons},'' \href{http://dx.doi.org/10.1007/BF01333110}{{\em Z. Phys.}
  {\bfseries 88} (1934) 612--625}.

\bibitem{Williams:1934ad}
E.~J. Williams, ``{Nature of the high-energy particles of penetrating radiation
  and status of ionization and radiation formulae},''
  \href{http://dx.doi.org/10.1103/PhysRev.45.729}{{\em Phys. Rev.} {\bfseries
  45} (1934) 729--730}.

\bibitem{Frixione:1993yw}
S.~Frixione, M.~L. Mangano, P.~Nason, and G.~Ridolfi, ``{Improving the
  Weizsacker-Williams approximation in electron - proton collisions},''
  \href{http://dx.doi.org/10.1016/0370-2693(93)90823-Z}{{\em Phys. Lett. B}
  {\bfseries 319} (1993) 339--345},
  \href{http://arxiv.org/abs/hep-ph/9310350}{{\ttfamily arXiv:hep-ph/9310350}}.

\bibitem{Galon:2016ngp}
I.~Galon, A.~Rajaraman, and T.~M.~P. Tait, ``{$H \rightarrow \tau^+ \tau^-
  \gamma$ as a probe of the $\tau$ magnetic dipole moment},''
  \href{http://dx.doi.org/10.1007/JHEP12(2016)111}{{\em JHEP} {\bfseries 12}
  (2016) 111}, \href{http://arxiv.org/abs/1610.01601}{{\ttfamily
  arXiv:1610.01601 [hep-ph]}}.

\bibitem{Rajaraman:2018uyb}
A.~Rajaraman, J.~N. Howard, R.~Riley, and T.~M.~P. Tait, ``{The \(\tau\)
  Magnetic Dipole Moment at Future Lepton Colliders},''
  \href{http://dx.doi.org/10.31526/lhep.2.2019.113}{{\em LHEP} {\bfseries 2}
  no.~2, (2019) 5}, \href{http://arxiv.org/abs/1810.09570}{{\ttfamily
  arXiv:1810.09570 [hep-ph]}}.

\bibitem{Buttazzo:2020ibd}
D.~Buttazzo and P.~Paradisi, ``{Probing the muon $g-2$ anomaly with the Higgs
  boson at a muon collider},''
  \href{http://dx.doi.org/10.1103/PhysRevD.104.075021}{{\em Phys. Rev. D}
  {\bfseries 104} no.~7, (2021) 075021},
  \href{http://arxiv.org/abs/2012.02769}{{\ttfamily arXiv:2012.02769
  [hep-ph]}}.

\bibitem{Han:2017yhy}
T.~Han and X.~Wang, ``{Radiative Decays of the Higgs Boson to a Pair of
  Fermions},'' \href{http://dx.doi.org/10.1007/JHEP10(2017)036}{{\em JHEP}
  {\bfseries 10} (2017) 036}, \href{http://arxiv.org/abs/1704.00790}{{\ttfamily
  arXiv:1704.00790 [hep-ph]}}.

\bibitem{Buttazzo:2026jfu}
D.~Buttazzo, G.~Levati, Y.~Ma, F.~Maltoni, P.~Paradisi, and Z.~Wang, ``{Probing
  $\tau$ lepton dipole moments at multi-TeV muon colliders},''
  \href{http://arxiv.org/abs/2611.21813}{{\ttfamily arXiv:2611.21813
  [hep-ph]}}.

\end{thebibliography}\endgroup

\end{document}